\documentclass[aps,prb,preprint,superscriptaddress,nofootinbib,longbibliography,floatfix]{revtex4-2}

\usepackage[utf8]{inputenc}
\usepackage[T1]{fontenc}
\usepackage{amsmath,amssymb,amsthm,mathtools}
\usepackage{bm}
\usepackage{graphicx}
\usepackage{xcolor}
\usepackage{hyperref}
\hypersetup{colorlinks=true,linkcolor=blue!60!black,citecolor=green!50!black,urlcolor=blue!60!black}
\usepackage{cleveref}

\graphicspath{{figures/}}

\newcommand{\adag}{\hat a^\dagger}
\newcommand{\aop}{\hat a}
\newcommand{\emat}{e_{\mathrm{mat}}}
\newcommand{\evar}{\tilde e}
\newcommand{\xmat}{\chi_{\mathrm{mat}}}
\newcommand{\mumat}{\mu_{\mathrm{mat}}}
\newcommand{\ket}[1]{|#1\rangle}

\begin{document}

\makeatletter\renewcommand\footnotesize{\@setfontsize\footnotesize\@xpt{12}}\makeatother

\title{Folds of one curve: the superradiant phase diagram of Dicke models with interacting matter\\[5pt]
\texorpdfstring{\normalsize\mdseries Onset, the order of transitions, and the stationarity curve,\\
from the Dicke--Ising model in general dimension to other cavity-coupled magnets}{Onset, the order of transitions, and the stationarity curve, from the Dicke-Ising model in general dimension to other cavity-coupled magnets}}
\author{Max H\"ormann}
\affiliation{Independent Researcher, Erlangen, Germany\\[2pt]\normalfont\texttt{max\_hoermann@web.de}}

\date{\today}

\begin{abstract}
We give an analytic account, in the thermodynamic limit, of Dicke models in which one cavity mode
couples collectively to \emph{interacting} matter. Integrating out the cavity yields an exact
self-consistent functional, $\evar(m)=\tfrac\lambda2 m^2+\emat(\lambda m)$: a classical penalty
against the bare-matter energy $\emat$ in the self-consistent field $h=\lambda m$, with
$\lambda=g^2/2\omega_c$ the collective coupling and $m$ the magnetisation. The photon supplies only
that scalar field, so it can create no phase the matter does not already possess. The functional's
whole phase diagram is in how its minima appear, split, and jump as the coupling is tuned. The states
that can hold its minimum form a single curve $\lambda(m)=\mumat^{-1}(m)/m$, single-valued because the
bare magnetisation rises monotonically with field, and connected as long as the matter orders
continuously. Because the curve is one connected object, the superradiant first-order transitions are
\emph{folds} of a single equation of state rather than crossings of disjoint sheets, and a fold can
straighten into a continuous line.

The remaining rules are local, each with a spectral counterpart. The onset of superradiance is set by
the leading singularity of the bare susceptibility $\xmat$ and shows up as the softening of the normal-phase (effective-Dicke)
lower polariton; its threshold is finite for a regular response, vanishes where the response diverges
(an immediate onset), or is approached through an essential singularity. On a superradiant vacuum the same reading, worked
out exactly at infinite dimension (\cref{app:dinf_dicke}), makes the superradiant--superradiant
transition's lower polariton soften with vanishing photon weight. Symmetry-protected pitchforks
of twin $\pm m$ branches occur only on the $m=0$ axis. Every other transition is a first-order fold, of one of two
kinds: a shape fold from a subcritically peaked response, or a Larkin--Pikin fold forced by
a divergent susceptibility at the matter criticality. The order of each transition is fixed by one
bare response: the Landau quartic where the susceptibility stays finite, or the susceptibility itself
where the matter turns critical, its divergence forbidding a minimum on the curve and forcing the
first-order fold (anchored at $\lambda_*=h_c/m_c$). A single-humped susceptibility caps the number of
stationary states, so no hidden minimum undercuts the local verdict, save a global Maxwell jump that
can preempt it. On stable branches each fold is a collective mode going soft through a Thouless
condition, wherever the coupled operator is not conserved.

For the Dicke--Ising model in general dimension the Landau coefficients are exact, giving in closed
form the second-order boundary and both zero-quartic fields (in the rescaled units $J\to J/d$ under which mean field is exact at $d=\infty$): the
ferromagnetic $\varepsilon=|J|/d$, an ordinary tricritical point, and the antiferromagnetic
$\varepsilon=2|J|/\sqrt{8d-3}$, where the second-order line instead ends in a first-order point.\footnote{These results were obtained by the author in November~2025 during the preparation of Ref.~\cite{leibig2025} in the group of K.~P.~Schmidt; see~\cref{sec:results_tri} for the full record and the relation to the work of J.~Koziol~\cite{Koziol2026}.} That
endpoint is preempted by a direct transition in every dimension; in one dimension the
antiferromagnetic superradiant (AS) phase then survives only as a narrow wedge near the quadruple
point, where the polarised and antiferromagnetically ordered phases, normal and superradiant (PN, PS,
AN, AS), all meet. A $1/d$ expansion maps all four phase boundaries to this order in $1/d$, including the AS phase, which admits no perturbative
reference, with the AS--polarised-superradiant (PS) transition first order for $d\le3$ (at or below the matter's upper critical dimension), where
the divergent $(d{+}1)$-Ising susceptibility forces a Larkin--Pikin fold, and tricritical points in the
$(d,\varepsilon)$ plane above. At the quadruple point the matter reduces to a detuned Rydberg-blockade chain; along its
undetuned ray, where the cavity perturbs the blockade manifold directly, the selection is disorder by
disorder, into the polarised superradiant state. That the antiferromagnetic superradiant phase
persists right up to the corner is then confirmed directly by strict-blockade infinite-system DMRG: it occupies
a finite wedge in one dimension, and across the AS--PS transition the magnetisation jumps by a finite amount that does not vanish as the corner is approached.

Deliberately varied magnets populate the same classification and map out its reach. The
frustrated triangular antiferromagnet (order by disorder, superradiant at arbitrarily weak coupling)
keeps its superradiant--superradiant line \emph{continuous} at all resolved scales, in the
$3$D-XY class, because its negative specific-heat exponent leaves the susceptibility
finite at criticality and \emph{no fold} is forced. The bond-alternating compass chain sits in the marginal
class: an essential-singularity onset of Berezinskii--Kosterlitz--Thouless type (disorder by disorder), with a
threshold $\lambda_c$ that vanishes only logarithmically as the bond asymmetry does. The
gapless Heisenberg and XX chains sit in the regular class, with a first-order onset that sends the
magnetisation straight to saturation; coupling through a \emph{conserved} operator, their onset is
spectrally silent, with no polariton and the photon pinned at $\omega_c$. The XX stationarity curve is
elementary, $\lambda(m)=\sin(\pi m/2)/m$, while the Heisenberg chain's curve follows from the Bethe
ansatz, and a transverse field opens it into a further exactly solved diagram whose
first-order line dies at a tricritical point of an unusual kind on the bare saturation line.
\end{abstract}

\maketitle

\clearpage
\makeatletter
\let\l@subsubsection\@gobble@tw@
\newcommand\appendixtocfix{\let\l@subsection\@gobble@tw@}
\renewcommand\l@f@section{\addpenalty{\@secpenalty}\addvspace{0.2em plus\p@}}
\makeatother
\tableofcontents
\clearpage

\section{Introduction}

A single cavity mode collectively coupled to interacting matter is, in the thermodynamic limit,
exactly reducible. Integrating out the photon leaves the matter in one self-consistent field
$h=\lambda m$, with $\lambda=g^2/2\omega_c$ the collective coupling and $m=\langle\sigma^z\rangle$ the
magnetisation that sources it, so the entire zero-temperature problem becomes the minimisation of one
classical functional of $m$, the single energy density
\begin{equation}
\evar(m)=\tfrac{\lambda}{2}m^2+\emat(\lambda m),
\label{eq:functional_intro}
\end{equation}
a classical parabola against the bare-matter energy $\emat$, which carries all the quantum mechanics. We
take this reduction as the foundation we build on, not as the
contribution (\cref{sec:decoupling}). What it makes plain is a structural point: the cavity adds no
degrees of freedom and only a $c$-number field, so it can create no phase the matter does not already
possess, and superradiance (at zero temperature, in the closed system) is nothing but the matter polarised along the cavity axis. The same field
can, however, change the order of the matter's transitions and shift their location. What it changes,
and to what, is fixed by the bare matter response alone.

A first-order transition is usually pictured as a contest: two phases, each with its own energy, meet
along a line and the lower one wins, so the order parameter jumps between separate states. There is a second picture. The van der Waals isotherm describes a first-order transition as a fold of one
connected curve, with liquid and gas a single equation of state joined through an unstable middle
branch that the equilibrium jump cuts across. The cavity--matter problem has this second structure.
The convex hull of its energy keeps the stable phases and the jump between them; we follow the whole
stationarity curve instead, the locus, i.e.\ the complete set, of every self-consistent state including the unstable branch the
hull leaves out, on which the superradiant first-order transitions are folds of one curve. That curve
is the bare matter's own response read back at self-consistency, so where it folds, and whether a fold
encloses a first-order jump or unbends into a continuous line, is set by the interacting matter.

The curve is single-valued because the bare magnetisation rises monotonically with the field. That it
is also one connected piece is not automatic: it requires the bare matter to order continuously, and
the cleanest example is the Ising magnet at zero longitudinal field. On its own this matter undergoes the
most ordinary continuous transition, the second-order breaking of a $\mathbb{Z}_2$ symmetry.
Placed in the cavity it turns first order~\cite{lee2004,Gammelmark2011,Rohn2020}. Because the bare transition stays continuous the
magnetisation still varies smoothly, so the first-order transition is a fold of one connected curve and not a collision of two phases.
Had the bare matter carried a first-order transition of its own, its
magnetisation would jump, the curve would split into disconnected pieces, and the separate-sheets
picture would be the right one. The single connected curve is therefore what the cavity makes of any
correlated matter with a continuous transition, and it lets the onset, the order of each transition,
and the number of phases be read from one object. On it the symmetry-protected pitchforks of twin $\pm
m$ branches sit only on the $m=0$ axis, where superradiance sets in, and every other transition is a
fold.

This is concrete in a specific setting. The Dicke model~\cite{Dicke1954}, many emitters coupled to one
photon mode, is a touchstone of collective light--matter physics since the identification of its
superradiant phase transition~\cite{HeppLieb1973,HeppLieb1973PRA,WangHioe1973}; cavity- and circuit-QED
platforms engineer the collective coupling~\cite{Kirton2018,Fink2009}, and the Dicke superradiant
transition, proposed for cavity QED~\cite{Dimer2007}, was realised with a superfluid gas in an optical
cavity~\cite{Baumann2010,Baumann2011}. Interest then turned to interacting matter placed in a cavity,
the paradigmatic case being the Dicke--Ising model~\cite{lee2004,Gammelmark2011,Zhang2014,Rohn2020}: an
Ising magnet with a longitudinal field $\varepsilon$ along its ordering axis, coupled collectively to
one cavity mode [\cref{eq:H} below] whose transverse field is supplied self-consistently by the cavity
displacement~\cite{Rohn2020,roman2022effective}. Its ingredients are within reach of circuit QED, where
Ising-coupled superconducting qubits share a resonator~\cite{Zhang2014}, and digital-analog quantum
simulators of the full model have been proposed~\cite{Shapiro2025}. For antiferromagnetic coupling the
diagram is rich, with normal and superradiant variants of both the polarised and the
antiferromagnetically ordered states (PN, PS, AN, AS) meeting at a quadruple
point~\cite{Zhang2014,langheld,leibig2025}, while the ferromagnet has only the PN/PS pair; the
diamagnetic $A^2$ term and the associated no-go theorems~\cite{Rzazewski1975} we do not treat here; for their effect on these phase diagrams see Ref.~\cite{langheld}.

What is known of this
diagram was assembled over two decades, and three questions about it had no analytic account, though numerics had begun to address them: what fixes the order of each transition in general dimension, whether the AS--PS transition is
first or second order, and how far the antiferromagnetic superradiant phase persists in one dimension.

The prior results group naturally under these three questions. On the order of the transitions, Lee and
Johnson~\cite{lee2004} found, with temperature as the main axis, that matter interactions can drive the
superradiant transition first order, and Gammelmark and M{\o}lmer~\cite{Gammelmark2011} mapped the
finite-temperature mean-field diagram with a tricritical point; at zero temperature Zhang \emph{et
al.}~\cite{Zhang2014} mapped the antiferromagnetic model at mean-field level, exact at $d=\infty$, where
the AS phase was first seen (a superradiant phase coexisting with lattice order had been predicted for
Rydberg matter in Ref.~\cite{ZhangRydberg2013}), and Rohn \emph{et al.}~\cite{Rohn2020} treated the
$\varepsilon=0$ chain in a displaced frame and likewise found a first-order transition. Its precise
location, however, requires the polarised superradiant branch taken self-consistently rather than at saturation; we place it at $\lambda_M\approx1.673\,J$. On the order of the AS--PS transition, quantitative diagrams
in one and two dimensions came with wormhole quantum Monte Carlo~\cite{langheld} and, by linked-cluster
methods plus DMRG, with Ref.~\cite{leibig2025}; both find the AS phase, the AS--PS boundary first order in one
dimension and, on the square lattice, first order except in a small-coupling window where the wormhole
data show apparent $3$D-Ising criticality. That window we show below to be precluded by a structural
constraint: the bare matter is either $(d{+}1)$-Ising, whose divergent susceptibility forces the fold
(\cref{sec:lp}), or it is first order, which makes the cavity transition first order as well. Either
way the antiferromagnetic AS--PS transition is first order throughout $d=1,2,3$, the apparent
criticality the finite-size shadow of a numerically hard-to-resolve small jump. On its persistence in one dimension, a recent study~\cite{mendoncca2025} has
questioned whether the AS phase exists there at all, although the wormhole and DMRG maps~\cite{langheld,leibig2025}
already find it a finite distance from the quadruple-point corner; its fate at the corner itself remained open. Analytically, only the onset had been treated: at finite
temperature by mean field~\cite{Gammelmark2011} and at zero temperature by the effective-Dicke mapping of
Ref.~\cite{Schellenberger2024} and the mean-field diagram of Zhang \emph{et al.}~\cite{Zhang2014}; the
closed-form ferromagnetic onset and tricritical point of \cref{sec:results_tri} we add here, also obtained
by Koziol~\cite{Koziol2026} from a Landau analysis of the same
functional.\footnote{For the provenance of these closed-form criteria relative to
Ref.~\cite{Koziol2026}---they were obtained by the author in November~2025---see the title-page note
and \cref{sec:results_tri}.} What no prior work assembled is the analytic account of where the lines lie
and what fixes the order of each transition. That is what this paper supplies, settling by DMRG that the AS phase exists in one dimension, with a first-order AS--PS transition, up to the quadruple-point corner.

The reduction behind \cref{eq:functional_intro} has a layered history, from the rigorous finite-temperature
separable-interaction theorems of the
1970s~\cite{Bogoliubov1972book,TindemansCapel1975,denOuden1976a,denOuden1976b,PerkCapel1977}, through
the cavity functionals~\cite{lee2004,Gammelmark2011} of the Dicke--Ising literature and its displaced
frame~\cite{Rohn2020}, whose Dicke--Lipkin--Meshkov--Glick form is due to Reslen \emph{et al.}~\cite{Reslen2005}
and which Rohn \emph{et al.}\ carried over to interacting matter, to the general, model-independent form of Rom\'an-Roche \emph{et
al.}~\cite{roman2022effective,roman2025linear,roman2025bound}, whose functional we adopt; the lineage,
and the precise sense in which the decoupling is exact, are set out in \cref{sec:decoupling}. We work at
the level of the ground state, asking which self-consistent matter state holds the global minimum and
thereby tracing the equilibrium phase boundaries. The same functional carries an exact spectral reading
where it is exact (\cref{sec:spectral}): a stable-branch lower polariton whose softening, a Thouless
instability, marks the superradiant onset and each fold of the curve, except where the cavity couples
through a conserved quantity, where the onset is thermodynamically present yet spectrally silent and the
photon line sits exactly at $\omega_c$. The full cavity-dressed excitation spectrum within the superradiant phase and on the
critical vacua, beyond the onset and fold softening just described and the exactly solvable $d=\infty$
case (\cref{app:dinf_dicke}), is left open (\cref{sec:conclusions}).

We preview the diagrams these rules yield before deriving them. The onset is classified by the leading
singularity of the bare susceptibility $\xmat=\mumat'$: a finite threshold $\lambda_c=1/\xmat(0)$; an
immediate onset; or an essential singularity. Its order is fixed by exact
closed-form Landau coefficients, in the rescaled units $J\to J/d$ under which mean field is exact at
$d=\infty$. The ferromagnetic tricritical point is
$\varepsilon_{\mathrm{tri}}^{\mathrm{ferro}}=|J|/d$ (a genuine, ordinary tricritical point,
$m\sim(\lambda-\lambda_c)^{1/4}$); the antiferromagnetic one is
$\varepsilon_{\mathrm{tri}}^{\mathrm{AF}}=2|J|/\sqrt{8d-3}$ ($J$ the Ising exchange, $d$ the spatial
dimension), where the branch instead leaves the axis unstably---a first-order endpoint that, as we
show, is never realised: the polarised state preempts it in every dimension, the AS phase surviving
only near the quadruple point. The superradiant--superradiant (AS--PS) transition of the Dicke--Ising
model can never be continuous in the physical dimensions $d=1,2,3$: the bare matter is in the
$(d{+}1)$-Ising class, whose susceptibility, with the same singular part as the specific heat, diverges
and forces a fold---the compressible-magnet mechanism of Larkin and Pikin~\cite{LarkinPikin1969}, in
cavity-QED form. At $d=3$ the $(d{+}1)=4$ matter sits exactly at the Ising upper critical dimension,
where the divergence is only logarithmic but still present, so the fold, and the first-order transition
with it, persist. Above it the divergence is gone, and the order is decided by the binding curvature: at large $d$, where the series is analytic, it is computed in closed
form, and the AS--PS line stays second order in its interior, bounded to this order in $1/d$ by two tricritical points near
its ends---a tricritical line in the $(d,\varepsilon)$ plane. A $1/d$ expansion maps all four boundaries to this order in $1/d$, including the AS phase, which admits no
perturbative reference and is instead expanded about the exact $d=\infty$ solution. And at the
quadruple point, where every expansion fails, the matter reduces to a detuned Rydberg-blockade chain
that we solve directly by infinite-system DMRG in the strict blockade manifold: the AS phase exists, as
a finite angular sector about $5^\circ$ wide (the rays $r\in(1,1.5)$), and at the scale-invariant corner
the AS--PS jump stays finite, not vanishing as the corner is approached, settling the
questioned existence in one dimension. Each of these first-order folds is, on its stable branch, a
collective mode going soft, so the lines carry a spectral signature; and a single-humped susceptibility
leaves no room for a further superradiant phase (\cref{sec:general}).

We test the framework on deliberately varied matter, each a different shape of the same curve. The
frustrated triangular antiferromagnet has a cusp onset (superradiance at any coupling) and a negative
specific-heat exponent $\alpha$, so that its susceptibility stays finite at criticality and a
\emph{continuous} superradiant--superradiant line survives there at all resolved scales, in the
$3$D-XY class---the one model in our gallery that escapes the Larkin--Pikin fold, precisely because
$\alpha<0$ leaves the susceptibility finite. The bond-alternating compass chain realises the marginal
row of the onset
classification in exactly solvable form, its threshold closing only logarithmically with the bond
asymmetry. The gapless Heisenberg and XX chains sit in the regular class and, coupled through a
\emph{conserved} quantity, show a superradiant onset that is thermodynamically present yet spectrally
silent, the photon line sitting exactly at $\omega_c$ through the transition; a transverse field on the
Heisenberg chain then opens an exactly solvable diagram whose first-order line ends, at the saturation
corner, in an exact tricritical point of an unusual kind. Throughout, the cavity couples transverse to
the matter order; it need not in general~\cite{Sur2026}. \Cref{sec:decoupling} derives the decoupling,
\cref{sec:general} the theory of the stationarity curve; \cref{sec:models} specifies the models,
\cref{sec:results} works out their diagrams, and \cref{sec:conclusions} lays out the open questions of Dicke-like models with interacting matter; the appendices carry the methods---the low-field linked-cluster expansion, the
$1/d$ expansion, the Larkin--Pikin cases, and the DMRG protocols---in reproducible detail.

\section{The general decoupling}
\label{sec:decoupling}

We work, throughout this section and the next, with an \emph{arbitrary} matter Hamiltonian
$\hat H_m$ collectively coupled to a single cavity mode,
\begin{equation}
\hat H = \omega_c\,\adag\aop + \frac{g}{\sqrt N}(\aop+\adag)\,\hat S_z + \hat H_m,
\qquad \hat S_z=\tfrac12\sum_{i=1}^N\sigma^z_i,
\label{eq:H_general}
\end{equation}
and specialise to a concrete model only in \cref{sec:models}. The photon couples to the
collective magnetisation $\hat S_z$; $\hat H_m$ is left unspecified. We use the per-site
magnetisation $m\equiv\langle\sigma^z\rangle$ (so $\langle\hat S_z\rangle/N=\tfrac12 m$) and the
coupling $\lambda\equiv g^2/(2\omega_c)$ (an energy; the dimensionless ratio is $g/\omega_c$); the combination that enters the functional is $h=\lambda m$ with penalty $\tfrac\lambda2 m^2$ (the paper-wide normalisations are collected in \cref{tab:conventions}).

\paragraph{The collective limit (Dicke $\to$ LMG).}
None of what follows in this section is new: every step is established literature, collected in
the lineage paragraph below; we spell the steps out because the precise form of
\cref{eq:functional} is the object the rest of the paper reads.
Completing the square in the photon---displacing $\aop\to\aop-\tfrac{g}{\omega_c\sqrt N}\hat S_z$ to
remove the linear coupling---eliminates the cavity exactly and leaves an attractive, all-to-all
interaction of the collective operator as its only trace,
\begin{equation}
\hat H\;\longrightarrow\;\hat H_m-\frac{g^2}{\omega_c N}\,\hat S_z^2 .
\label{eq:Sz2}
\end{equation}
The cleanest place to see what this means is the interaction-free corner: take $\hat H_m$ to be a
single-site field transverse to the cavity axis (along $\sigma^x$), with no two-body terms, and \cref{eq:H_general} is the bare Dicke model, while
\cref{eq:Sz2} is the Lipkin--Meshkov--Glick model---a single large spin with an all-to-all
$\hat S_z^2$ interaction in a field (the equivalence noted early by Reslen
\emph{et al.}~\cite{Reslen2005}). For such an infinite-range interaction the coherent (product)
mean field is exact in the thermodynamic limit, as established for the
Dicke model by Wang and Hioe~\cite{WangHioe1973,Hioe1973} (and rigorously by Hepp and Lieb~\cite{HeppLieb1973}), with the $O(1/N)$ finite-size corrections
worked out for the Lipkin--Meshkov--Glick model
in Refs.~\cite{Leyvraz2005,Ribeiro2008,Dusuel2004,Dusuel2005}. The same holds for \emph{any}
$\hat H_m$ (\cref{sec:models})---its own interactions may even be long-ranged~\cite{Mori2010}---because only the collective $\hat S_z^2$
is treated at mean-field level, never the matter itself. The mechanism is the textbook exactness of
mean field for an all-to-all interaction. The photon mediates only the single collective $\hat S_z^2$,
so a Hubbard--Stratonovich field reduces the problem to a one-dimensional integral over the
self-consistent field, dominated by its saddle as $N\to\infty$ (Laplace); the decoupling error is the
collective variance $\langle(\hat S_z-\langle\hat S_z\rangle)^2\rangle$, which is $O(N)$ whenever the
matter clusters in the cavity-coupled channel---an $O(1)$, subextensive correction to the energy, so
the energy \emph{density} is exact up to $O(1/N)$. Two things this does \emph{not} require:
analyticity of $\emat$---Laplace needs only that it be continuous, which concavity guarantees, so an
extensively degenerate ground manifold (merely a cusp in $\emat$, \cref{sec:landau}) is harmless; and
a gap---the argument is insensitive to whether the matter is gapped, gapless, or degenerate, failing
only if the cavity-coupled $\sigma^z$ channel itself develops macroscopic [$O(N^2)$]
\emph{connected} fluctuations about the mean (a cat of two macroscopic $S_z$ values)---the bare matter's own spontaneous long-range order
in $\sigma^z$, as distinct from the self-consistent polarisation of the superradiant phase. The frustrated magnets of \cref{sec:model_tri,sec:results} are degenerate in
the order-parameter channel, not the cavity one.

\paragraph{The matter functional.}
That single mean-field step is the factorisation of the collective term: the standard decoupling of
the squared collective operator,
\begin{equation}
\hat S_z^2\;\longrightarrow\;2\langle\hat S_z\rangle\,\hat S_z-\langle\hat S_z\rangle^2
=\tfrac N2\,m\!\sum_i\sigma^z_i-\tfrac{N^2}{4}m^2 ,
\label{eq:Sz2_MF}
\end{equation}
is exact for the infinite-range $\hat S_z^2$ as $N\to\infty$. Inserting \cref{eq:Sz2_MF} into
\cref{eq:Sz2} replaces the cavity by a uniform, self-consistent field $h=\lambda m$ on the matter
plus a constant penalty $\tfrac\lambda2 N m^2$ (with $\lambda=g^2/2\omega_c$ as above), so the energy
density is the single scalar functional of one mean field,
\begin{equation}
\boxed{\;\evar(m) = \tfrac{\lambda}{2}\,m^2 + \emat(\lambda m)\;},
\qquad
\emat(h)=\lim_{N\to\infty}\tfrac1N\,E_{\mathrm{GS}}\!\left[\hat H_m - h\textstyle\sum_i\sigma^z_i\right],
\label{eq:functional}
\end{equation}
to be minimised over $m$, now a free variational parameter whose minimiser is the self-consistent
magnetisation---the exact matter functional in the general form of Rom\'an-Roche and
co-workers~\cite{roman2022effective}, resting on the separable-interaction exactness theorems
reviewed below. Here $\emat(h)$ is the ground-state energy density of the bare matter
in a uniform field $h$, with bare magnetisation and susceptibility
\begin{equation}
\mumat(h)\equiv-\emat'(h)=\langle\sigma^z\rangle,\qquad
\xmat(h)\equiv-\emat''(h).
\label{eq:mu_chi}
\end{equation}
The two terms of \cref{eq:functional} are in competition: the penalty $\tfrac\lambda2 m^2$ is the
energy cost of displacing the photon---what keeps a normal phase $m=0$ stable at weak coupling---while
$\emat(\lambda m)$ is the energy the matter gains by polarising in the field $h=\lambda m$ the photon
mediates. The two terms are sharply asymmetric: the penalty $\tfrac\lambda2 m^2$ is a classical
$c$-number scalar (a parabola $\propto g^2$, carrying no operators and no light--matter
entanglement), while \emph{all} the quantum mechanics---the interactions, correlations, and
criticality of the matter---resides in $\emat(\lambda m)$. In equilibrium the cavity contributes only
that scalar; what makes the phase diagram rich is how this trivial term competes with, and folds, the
matter's own quantum response. The reduction is therefore exact but \emph{conditional}: every
approximation that survives it is an approximation of $\emat(h)$ and nothing else. Where $\emat(h)$
is known in closed form---the solvable chains of \cref{sec:results_iso}---the superradiant phase
diagram inherits that exactness; where it is not, the sole error is in $\emat(h)$ itself, computed
by the controlled matter approximations of \cref{sec:results} and flagged at the end of this
section. The functional itself has a lineage, and its rigorous core is half a century old. That a separable
(all-to-all) coupling decouples \emph{exactly} in the thermodynamic limit---the free energy equal to
that of a matter Hamiltonian \emph{linear} in the coupled operator, the field fixed by min--max
self-consistency---was proven at finite temperature in the 1970s: by
Bogoliubov~Jr.~\cite{Bogoliubov1972book} and, for an arbitrary interacting short-range reference, by
the Leiden school~\cite{TindemansCapel1975,denOuden1976a,denOuden1976b}, who also recast the exact
free energy as a convex-envelope construction~\cite{PerkCapel1977}; the bridge to Dicke-type models
was made explicit early on~\cite{Brankov1975,Gibberd1974}. In the cavity setting the interacting-matter problem was taken up at
finite temperature by Lee and Johnson~\cite{lee2004}, who found that matter
interactions can render the superradiant transition first order, and by Gammelmark and
M\o lmer~\cite{Gammelmark2011}, who mapped the finite-temperature mean-field diagram of the
cavity-coupled Ising chain---first- and second-order onsets with a tricritical point; at $T=0$ the displaced-frame transformation of Rohn \emph{et al.}~\cite{Rohn2020} removes the photon
but leaves the residual collective $\hat S_z^2$ of \cref{eq:Sz2} still to be treated correctly, that
is, self-consistently; at zero and finite temperature such a self-consistent treatment of the
cavity-coupled Ising case was carried out in Ref.~\cite{roman2021photon}; and a complementary route
reaches a matter-only description differently, Lenk \emph{et al.}~\cite{lenk2022collective} expressing
the collective light--matter response through the nonlinear response functions of the uncoupled
matter. The general, model-independent form---valid for arbitrary $\hat H_m$ at any
temperature~\cite{roman2022effective}, proven exact via a generalised Hubbard--Stratonovich
transformation~\cite{roman2023exact}, and extended to the full dynamical response in
Refs.~\cite{roman2025linear,roman2025bound}---is due to Rom\'an-Roche and co-workers; it is the
modern, cavity-native form of the separable-interaction theorems above, and the foundation on which
everything below is built.

\paragraph{Stationarity is self-consistency.}
Minimising \cref{eq:functional} gives the self-consistency condition $m=\mumat(\lambda m)$---the
cavity field $\lambda m$ polarises the matter, whose magnetisation sources the field in turn. Its
global structure is the subject of \cref{sec:general}, where the stationarity is developed in full.

\paragraph{A pure matter Hamiltonian.}
The content of \cref{eq:functional} is that, in the thermodynamic limit, the joint light--matter
ground state is a displaced photon vacuum times the ground state of a \emph{pure matter
Hamiltonian}---the bare chain $\hat H_m-h\sum_i\sigma^z_i$ in the field $h=\lambda m$. No states
beyond those of $\hat H_m$ appear: once the photon sits at its coherent displacement, the
light--matter coupling is, in effect, nothing but a self-consistent field term on the matter,
changing only the energies and the value of the field, not the Hilbert space. In particular it creates no phase the matter does not already
possess---every superradiant phase is just the matter polarised along $\sigma^z$. Because the collective
coupling is all-to-all, this exactness extends beyond the global minimum to every \emph{stable}
self-consistent solution: each metastable superradiant branch is, as much as the true ground state,
the exact ground state of the bare matter in its own field $h=\lambda m$, reproducing as
$N\to\infty$ the exact energy density and magnetisation of a (meta)stable state of the full
model---the self-consistent minima are the model's actual states, not a bookkeeping construction. The genuinely
quantum cavity effects, light--matter entanglement and photon squeezing, are $O(1/N)$ corrections
to this saddle point and so enter only at finite $N$ (\cref{sec:conclusions}).

\paragraph{Two distinct mean-field limits.}
One caveat for later. The cavity saddle point above is exact at $N\to\infty$ for \emph{any}
matter, and is the only mean-field step taken here. It must not be conflated with a second,
independent mean-field limit---matter mean field becoming exact as $d\to\infty$ with $J\to J/d$
on a hypercubic lattice---which we invoke purely as a calculational tool in \cref{sec:results}.

\section{The stationarity curve and its landscape}
\label{sec:general}

Everything in this paper is a property of the single functional \cref{eq:functional}, worked out
here for arbitrary matter and specialised to concrete models only in \cref{sec:models,sec:results}.
Its whole content is carried by one object---the \emph{stationarity curve} $\lambda(m)$ in the
coupling--magnetisation plane. Concavity of the bare-matter energy makes the curve a single-valued
graph (\cref{sec:concavity}). A symmetry-protected \emph{pitchfork}---two mirror
branches $\pm m$ growing out of the symmetric solution---can occur only on the $m=0$ axis: that is
the onset of superradiance. Everywhere else the curve cannot branch---it can only \emph{fold}: it
bends back, two phases coexist, and the global state jumps between them---a first-order transition.
Which way the curve bends, and so which transitions are continuous and which first order, is fixed
by a single local sign (\cref{sec:folds}); the whole curve also carries a spectral reading---folds
are softening polaritons (\cref{sec:spectral}). A bare-matter critical point pins one further point on
the curve, and whether the curve passes it as a minimum or a maximum is decided by the divergence of
the matter susceptibility there---the Larkin--Pikin mechanism (\cref{sec:anchor,sec:lp}). A short
completeness argument shows the curve hides no further phases (\cref{sec:completeness}). The one
exception: a bare matter with its own first-order transition breaks the curve into disconnected
pieces (\cref{sec:folds}).

\subsection{Concavity: the response is a monotone, bounded curve}
\label{sec:concavity}

The one input we need is almost obvious: \emph{turning up the field raises the magnetisation}.
Increasing the field $h$ conjugate to $\sigma^z$ can only increase the bare magnetisation
$\mumat(h)=-\emat'(h)=\langle\sigma^z\rangle$ of \cref{eq:mu_chi}, so the response is a single,
bounded, monotone curve: $\mumat$ is non-decreasing in $h$, strictly increasing wherever the bare
susceptibility $\xmat(h)=-\emat''(h)=\mumat'(h)$ is nonzero, and saturating at
$|\mumat|\le\mu_{\mathrm{sat}}$; equivalently $\xmat(h)\ge0$. The reason is that $\emat(h)$ is
\emph{concave}: the ground-state energy in a field is the minimum of a family of straight lines in
$h$---one per state, with slope $-\langle\sigma^z\rangle$---and a minimum of straight lines always
bends down.\footnote{$\emat(h)=\inf_{|\psi\rangle}\langle\psi|\hat
H_m-h\sum_i\sigma^z_i|\psi\rangle/N$ is an infimum of functions \emph{affine} in $h$, hence concave;
therefore $\xmat=-\emat''\ge0$ and $\mumat$ is monotone and bounded. The argument needs only that
$\emat$ is a minimum of an $h$-affine family over \emph{some} set of states, so it holds for the
exact ground state and in any product-state (mean-field) restriction alike.} This concavity is what
makes the stationarity curve of \cref{sec:folds} \emph{single-valued}: it pins one coupling
$\lambda$ to each magnetisation $m$, so the curve is a graph over $m$. (A fold does not violate
this---the curve may still bend back so that one $\lambda$ meets \emph{several} $m$, the coexisting
phases; what is excluded is the reverse, several $\lambda$ at one $m$.) Single-valuedness in fact
needs $\mumat$ \emph{strictly} monotone---$\xmat>0$, not merely $\xmat\ge0$---since an interior
magnetisation plateau ($\xmat=0$ over a field interval) would map an interval of $\lambda$ to a
single $m$. None occurs here: the susceptibility
$\xmat(h)=\tfrac2N\sum_{n\neq0}|\langle n|\hat M|0\rangle|^2/(E_n-E_0)$, with $\hat M=\sum_i\sigma^z_i$
and $|n\rangle$ the eigenstates of the bare matter in the field $h$, vanishes only when the ground
state is an eigenstate of $\hat M$; for matter whose order lies along $\sigma^x$ this happens only at
full $\sigma^z$ saturation, so $\xmat>0$ strictly on the interior and the only
$\xmat=0$ point is the benign saturation edge $m=\mu_{\mathrm{sat}}$---benign for
single-valuedness, though the response may still diverge as the edge is \emph{approached} from
below (\cref{sec:results_iso}). One bookkeeping note: throughout this paper $\xmat$ is the
\emph{thermodynamic} curvature $-\emat''$. The spectral representation above is its Kubo form,
valid when the coupled operator is not conserved---the generic case here. It is one susceptibility
in two readings, not a new quantity. For the conserved
chains of \cref{sec:results_iso} the Kubo sum vanishes identically, yet the thermodynamic
$\xmat$ stays finite and positive, assembled from level crossings between magnetisation sectors
rather than from matrix elements.

\subsection{The self-consistent landscape}
\label{sec:landscape}

For fixed coupling, \cref{eq:functional} is a one-parameter family of scalar landscapes
$\evar(\,\cdot\,;\lambda)$, and the physical state is its global minimiser; a phase boundary is
where that minimiser jumps or splits. The first two derivatives are
\begin{align}
\evar'(m)&=\lambda\big[m-\mumat(\lambda m)\big],\label{eq:devar}\\
\evar''(m)&=\lambda\big[1-\lambda\,\xmat(\lambda m)\big],\label{eq:ddevar}
\end{align}
so that stationarity, $\evar'(m)=0$, is exactly the self-consistency condition $m=\mumat(\lambda m)$:
the self-consistent states are the intersections of the diagonal with the rescaled response
$\mumat(\lambda m)$, and the curvature \cref{eq:ddevar} says whether an intersection is a stable
minimum or a barrier maximum.

\subsection{The onset, the curve, and its folds}
\label{sec:landau}
\label{sec:folds}

\begin{figure}[tbp]
\centering
\includegraphics{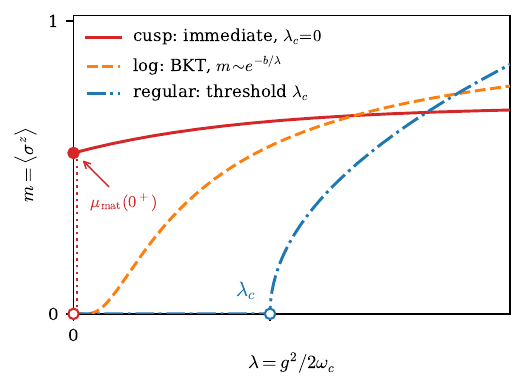}
\caption{The three onset classes set by the leading singularity of $\emat(h)$ (schematic). A
\emph{cusp} (divergent $\xmat(0)$) gives an immediate onset at $\lambda_c=0$: the magnetisation
jumps from $m=0$ (open circle at the origin) to the bare value $\mumat(0^+)$ (filled circle,
dotted riser); a \emph{log}-divergent
$\xmat(0)$ gives a Berezinskii--Kosterlitz--Thouless-type essential singularity $m\sim e^{-b/\lambda}$
with no sharp threshold; a \emph{regular} (finite $\xmat(0)$) response gives a finite threshold
$\lambda_c=1/\xmat(0)$ (open circle).}
\label{fig:onset}
\end{figure}

At weak coupling the photon penalty $\tfrac\lambda2 m^2$ of \cref{eq:functional} holds the system
in a normal phase $m=0$, exactly as in the bare Dicke
model. Superradiance is the onset of $m\neq0$ as the coupling grows, and \emph{how} it sets in is
fixed by the leading singularity of the matter response $\emat(h)$ as $h\to0$. Three cases cover the
matter of this paper, spanning gapped, degenerate, and marginal responses:
\begin{center}
\begin{tabular}{lll}
\hline
$\emat(h)-e_0$ & $\xmat(0)$ & onset \\
\hline
$-c\,|h|$ (cusp) & divergent & immediate, $\lambda_c=0$ \\
$-\tfrac12 A\,h^2\ln\tfrac1{|h|}$ & log-divergent & $m\sim e^{-b/\lambda}$ (BKT-type) \\
$-\tfrac12\xmat(0)\,h^2$ (regular) & finite & finite threshold $\lambda_c=1/\xmat(0)$ \\
\hline
\end{tabular}
\end{center}
A \emph{regular} response---a finite zero-field susceptibility, with $\emat$ even and analytic in
$h$ (the Ising axis is orthogonal to the cavity axis, so $\langle\sigma^z\rangle$ is odd under the
residual spin-flip symmetry)---is the familiar Dicke case: $\evar$ is even in $m$, the normal branch
$m=0$ is stationary at \emph{every} coupling, and it destabilises only at a finite threshold. The
other two rows break this picture at $m=0$ itself. A \emph{linear cusp} makes the functional
$\evar(m)=\tfrac\lambda2 m^2-c\lambda|m|+\dots$ kink \emph{downward} at the origin: $m=0$ is no
longer stationary [$\evar'(0^+)=-c\lambda<0$], the system slides off the axis for every $g>0$, and
the magnetisation jumps to the bare value $\mumat(0^+)$ (\cref{fig:onset}). A \emph{marginal},
logarithmically soft response is the borderline: $m=0$ stays stationary, but the threshold collapses
to zero through a Berezinskii--Kosterlitz--Thouless (BKT)-\emph{type} essential singularity
$m\sim e^{-b/\lambda}$---the functional form of a BKT onset---turning
on continuously, with no power law and no jump. Which case a given model realises is taken up with the
models in \cref{sec:results}.

The rest of this subsection develops the \emph{regular} row, on a fixed route: first the global
object---the stationarity curve, and the angle that reads stability off it; then why the curve can
fold but never branch; then the onset, as the one symmetry-protected exception at the axis, where
the Landau analysis and the bend of the curve turn out to be the same statement; and finally the
connectedness of the curve. Its microscopic, spectral reading then gets a subsection of its own
(\cref{sec:spectral}). The cusp and marginal rows return
with the frustrated and compass magnets of \cref{sec:results}.

\emph{The stationarity curve, and the angle that reads stability.} Solving the self-consistency
$m=\mumat(\lambda m)$ for the coupling gives a single curve in the $(m,\lambda)$ plane,
\begin{equation}
\boxed{\;\lambda(m)=\frac{\mumat^{-1}(m)}{m}\;},
\label{eq:statcurve}
\end{equation}
single-valued because $\mumat$ is strictly monotone (the concavity of \cref{sec:concavity});
coexisting phases can appear only where it \emph{folds}---bends back, so that one coupling supports
several self-consistent magnetisations. A fold is a turning point, $\lambda'(m)=0$. Differentiating
\cref{eq:statcurve} (with $h=\mumat^{-1}(m)$, so $h=\lambda m$ and $dh/dm=1/\xmat$),
\begin{equation}
\lambda'(m)=\frac{d}{dm}\frac{h}{m}=\frac{1-\lambda\xmat}{m\,\xmat},
\label{eq:lambdaprime}
\end{equation}
which, compared with the curvature \cref{eq:ddevar}, gives the identity
\begin{equation}
\evar''(m)=\lambda\,m\,\xmat(\lambda m)\,\lambda'(m).
\label{eq:eddot_slope}
\end{equation}
Here $\evar''$ is evaluated at the self-consistent coupling $\lambda=\lambda(m)$: the symbol
$\lambda$ is the control parameter in \cref{eq:ddevar} but the curve value in \cref{eq:statcurve},
and the two coincide exactly on the curve, where this identity lives.
Since $\lambda,m,\xmat>0$, the curvature of the landscape and the slope of the curve carry the same
sign, and this is sharpest read as an angle. By the $m\to-m$ symmetry we may take $m\ge0$ and
measure the curve's tangent by the angle $\theta$ it makes with the coupling axis of the
phase diagram as $m$ increases ($\lambda\propto g^2$, so the same reading applies verbatim to the
$g^2$ axes of the figures). A stationary point is then a \emph{minimum} for
$\theta\in(0^\circ,90^\circ)$, where the tangent runs to larger $\lambda$; a \emph{maximum} for
$\theta\in(90^\circ,180^\circ)$, running to smaller $\lambda$; and a \emph{fold} exactly at the vertical
tangent $\theta=90^\circ$, where $\lambda$ turns around ($\lambda'=\evar''=0$): the \emph{spinodal}, the
point at which a metastable branch loses its local stability (for short-range interactions the
mean-field spinodal is not sharp---metastable states decay by nucleation before it is
reached~\cite{Binder1987}; here the collective coupling is infinite-ranged and the spinodal is
exact). A vertical $\lambda=\text{const}$
through a back-bent region therefore meets the curve min--max--min (two stable phases flanking the
barrier), and the global state jumps between the outer arms where their energies cross, by a finite
$\Delta m$ (the Maxwell construction---the equal-energy rule; in the separable-interactions
literature this equilibrium statement is the rigorous convex-envelope
theorem~\cite{PerkCapel1977}).

\emph{Why the curve never splits off the axis.} Read the stationary set as the level curve
$F\equiv\partial_m\evar=0$ in the $(m,\lambda)$ plane. The level curve runs perpendicular to the
gradient $\nabla F=(\partial_m F,\partial_{\lambda}F)$, so its tangent is well defined---one smooth
branch---wherever $\nabla F\neq0$; two branches could meet only where the gradient vanishes
entirely. But $\partial_{\lambda}F\propto m\,\xmat$ carries a factor of $m$: off the axis it is nonzero
(at a fold, $\evar''=0$ forces $\xmat=1/\lambda>0$), so away from $m=0$ the curve folds but never
splits. Both components vanish together only on the axis, where $\{m=0\}$ is stationary at every
coupling. The finite-$m$ folds are in this sense the generic version of the onset, with the
symmetry protection removed---and the onset itself is next. The curve has no loose ends: it runs
from the axis out to the saturation edge---reached at a finite coupling when the saturation
field is finite (the isotropic chains of \cref{sec:results_iso}, where the saturated branch then
takes over), and only as $\lambda\to\infty$ otherwise.

\emph{The onset is the curve leaving the axis---and $a_4$ is the direction it takes.} On the axis
the normal phase is stationary at every coupling, and it destabilises where the curvature
$\evar''(0)=\lambda[1-\lambda\xmat(0)]$ passes through zero---the response catching up with the
penalty, at
\begin{equation}
\boxed{\;\lambda_c=\frac1{\xmat(0)}\;}
\label{eq:lambdac_general}
\end{equation}
(in bare units, the superradiance coupling $g_c^2=2\omega_c/\xmat(0)$---quoted once here; the rest
of the paper stays in $\lambda$). There the symmetry-protected pitchfork happens: the superradiant
branch leaves the axis vertically ($\lambda(m)$ is even), and what detaches is read off an ordinary
Landau analysis. For the regular row $\evar$ is analytic in the single variable $z\equiv m^2$ (the
marginal row's $m^2\ln\tfrac1{|m|}$ is not---no Landau series exists there, and its onset was
settled directly above):
\begin{equation}
\evar=a_0+a_2\,z+a_4\,z^2+a_6\,z^3+\dots,\qquad
a_2=\tfrac\lambda2\big(1-\lambda\,\xmat(0)\big),\quad
a_4=-\tfrac1{24}\,c_4\,\lambda^4,
\label{eq:landau}
\end{equation}
with $c_4=\mumat'''(0)$; the normal phase is the boundary $z=0$ of the allowed range $z\ge0$, a
minimum while $a_2>0$. The quartic $a_4$ is, geometrically, the leading curvature of the
stationarity curve at the origin---the even $m^2$ expansion seen sideways, as a plane curve.
Expanding \cref{eq:statcurve} about the origin,
\begin{equation}
\lambda(m)=\lambda_c-\frac{c_4}{6\,\xmat(0)^4}\,m^2+O(m^4),
\label{eq:lambda_origin}
\end{equation}
so the Landau alternative at the onset and the bend of the curve are one and the same
fact.\footnote{At the onset this is the quartic-sign criterion of Landau theory, in the form
experimental magnetism uses daily: a negative initial slope of the Arrott isotherm $H/M$ vs.\
$M^2$ signals a first-order transition (the Banerjee criterion~\cite{Arrott1957,Banerjee1964}). \cref{eq:lambda_origin} is that line for the bare
matter at zero temperature---its ordinate $h/m$ \emph{equals} the coupling by
self-consistency---and the threshold $\lambda_c=1/\xmat(0)$ is Arrott's zero-intercept criterion
for the critical isotherm. What is there a mean-field diagnostic of a thermal transition is here
exact for the functional, the angle reading of \cref{eq:eddot_slope} extending it from the onset
to the whole curve.} For
$a_4>0$ ($c_4<0$) the curve rises from the axis and the interior stationary point $z_*=-a_2/2a_4$
grows continuously out of $z=0$ as $a_2$ turns negative---a pitchfork, second order. For $a_4<0$
($c_4>0$) the curve bends \emph{back} toward weaker coupling: the branch that detaches is a
\emph{maximum} of $\evar(z)$---a barrier---born together with a true minimum beyond it, where the
finite-spin saturation turns the energy back up, while $z=0$ is still locally stable; the two minima are separated
by the barrier and the system jumps across it, first order. The tricritical point $a_2=a_4=0$ is
exactly where the fold is born and the superradiant extremum detaches from the axis; there the
$m^2$ bend vanishes and the next coefficient takes over---the branch leaves flatter still
($\propto m^4$), toward larger $\lambda$ for $a_6>0$ (a genuine, \emph{ordinary} tricritical point) or
toward smaller $\lambda$ for $a_6<0$, in which case $a_4=0$ marks a first-order point itself, with
$m=0$ locally unstable (\cref{app:a4}). The vocabulary is that of catastrophe theory---fold and
cusp of one potential family~\cite{Gilmore1981}; that the symmetric $m^6$ Landau family realises
a higher catastrophe, the butterfly, was noted by Schulman~\cite{Schulman1973}, and Gilmore carried the
structural-stability analysis out for Dicke-like models themselves~\cite{Gilmore1977}. What the
curve adds is the matter input: \emph{which} catastrophe occurs, and where, is fixed here by
exact response coefficients of correlated matter, not by an unfolding's free parameters.

Because \cref{eq:functional} is exact in the thermodynamic limit, these are \emph{exact locations}:
$a_2=0$ and $a_4=0$ are fixed by single low-order responses of the bare matter ($\xmat(0)$ and
$c_4$), independent of every higher-order term, even for interacting, spatially correlated matter.
Both the location and the exponents are exact: the location is pinned by these low-order responses,
and non-universal, while the exponents are the universal mean-field ones; both follow from the one
feature that makes the bare Dicke transition exact: $m$ is a single collective mode whose fluctuations vanish as $1/N$ away from
criticality (anomalously, more slowly, on the transition lines themselves). (The
marginal row shows how this bookkeeping degenerates gracefully between the gapped and the critical
normal phase: there $\xmat(0)$ diverges, so $a_2<0$ at every coupling---the exact location
$\lambda_c=1/\xmat(0)$ is simply zero---and the minimum sits at the essentially small
$m\sim e^{-b/\lambda}$; the location statement survives, only the expansion does not.) The
critical behaviour is therefore always of \emph{mean-field} type, but \emph{which} class depends on
the case: the second-order line carries the bare Dicke exponents
($m\sim(\lambda-\lambda_c)^{1/2}$), a genuine tricritical point ($a_2=a_4=0$, $a_6>0$) the
mean-field tricritical ones ($m\sim(\lambda-\lambda_c)^{1/4}$), and a marginal response an essential
singularity $m\sim e^{-b/\lambda}$ with no power law at all. The $m=0$ end of the curve is also
where the analysis is easiest: the expansion starts from the matter's zero-field ground
state---gapped and non-degenerate---so perturbation theory in the field converges and proceeds
outward along the nascent superradiant branch, toward the fold, its low-order
coefficients well-defined convergent responses (closed-form for the solvable matter of
\cref{sec:results}). A fold at finite $m$ sits at finite field, beyond the reach of any low-order
expansion about $h=0$; and for a degenerate or gapless reference no regular expansion exists at all,
the onset being pushed to $\lambda_c=0$ (the cusp and marginal rows above).

One caveat, and it is worth keeping the two levels strictly apart. Everything above is \emph{local}:
$a_2$ and $a_4$ describe the landscape near $m=0$---when the origin destabilises, and what detaches
from it. Whether the \emph{global} minimum is still there is a separate question: a first-order jump
to a finite-$m$ minimum can preempt the local instability, the system leaving $m=0$ while it is
still locally stable. The local expansion never decides this; the global curve does.

\emph{One connected curve: the van der Waals reading.} The curve is not only single-valued but
\emph{connected}: $\lambda(m)$ is continuous wherever $\mumat$ is, and that is everywhere unless the
bare matter has its own first-order field-driven transition. The two superradiant phases flanking a
fold are therefore not two unrelated ground states crossing, as disconnected free-energy sheets do
at a level crossing; they are the two arms of \emph{one} curve, joined through the unstable branch
between them---the cavity analogue of the van der Waals isotherm, whose liquid and gas are arms of
one equation of state. The geometry itself has a classical ancestor: for an all-to-all spin toy the
same S-shaped curve was drawn---spinodal and equal-area rule included---in the separable-interactions
literature~\cite{denOuden1976a}. There, however, the folds were injected by hand, through a quartic
collective term acting on \emph{free} spins; here the collective term is the bare parabola of
\cref{eq:functional}, and every feature of the curve is supplied by the response of the interacting
matter---the curve probes the matter, not the coupling polynomial. That folds-not-crossings
distinction matters because a fold can \emph{unbend}: the back-bend straightens out without two
minima merging. Above the upper
critical dimension the first-order line can give way to a continuous one (reached by tuning the
dimension and $\varepsilon$ into the regime where the fold unbends; \cref{sec:lp}), an option
a level crossing of disconnected branches does not have. Where it does, the matter's own order
parameter (the staggered magnetisation of an antiferromagnet, say) vanishes there continuously;
because the two phases stay symmetry-distinct, the line cannot terminate the liquid--gas way,
with two minima merging, so the endpoint is a tricritical point. The same mechanism---an order
parameter slaving a non-critical variable, here the photon displacement, there the lattice
strain---drives the compressible magnet of Bean and Rodbell~\cite{BeanRodbell1962}. Here the slaver
is the matter's own order parameter, the staggered magnetisation of the antiferromagnet, and the
slaved variable is the photon displacement $m$ itself; this is the fold (superradiant--superradiant)
context, distinct from the symmetry-protected pitchfork onset. The chain is
worth stating once: \emph{continuity of the bare matter $\Rightarrow$ one connected curve
$\Rightarrow$ folds, not crossings $\Rightarrow$ a superradiant first order that can turn
continuous.}

\emph{First-order matter: the disconnected exception.} If instead the \emph{bare} transition is
first order, $\mumat$ jumps, and the gap in its range cuts the stationarity curve into two
disconnected pieces---one per matter phase, each with its own normal phase and onset, the physical
curve being their lower-energy envelope. The reason is generic: at the crossing the two matter
phases carry \emph{different} matter magnetisations $\mumat$ (so $\mumat$ jumps across the bare
transition), so the state cannot pass smoothly from one to the
other---and with the connectedness goes the option of unbending. The long-range Ising devil's
staircase is such a case: the cavity melts its plateaux on disconnected sheets~\cite{Koziol2025}.
Whether the melting on each sheet is continuous or first order is out of scope here; in the former
case a single-valued stationarity curve is assigned to each sheet, in the latter a split into yet more
sheets occurs.

\subsection{The microscopic and spectral reading}
\label{sec:spectral}

The threshold \cref{eq:lambdac_general} and the quartic $a_4$ of \cref{eq:landau} have a
microscopic reading, valid for any gapped matter. At $m=0$ the relevant excitations are single spin flips in the
$\sigma^z$ channel; the cavity couples to their uniform ($q=0$) superposition, so at this level the
coupled problem \emph{is} an effective Dicke model~\cite{Schellenberger2024}, built on the bare flip gap and the matrix
elements that compose $\xmat(0)$, and the onset is the softening of its lower polariton---the lower
of the two hybrid light--matter normal modes---at exactly $\lambda_c=1/\xmat(0)$, the textbook Dicke
result~\cite{Emary2003}. What this single mode cannot decide is the direction the new branch
takes---the sign of $a_4$. That is set by processes involving \emph{pairs} of nearby
flips (physics beyond independent flips, to which a mode built on them is blind), computed for the
Dicke--Ising model in \cref{sec:results_tri} and \cref{app:a4}. When the pair processes win and
$a_4<0$, the polariton still softens at $\lambda_c$---the gap closes regardless---but the branch
that detaches there is the barrier, not a minimum: the uniform-mode description correctly locates
the instability and mispredicts the character of what emerges.

The same reading extends along the whole connected curve, through the cavity-QED linear-response
theory of Rom\'an-Roche \emph{et al.}~\cite{roman2025linear,roman2025bound}: around any
self-consistent state (the self-consistent stationary solution) on the connected curve---stable, metastable, or barrier---the cavity
hybridises only with the \emph{uniform} ($q=0$) matter response, described by the dynamic
susceptibility $\xmat^R(\omega)$ of the bare matter in its static field $h=\lambda m$ [its static
limit is the $\xmat(\lambda m)$ of \cref{eq:mu_chi}], and eliminating the photon at Gaussian level
gives the collective-mode condition $\Omega^2=\omega_c^2\big[1-\lambda\,\xmat^R(\Omega)\big]$---the
poles of the cavity-dressed photon propagator fixed by the matter's dynamic
response. The same linear pole condition is reached from the other direction by Lenk
\emph{et al.}~\cite{lenk2022collective}, who build the collective modes directly from the uncoupled
matter's nonlinear response functions rather than from the equilibrium functional; the two are
complementary derivations of one condition, the functional route additionally fixing \emph{which}
self-consistent state on the curve each mode is built around. For
matter gapped in this channel the lowest solution obeys $\Omega^2=Z\,\evar''(m)$ in the soft limit,
with $Z>0$ (the dispersion of $\xmat^R(\omega)$ only renormalises the prefactor---by spectral
positivity it cannot move the zero\footnote{In the Lehmann representation
$\xmat^R(\omega)=\sum_{n\neq0}2E_{n0}|\langle n|\hat O|0\rangle|^2/(E_{n0}^2-\omega^2)$ of a gapped
ground state ($E_{n0}=E_n-E_0>0$, $\hat O=\tfrac1{\sqrt N}\sum_i\sigma^z_i$, the standard uniform ($k=0$) observable), $\xmat^R$ is monotone increasing
in $\omega^2$ below the gap---$\mathrm{d}\xmat^R/\mathrm{d}(\omega^2)=\sum_{n\neq0}2E_{n0}|\langle
n|\hat O|0\rangle|^2/(E_{n0}^2-\omega^2)^2>0$---so
$Z=\omega_c^2/[1+\omega_c^2\lambda\,\mathrm{d}\xmat^R/\mathrm{d}(\omega^2)|_0]>0$: the dispersion
stiffens the mode but cannot shift its zero from $\evar''=0$. The argument fails precisely when the
gap closes, at the bare critical point, where the continuum reaches zero frequency---the exclusion
noted next.}): stable branches carry a true polariton, the barrier branch an
imaginary frequency, and a fold is exactly where the mode goes soft, $\Omega\to0$ at
$\evar''=0$---Thouless's criterion, the collective soft mode coinciding with the static
instability~\cite{Thouless1961}. In particular, at the onset $m=0$ this collapses onto the Landau
analysis: $\evar''(0)=2a_2$ by \cref{eq:landau,eq:ddevar}, so $\Omega^2(0)=2Z\,a_2$---the
lower-polariton frequency squared is, up to the positive factor $Z$, the Landau mass $2a_2$ itself;
static instability ($a_2=0$) and soft mode ($\Omega\to0$) are then one event, not two statements
that happen to coincide. In the collective channel this is exact at leading order in $1/N$ away from
criticality (on the transition lines the collective gap scales anomalously and the fluctuations are
no longer simply $1/N$), by the same counting that makes the saddle point exact, with two exclusions.\footnote{A third,
minor exclusion: the reading does
not apply on the metastable continuations of the \emph{bare} matter across its own first-order transition---those are not
ground states of any displaced frame, and their response is not a ground-state susceptibility.} It
needs a non-conserved coupling operator: for the XX and Heisenberg chains of \cref{sec:results_iso},
whose total $\sigma^z$ commutes with the Hamiltonian, the dynamic response vanishes and the onset
proceeds by level crossings, with no soft mode. And it fails at the bare-matter critical point
itself when the response diverges there ($\alpha\ge0$, the same divergence the Larkin--Pikin
criterion of \cref{sec:lp} runs on): the critical continuum then reaches down to zero frequency, the
pole moves off the real axis, and the sharp mode is replaced by a broad, damped
response---while the folds, sitting strictly off-critical, are safe. That broadening is the
divergent-$\xmat(0)$ case. The controlling exponent in both cases is the matter's specific heat $\alpha$: because the
cavity-coupled operator is conjugate to the field that drives the matter transition,
$\xmat(0)=-\emat''$ inherits the specific-heat singularity $\sim|t|^{-\alpha}$---divergent for
$\alpha\ge0$, finite for $\alpha<0$. A \emph{continuous} transition with $\alpha<0$ therefore keeps
$\xmat(0)$, and the curvature $\evar''$, finite---no Larkin--Pikin instability---yet its gap still
closes, and there the static value and the dynamics part company: $\xmat(0)$ stays finite, but the
\emph{dynamic} coefficient $\mathrm d\xmat^R/\mathrm d(\omega^2)$ diverges on its own---it weights the
same spectral density by an extra $1/E^2$, two more powers than $\xmat(0)$---so $\xmat^R(\omega)$ is
non-analytic at $\omega=0$. The static curvature does not soften ($\evar''>0$: no fold, no Thouless instability), so any softening
would have to come from the dynamics. Formally it does: with $\mathrm d\xmat^R/\mathrm d(\omega^2)$
diverging, the photon residue $Z=\omega_c^2/[1+\omega_c^2\lambda\,\mathrm d\xmat^R/\mathrm d(\omega^2)]$
collapses to zero, $Z\sim|t|^{2\nu z+\alpha}$ ($t$ the distance to criticality, $\nu z$ the matter-gap
exponent), and $\Omega^2=Z\evar''$ then sends the lower polariton soft. At $d=\infty$, where the AS--PS line is approached off the staggered ($q=\pi$) point so the uniform
$\xmat(0)$ stays finite and the matter response is a single sharp pole rather than a continuum, it does exactly this: the AS--PS lower polariton
softens, $\Omega_1\propto|t|^{1/2}$, shedding its photon weight $Z\propto|t|$ (the mean-field
$2\nu z+\alpha=1$ case, \cref{app:dinf_dicke}). Away from $d=\infty$, though, that reduction is only the
small-$\omega$ truncation of the pole condition, and its validity is exactly what is in doubt: the soft
pole it would place overlaps the gapless continuum the cavity hybridises with---the matter orders at a
finite wavevector and reaches the uniform $q=0$ channel only through the superradiant canting (with
vanishing weight)---so the truncation runs past its own range. Whether the lower polariton then survives
as a sharp resonance, softening with the matter's exponent, or its weight and frequency dissolve into the
continuum, is an open question (\cref{sec:conclusions}); only where the response itself diverges
($\alpha\ge0$, so $\evar''\to0$) is the loss of a sharp mode certain. This is realised on the continuous $3$D-XY superradiant
line of the triangular antiferromagnet (\cref{sec:results_triangular}), where $\alpha\approx-0.015$
keeps $\xmat(0)$ finite---the finite peak of \cref{fig:tri_sc}---while
$\mathrm d\xmat^R/\mathrm d(\omega^2)$ diverges. The unstable branch joining the
two spinodals is an excited-state quantum phase transition~\cite{Caprio2008}.

Read spectrally, the response classes of \cref{sec:folds} sort the fate of the collective mode:
\begin{center}
\begin{tabular}{l l p{0.42\linewidth}}
\hline
matter, $q=0$ channel & $\xmat^R(\omega\!\to\!0)$ & collective mode \\
\hline
regular, non-conserved & finite & sharp polariton; soft at onset ($\Omega^2=2Z\,a_2$) and at each fold ($\Omega^2=Z\,\evar''$) \\
conserved $\sum_i\sigma^z_i$ & $\equiv0$ & none; onset by level crossing \\
cusp & divergent & none at $m=0$; pole only on the superradiant branch \\
marginal (log) & log-divergent & overdamped at $\lambda=0$; sharp for $\lambda>0$ \\
\hline
\end{tabular}
\end{center}
(The column is the Kubo/dynamic response $\xmat^R(\omega\!\to\!0)$---what governs the
collective mode; for the conserved chains it vanishes identically, while the \emph{thermodynamic}
$\xmat(0)=-\emat''$ that fixes the stationarity curve stays finite and positive, the same distinction
drawn in \cref{sec:concavity}.)
Conservation is an axis of its own, orthogonal to the cusp/marginal/regular split: degeneracy and
gaplessness matter spectrally only where they sit in the uniform channel the cavity couples
to---the gapless $q=\pi$ modes of the Heisenberg and XX chains, for instance, leave the static $q=0$
susceptibility finite and the threshold $\lambda_c=1/\xmat(0)$ ordinary, while the absence of a soft
mode at that onset is a separate matter of conservation, not of the ordering wavevector.

\subsection{The critical-point anchor}
\label{sec:anchor}

A bare-matter critical point pins one further point on the curve exactly. Let $h_c$ be the bare
critical field and $m_c=\mumat(h_c)$ the magnetisation there---both cheap inputs: $h_c$ is the bare
matter's own critical field, and $m_c=-\emat'(h_c)$ is its ground-state magnetisation at that field,
available by Hellmann--Feynman from a single ground state. The critical magnetisation is then
self-consistent at the single coupling
\begin{equation}
\lambda_*=\frac{h_c}{m_c},
\label{eq:lambdastar}
\end{equation}
read straight off the bare matter---so the cavity critical line is located with no further
calculation. And because the field $h=\mumat^{-1}(m)$ rises strictly monotonically along the curve,
the curve crosses the bare critical field exactly once: past the anchor, the matter sits in its
other phase for good. Two different points of the curve must not be conflated here: the \emph{anchor}
$(m_c,\lambda_*)$, where the bare matter is critical, and a \emph{fold}, where $\lambda'(m)=0$ at
finite $\xmat=1/\lambda$. When the susceptibility diverges at the anchor the two never coincide---the
divergence in fact pushes the folds \emph{away} from it, since the spinodal condition
$\xmat(\lambda m)=1/\lambda$ is then met strictly off-critical. Whether the curve passes the anchor
as a minimum, leaving the transition continuous, or as a maximum, forcing it first order, is the
Larkin--Pikin question taken up next.

\subsection{Larkin--Pikin: a divergent susceptibility forces the transition first order}
\label{sec:lp}

Whether the cavity transition through the anchor is first or second order comes down to one number:
how large $\xmat$ is at the bare critical point $h_c$. Split $\emat=e_{\mathrm{reg}}+e_{\mathrm{sing}}$
into regular and singular parts at $h_c$, with $\chi_{\mathrm{reg}}$ the finite part of the response,
and expand about the anchor $\lambda_*$ of \cref{eq:lambdastar} with $\delta=m-m_c$ (a local
expansion variable, not the quadruple-point detuning) and $u=\lambda_*\delta$,
\begin{equation}
\Delta\evar=\tfrac{K_2}{2}\,\delta^2+e_{\mathrm{sing}}(\lambda_*\delta),\qquad
K_2=\lambda_*\big(1-\lambda_*\chi_{\mathrm{reg}}\big),
\end{equation}
with $e_{\mathrm{sing}}\sim-A|u|^{2-\alpha}$ (power, $\alpha>0$) or $B\,u^2\ln|u|$ (log, $\alpha=0$).
When $\xmat$ diverges the singular term dominates the regular $\delta^2$, so $m_c$ is a local
\emph{maximum} and the cavity transition is forced first order. The sharp condition is that $\xmat$
\emph{diverge}---not merely $\alpha>0$; for finite $\xmat$ ($\alpha<0$) the singular part is regular
and one recovers the quantitative classical criterion
$\xmat(h_c)<m_c/h_c=1/\lambda_*$~\cite{LarkinPikin1969} (the three cases are derived in
\cref{app:lp}).

This is the cavity-QED form of the Larkin--Pikin mechanism: the cavity displacement plays the role
of the elastic strain in a compressible magnet, with the same local stability conclusion. For
separable couplings the dichotomy is in fact a rigorous finite-temperature theorem, proven by Capel,
den Ouden, and Perk~\cite{CapelDenOudenPerk1979}: a reference critical point with divergent second
derivatives is unstable under arbitrarily small separable perturbations, leaving either renormalised
exponents (Fisher renormalisation~\cite{Fisher1968}) or a first-order transition terminating in
new mean-field critical points---the present
section is, in that sense, the $T=0$ quantum case, in which the single global mode (its fluctuations
$1/N$-suppressed) realises the latter, first-order branch rather than the Fisher-renormalised one. For elastic media the first-order outcome at
a pressure-tuned quantum critical point (the coupling strengthening as $T_c\to0$) was
anticipated at the Landau level by Gehring~\cite{Gehring2008}, and the $T=0$ scaling
theory was developed as ``quantum annealed criticality'' by Chandra
\emph{et al.}~\cite{Chandra2018QAC,Chandra2020QAC}, who found, consistent with the analysis
here, that \emph{above} the upper critical dimension zero-point fluctuations can restore the
criticality. That is precisely the
unbending of the fold noted in \cref{sec:folds}.

A field-driven $\mathbb Z_2$ matter transition---for instance the antiferromagnetic ordering of an
Ising chain in its self-consistent field, which breaks the sublattice (translational) $\mathbb Z_2$---is
in the $(d{+}1)$-Ising class, whose susceptibility diverges in every physical dimension ($d\le3$;
the cases are listed in \cref{app:lp}), so the
superradiant--superradiant line is first order---as we find in one dimension along its whole length up to the quadruple point
(\cref{sec:results}). (This takes the bare field-driven antiferromagnetic melting to be a
continuous, second-order $\mathbb Z_2$-breaking transition in the $(d{+}1)$-Ising class. Its
single-component order parameter fixes the class, and the continuity is consistent with the
finite-size scaling of Kaneko \emph{et al.}~\cite{Kaneko2021} on the square lattice with a
longitudinal field and is found on the chain~\cite{ovchinnikov2003antiferromagnetic,Bonfim2019}; were
it instead first order, the curve would split into disconnected sheets and the transition stay first
order regardless.) This follows from the divergence, not from a universal law: above the upper
critical dimension the response is mean-field with finite $\xmat$, the classical criterion applies, and
the transition can stay second order---as in the $d\to\infty$ limit, matching Zhang
\emph{et al.}~\cite{Zhang2014}.

Two routes thus escape the first-order outcome. One is dimensional, just seen: above the upper
critical dimension the response is finite and the classical criterion applies. The other is a
property of the matter critical point itself: $\xmat$ stays
\emph{finite} there---a negative specific-heat exponent, $\alpha<0$. This is the \emph{generic}, though
not universal, situation for a continuous-symmetry-breaking ($O(n\!\ge\!2)$) order parameter in the
relevant dimensions---the triangular-lattice antiferromagnet in the $3$D-XY
class~\cite{IsakovMoessner2003} (worked out in \cref{sec:results}) is one instance---in contrast to the
discrete Ising case ($n=1$, $\alpha>0$), which is forced first order. But it is not special to
$O(n\!\ge\!2)$: a Berezinskii--Kosterlitz--Thouless transition (whose essential singularity leaves
$\xmat$ finite), and in principle any other critical point with $\alpha<0$---including more exotic ones
such as deconfined or spin-liquid criticality---escape on the same footing. What forces the first order
is the \emph{divergence} of $\xmat$, and only that.

\begin{figure}[tbp]
\centering
\setlength\abovecaptionskip{5pt}
\includegraphics{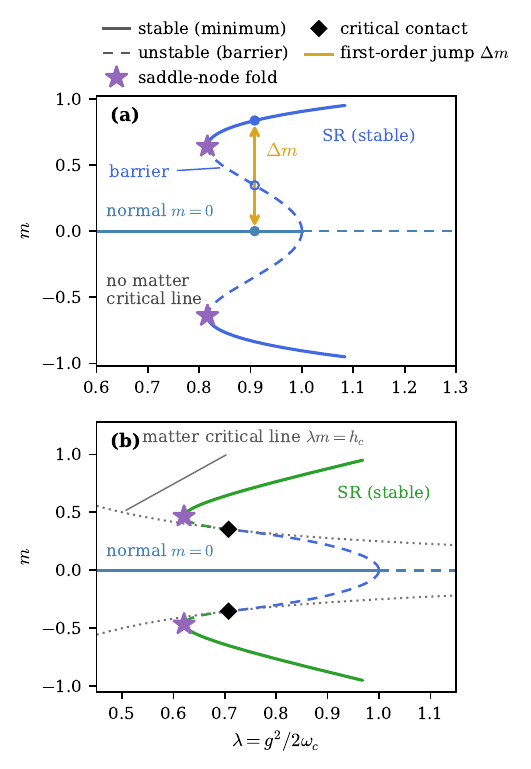}
\caption{The single-valued stationarity curve $\lambda(m)=\mumat^{-1}(m)/m$ (\cref{eq:statcurve})
and its two kinds of fold (schematic). \textbf{(a)} Shape fold: a finite but subcritically peaked $\xmat$
($a_4<0$) makes $\lambda(m)$ bend back with \emph{no} matter critical line. \textbf{(b)}
Larkin--Pikin fold: a $\xmat$ diverging on a matter critical line forces the back-bend; the
critical line $\lambda m=h_c$ (dotted) touches the curve tangentially (black diamond) on the
unstable dashed segment below the fold, since $d\lambda/dm=(\xmat^{-1}-\lambda)/m\to-\lambda/m$ as
$\xmat\to\infty$. In both, a
vertical $\lambda=\text{const}$ cuts the curve three times---stable normal ($m=0$), barrier
maximum, stable superradiant (dots in (a); the open dot is the barrier)---and the global minimiser
jumps by a finite $\Delta m$ (Maxwell
construction). \emph{Colour and marker convention, used in most of the stationarity-curve
figures:} branch \emph{colour} encodes the matter order---ordered (antiferromagnetic or
ferromagnetic, blue; the $m{=}0$ normal line steel-blue) versus polarised (green); \emph{line
style} encodes local stability (solid minimum, dashed barrier); a saddle-node fold is a purple
star, a continuous critical point a filled diamond, a tricritical
point a black star, and the first-order Maxwell jump is goldenrod (the $m{=}0$ onset $\lambda_c$ a
black dotted line).}
\label{fig:folds}
\end{figure}

\emph{Two kinds of fold.} The mechanism sorts the folds of \cref{sec:folds} into two kinds
(\cref{fig:folds}). \emph{(i) Shape fold:} $\xmat$ is finite everywhere but its shape---a broad
subcritical peak, equivalently $a_4<0$ in \cref{eq:landau}---makes $\lambda(m)$ bend back. There is
no matter critical line; the first-order Dicke transition is purely a property of the response
shape. \emph{(ii) Larkin--Pikin fold:} $\xmat$ \emph{diverges} on a matter critical line, and the
divergence just described forces the back-bend. Both are folds of the same curve \cref{eq:statcurve},
differing only in whether the susceptibility that produces them is finite or critical. An exactly
solvable instance of the second is the transverse-field Ising chain at zero longitudinal field,
whose uniform response diverges logarithmically at its critical field and folds the curve into a
first-order superradiant onset---the guided example of \cref{sec:results_guide}. In a
perpendicular-field extension of the same chain, Rao \emph{et al.}~\cite{Rao2025unilateral}
found the matching endpoint phenomenon. There the same logarithmic singularity enters through the
photon amplitude in quadrature, appearing as its fourth power times a logarithm, and pins a finite
jump at the endpoint of the Ising critical line with one-sided critical signatures. In the present
classification this is a fold pinned by a log-divergent response---the marginal ($\alpha=0$, logarithmic) case (Case B of \cref{app:lp}).

\subsection{Completeness: no hidden branch}
\label{sec:completeness}

\begin{figure*}[tbp]
\centering
\includegraphics[width=\linewidth]{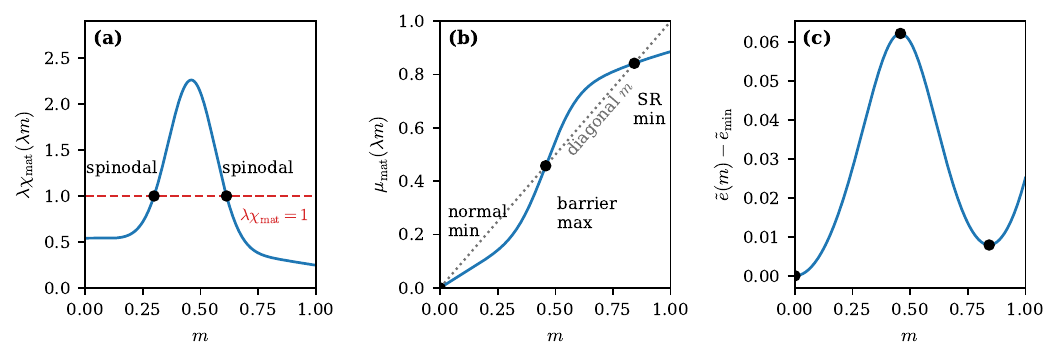}
\caption{The landscape is complete: no hidden superradiant branch (schematic, generic single-peaked $\xmat$). \textbf{(a)} A
single matter transition gives a \emph{single}-humped susceptibility; the level $1/\lambda$ (here
$\lambda\xmat=1$) cuts it \emph{at most twice}---at most two spinodals ($\evar''=0$)---regardless of the peak height.
\textbf{(b)} The self-consistent states are where the response $\mumat(\lambda m)$ meets the diagonal
$m$; with at most two spinodals between them, there are \emph{at most three}: normal minimum, barrier
maximum, superradiant minimum. \textbf{(c)} In the folded (first-order) regime the resulting landscape
$\evar(m)$ is min--max--min (shown for $m\ge0$; $\evar$ is even). The peak \emph{height} in (a) sets the order
(Larkin--Pikin, \cref{sec:lp}); the single \emph{hump} sets the count. A fourth phase would require a
second $\xmat$ peak, i.e.\ a second matter transition.}
\label{fig:completeness}
\end{figure*}

The Larkin--Pikin criterion is \emph{local}: it reads the curvature at the one critical point $m_c$.
Could the landscape hide a further superradiant phase---a second superradiant minimum somewhere we
have not looked? It cannot. The \emph{order} needed a number, the size of $\xmat(h_c)$ against
$1/\lambda_*$; the \emph{count} needs only the shape of $\xmat$. Differentiating \cref{eq:ddevar}
once more,
\begin{equation}
\evar'''(m)=-\lambda^3\,\xmat'(\lambda m),
\label{eq:eddd}
\end{equation}
so the inflections of the landscape sit exactly at the extrema of the susceptibility---and since the
field $h=\mumat^{-1}(m)$ rises strictly monotonically along the curve (\cref{sec:concavity}), the
shape of $\xmat$ in the field \emph{is} its shape along the curve. If $\xmat(h)$ falls monotonically
from $h=0$, $\evar''$ crosses zero at most once: at most one superradiant minimum beside the normal
one. If $\xmat(h)$ rises monotonically to a single peak and falls monotonically beyond it---a
\emph{single hump}---then by \cref{eq:eddd} $\evar''$ dips below zero at most once: the level
condition $\xmat(\lambda m)=1/\lambda$ has at most two roots, whatever the peak height
(\cref{fig:completeness}a). Between consecutive stationary points the curvature must vanish (Rolle),
so two such roots admit at most \emph{three} stationary points, and since $\evar\to+\infty$ at large
$m$ (the photon penalty) they can only be min--max--min---the normal minimum, the barrier, and
\emph{one} superradiant minimum (\cref{fig:completeness}b,c); past the pitchfork, where $m=0$ has
turned maximum, there are just two. A second superradiant phase---a third minimum---would require
$\evar''$ to dip twice, that is, a \emph{second} peak of $\xmat$. Matter with several consecutive
\emph{continuous} field-driven transitions extends the count phase by phase: each adds one $\xmat$
peak and at most one further min--max pair.

What the single hump rests on, stated as assumptions: (i) concavity of $\emat$---proven
(\cref{sec:concavity}); (ii) the matter orders \emph{continuously}---a first-order bare transition
is the disconnected case of \cref{sec:folds}, where the count proceeds piece by piece; (iii) $\xmat$
has no second, smooth, \emph{non-critical} hump. The critical peak itself is not assumed---a single
matter transition \emph{produces} it ($\xmat$ diverging, or cusping, at criticality); the only
genuine assumption is (iii), equivalently that the bare matter has a \emph{single} field-driven
transition, with $\xmat$ monotone within each phase and no interior plateau ($\xmat>0$ strictly,
\cref{sec:concavity}). Granted the single hump, a second superradiant phase would need a second
\emph{matter} transition, which the cavity never supplies---the precise sense in which it creates no
phase the matter does not already have (\cref{sec:decoupling}). Assumption (iii) is a genuine restriction---it excludes matter with several field-driven
transitions or magnetisation plateaus---and it is not proven here but \emph{checked} for each
model, from its computed $\xmat(h)$, in \cref{sec:results}.

This is the guarantee the local criterion lacked: \emph{whatever} order the magnitude assigns at
$m_c$, the Larkin--Pikin maximum is the \emph{only} maximum, the Maxwell construction between the
normal and superradiant branches is exhaustive, and the onset destabilises into a \emph{single}
superradiant phase whose order is settled by $a_4$ and Larkin--Pikin alone. The same susceptibility
curve carries both pieces of information at once: its peak \emph{height} fixes the order, its single
\emph{hump} fixes the count. Where the matter response is marginal or critical---above all the
superradiant--superradiant transition near the quadruple point of \cref{sec:results}, where
$\lambda\to0$ disables perturbation theory---the order there follows from the Larkin--Pikin mechanism (the divergent susceptibility, \cref{sec:lp}), with the finite jump confirmed by direct numerics; with that, these handles fix the onset, the order, and the count of every phase. That is the payoff of reducing the whole problem
to one stationarity curve: the cheap local quantities are handles on one global object.

\medskip
\noindent\fbox{\parbox{0.96\columnwidth}{\small
\emph{The handles, for use in \cref{sec:models,sec:results}:} onset at $\lambda_c=1/\xmat(0)$
[\cref{eq:lambdac_general}], its order from the sign of $a_4$ [\cref{eq:landau}]; bare criticality
pinned at $\lambda_*=h_c/m_c$ [\cref{eq:lambdastar}], its order from $\xmat(h_c)$: divergent
$\Rightarrow$ first order, finite $\Rightarrow$ classical criterion $\xmat(h_c)\gtrless1/\lambda_*$
(\cref{sec:lp}); jumps located by the Maxwell equal-energy rule (\cref{sec:folds}); the phase count
by the single hump of $\xmat$ (\cref{sec:completeness}).}}

\section{The models}
\label{sec:models}

Because the whole superradiant phase diagram is the bifurcation structure of the one functional
$\evar(m)=\tfrac\lambda2 m^2+\emat(\lambda m)$ (\cref{sec:decoupling,sec:general}), a cavity-coupled
magnet is specified entirely by its bare-matter response $\emat(h)$---the ground-state energy density
in the self-consistent field $h=\lambda m$. The models below are, in that sense, a gallery of
responses. For each we give the cavity Hamiltonian, its cavity-decoupled matter
$\hat H_m-h\sum_i\sigma^z_i$, and $\emat$, recalling how the bare matter is solved, its universality
class, and whether a field selects order by disorder---and defer the cavity consequences (which
response class of \cref{sec:landau} each realises, and the resulting phase structure) to
\cref{sec:results}. (For the non-Ising chains of \cref{sec:model_compass,sec:model_iso} and the
frustrated antiferromagnet of \cref{sec:model_tri} we keep the paper-wide normalisation $\omega_c=1$, $\lambda=g^2/2\omega_c$, the
cavity coupled to $\hat S_z=\tfrac12\sum_i\sigma^z_i$ as in \cref{eq:H_general}, and $m=\langle\sigma^z\rangle$
the cavity-axis magnetisation.)

\subsection{The Dicke--Ising model}
\label{sec:model_DI}

The central model couples the cavity to an Ising magnet on a hypercubic lattice (coordination
$z=2d$) in a longitudinal field,
\begin{equation}
\hat H = \omega_c\,\adag\aop + \frac{g}{\sqrt N}(\aop+\adag)\,\hat S_z
+ J\!\sum_{\langle ij\rangle}\!\sigma^x_i\sigma^x_j - \varepsilon\sum_i\sigma^x_i ,
\qquad \hat S_z=\tfrac12\sum_i\sigma^z_i ,
\label{eq:H}
\end{equation}
with $\varepsilon$ a longitudinal field along the Ising ($\sigma^x$) axis; we treat both the
antiferromagnet ($J>0$) and the ferromagnet ($J<0$). Integrating
out the cavity (\cref{sec:decoupling}) leaves the bare chain in the self-consistent transverse field
$h=\lambda m$,
\begin{equation}
\hat H_{\mathrm{mat}}(h) = J\!\sum_{\langle ij\rangle}\!\sigma^x_i\sigma^x_j - \varepsilon\sum_i\sigma^x_i
- h\sum_i\sigma^z_i ,
\label{eq:H_DI_mat}
\end{equation}
whose ground-state energy density is $\emat(h)$ and whose magnetisation $m=\langle\sigma^z\rangle$
along the cavity axis is the superradiant order parameter; the conventions used throughout are
collected in \cref{tab:conventions}.

\begin{table}[tb]
\caption{Conventions used throughout. All couplings are in Pauli ($\sigma$) units unless a value
is tagged ``$S=\tfrac12$''; $\omega_c=1$ in all numerical values.}
\label{tab:conventions}
\begin{ruledtabular}
\begin{tabular}{lp{0.70\textwidth}}
$\sigma^a$ & Pauli matrices; for $S=\tfrac12$ spins $\sigma^a=2S^a$. \\[2pt]
$\lambda\equiv g^2/2\omega_c$ & collective coupling (energy units). \\[2pt]
$m\equiv\langle\sigma^z\rangle$ & cavity (superradiant) magnetisation, cavity axis $\sigma^z$;
\cref{app:1d} and the triangular figures rotate the cavity onto $\sigma^x$---the same physical
$m$. \\[2pt]
$\varepsilon$ & longitudinal Ising field, $-\varepsilon\sum_i\sigma^x_i$ (no factor $\tfrac12$). \\[2pt]
$h=\lambda m$ & self-consistent cavity field on the matter ($h_z$ in the blockade chain). \\[2pt]
$\delta\equiv\varepsilon-2|J|$ & quadruple-point (QP) detuning; ray $r=\lambda/2|\delta|$ (one flip
costs $2\delta$); reused as a local expansion variable in \cref{app:lp}. \\[2pt]
QP rays & quoted in the $(\delta,\lambda)$ plane; the $(\delta,h_z)$ and $(\delta,g^2)$ variants
are linked by $h_z=\lambda m$, $g^2=2\lambda$ (\cref{app:idmrg}). \\
\end{tabular}
\end{ruledtabular}
\end{table}

The symmetry, and hence the phase table, differs between the two signs of $J$. For the
\emph{antiferromagnet} ($J>0$) there are \emph{two} independent $\mathbb Z_2$ symmetries---the
sublattice (staggered) $\mathbb Z_2$ of the matter order $\langle\sigma^x_A-\sigma^x_B\rangle$, and the
photon-parity $\mathbb Z_2$ under which the superradiant magnetisation $m$ is odd---so the model is
$\mathbb Z_2\times\mathbb Z_2$, and switching each order on or off independently gives the $2\times2$
table of phases,
\begin{center}
\begin{tabular}{lcc}
\hline
 & $m=0$ (normal) & $m\neq0$ (superradiant)\\
\hline
no antiferromagnetic order & PN & PS\\
antiferromagnetic order & AN & AS\\
\hline
\end{tabular}
\end{center}
polarised normal (PN), polarised superradiant (PS), antiferromagnetic normal (AN), and antiferromagnetic
superradiant (AS), in which superradiance coexists with staggered (antiferromagnetic) order; the four
boundaries cross at the \emph{quadruple point} $\varepsilon=2|J|$, $g=0$
(\cref{sec:model_QP}). The two-$\mathbb Z_2$ structure is itself a geometric choice: in the
perpendicular-field chain of Rao \emph{et al.}~\cite{Rao2025unilateral} cavity and staggered
order share a \emph{single} $\mathbb Z_2$, the AS phase is forbidden by symmetry, and the
critical line ends on the first-order boundary instead. The \emph{ferromagnet} ($J<0$) has no staggered order---only the photon-parity
$\mathbb Z_2$ survives---so its diagram is the single normal/superradiant column PN/PS, the two phases
separated by a line that turns first order below the tricritical point $\varepsilon_{\mathrm{tri}}^{\mathrm{ferro}}=|J|/d$ (\cref{sec:results_tri}). Both signs are treated in
\cref{sec:results}, the low-field analysis being organised by the reference vacuum---polarised (the
ferromagnet, and the antiferromagnet at strong field $\varepsilon>2|J|$) or staggered (the antiferromagnet
at weak field)---which is what gives the distinct ferromagnetic and antiferromagnetic tricritical
structures there. What is known of the bare matter \cref{eq:H_DI_mat}: the onset of
antiferromagnetic order under the transverse field $h=\lambda m$ is a $\mathbb Z_2$-symmetry-breaking transition in the $(d{+}1)$-dimensional Ising
universality class (the quantum--classical mapping~\cite{Suzuki1976})---the
divergent-susceptibility case of \cref{sec:lp}. In one dimension at
$\varepsilon=0$ the matter is the transverse-field Ising chain---exactly solvable in closed
form~\cite{Pfeuty1970}, the guided example of \cref{sec:results_guide}. The full Dicke--Ising phase diagram has been mapped numerically by wormhole quantum Monte
Carlo~\cite{langheld} and by linked-cluster plus DMRG~\cite{leibig2025}.

\subsection{The quadruple point: a detuned Rydberg-blockade chain}
\label{sec:model_QP}

At the quadruple point of the Dicke--Ising diagram ($\varepsilon=2|J|$, $g=0$) the matter
reorganises. Writing $\delta=\varepsilon-2|J|$, at small $\delta$ and weak coupling the low-energy
states are the \emph{independent-set} configurations---no two neighbouring spins up---a manifold of
dimension $\sim\varphi^N$ in one dimension ($\varphi$ the golden ratio) and $\sim1.503^N$ in two~\cite{BaxterEntingTsang1980}. In one
dimension this constrained manifold is the \emph{Fibonacci Hilbert space}: the number of independent
sets is a Fibonacci number, exactly the dimension of the fusion space of Fibonacci anyons, whose fusion
rule $\tau\times\tau=\mathbf 1\oplus\tau$ generates the golden-ratio
growth~\cite{Feiguin2007,chepiga2019dmrg}. (The Hamiltonian \cref{eq:IS} below is the hard-boson chain,
not the antiferromagnetic golden chain, so its criticality is the $2$D-Ising point, $c=\tfrac12$,
not the golden chain's tricritical Ising, $c=\tfrac7{10}$.)
Projecting the cavity perturbation onto this manifold gives a detuned, self-consistent
Rydberg-blockade (PXP) chain,
\begin{equation}
\hat H_{\mathrm{IS}} = 2\delta\sum_i n_i - h_z\sum_i \tilde X_i,\qquad
\tilde X_i = P_{i-1}X_iP_{i+1},
\label{eq:IS}
\end{equation}
where $P_j=(1-n_j)$ projects site $j$ onto the empty state, $n_i$ counts up-spins along the Ising ($\sigma^x$)
axis, $\tilde X_i$ flips spin $i$ where both neighbours are empty---the restriction to the
manifold of the cavity-axis flip, whence the name---and the field $h_z=\lambda m$ (the cavity
field of \cref{eq:functional}, restricted to the manifold) is again set self-consistently (the
factor of two in $2\delta$ is the cost of creating one up-spin in the blockade). What is known:
this is exactly the hard-boson model of Fendley \textit{et al.}~\cite{fendley2004competing} at
zero next-neighbour repulsion, and the bare chain undergoes a single second-order transition in
the $2$D-Ising universality class~\cite{fendley2004competing,chepiga2019dmrg}. Along
the perpendicular ray $\delta=0$ it reduces to the pure PXP chain
$-h_z\sum_i\tilde X_i$---the disorder-by-disorder ray analysed in \cref{sec:results_qp}, where
this corner is examined directly and the order of its transitions is settled.

\subsection{Frustrated antiferromagnets: the triangular lattice}
\label{sec:model_tri}

Putting a geometrically frustrated Ising antiferromagnet on the cavity at $\varepsilon=0$,
\begin{equation}
\hat H = \omega_c\,\adag\aop + \frac{g}{\sqrt N}(\aop+\adag)\,\hat S_z
+ J\!\sum_{\langle ij\rangle}\!\sigma^x_i\sigma^x_j \qquad(\text{triangular lattice}),
\end{equation}
the cavity-decoupled matter is the \emph{transverse-field} Ising antiferromagnet
$J\sum_{\langle ij\rangle}\sigma^x_i\sigma^x_j - h\sum_i\sigma^z_i$, with the cavity field $h=\lambda m$
playing the role of the transverse field. What is known: the classical ($h=0$) frustrated ground
manifold is extensively degenerate, with a finite residual entropy. A transverse field lifts this
degeneracy by \emph{order by disorder}---selection by quantum fluctuations rather than by an added
interaction~\cite{MoessnerSondhiChandra2000}---and on the triangular lattice it selects the
three-sublattice clock-ordered ($\sqrt3\times\sqrt3$)
state~\cite{MoessnerSondhiChandra2000}. As the field grows, that clock order melts in a
transition belonging to the $3$D-XY universality
class~\cite{IsakovMoessner2003,Humeniuk2016}. On the kagome lattice the same mechanism instead selects a \emph{disordered} paramagnet---disorder by
disorder~\cite{MoessnerSondhiChandra2000,Priour2001}.

\subsection{The compass chain}
\label{sec:model_compass}

The bond-alternating compass chain is, written in spin operators,
\begin{equation}
\hat H_{\mathrm{compass}} = J_1\!\!\sum_{\text{even bonds}}\!\!\sigma^x_i\sigma^x_{i+1}
+ J_2\!\!\sum_{\text{odd bonds}}\!\!\sigma^y_i\sigma^y_{i+1} ,
\end{equation}
alternating $\sigma^x\sigma^x$ and $\sigma^y\sigma^y$ bonds along the chain; coupled to the cavity it
acquires the $\omega_c\adag\aop+(g/\sqrt N)(\aop+\adag)\hat S_z$ terms, and the decoupled matter is
$\hat H_{\mathrm{compass}}-h\sum_i\sigma^z_i$. What is known~\cite{brzezicki2007,eriksson2009}: writing the bond asymmetry as
$J_1=1+\Delta$, $J_2=1-\Delta$, the symmetric point $\Delta=0$ is gapless and maximally frustrated;
an extensively degenerate ground manifold---exact zero modes, inert in the cavity-coupled
channel---persists at every $\Delta$, while $\Delta\neq0$ opens a gap in the dispersive band the
cavity couples to. It is nonetheless
exactly solvable---a Jordan--Wigner transformation maps the spins to free fermions, so $\emat(h)$ is
obtained in closed form. Because the chain contains no term odd under $\sigma^z\to-\sigma^z$,
$\emat(h)$ is \emph{even} in $h$, with no linear cusp.

\subsection{Isotropic chains: Heisenberg and XX}
\label{sec:model_iso}

The antiferromagnetic Heisenberg chain and the XX chain,
\begin{equation}
\hat H_{\mathrm{Heis}} = J\sum_i \vec S_i\!\cdot\!\vec S_{i+1}-h_x\sum_i\sigma^x_i,\qquad
\hat H_{\mathrm{XX}} = \tfrac12\sum_i\big(\sigma^x_i\sigma^x_{i+1}+\sigma^y_i\sigma^y_{i+1}\big),
\end{equation}
coupled to the cavity through $\hat S_z=\tfrac12\sum_i\sigma^z_i$, decouple to the chain in a uniform field
$\hat H_{\mathrm{Heis/XX}}-h\sum_i \sigma^z_i$. Only the Heisenberg Hamiltonian carries a transverse field $h_x$; the XX chain stays at $h_x=0$, for the reason taken up at the end of this subsection. What is known: the Heisenberg chain is solved by the Bethe
ansatz~\cite{Bethe1931} and the XX chain by a Jordan--Wigner transformation to free fermions, the latter giving
$m(h)$ in closed form (saturating at $h=1$; \cref{sec:results_iso}). Both chains are gapless---but the gaplessness lives at nonzero wavevector (the antiferromagnetic point
for Heisenberg, the Fermi points for XX) and in the spin correlations, not in the uniform ($q=0$)
channel the cavity couples to, so the
\emph{uniform} susceptibility is \emph{finite}: $\xmat(0)=4/\pi^2$ for Heisenberg and $2/\pi$ for XX.
The field $h_x$---transverse to the cavity axis, the role the longitudinal field $\varepsilon$
plays for the Ising matter---and the self-consistent cavity field $h$ combine, because the exchange
is SU(2)-symmetric, into a single tilted field along $\hat n\propto(h_x,0,h)$. This breaks the SU(2)
of the exchange down to the U(1) of rotations about $\hat n$, whose component stays conserved, so the
Bethe solution carries over unchanged (\cref{sec:results_iso}); the XX chain, lacking the SU(2),
admits no such reduction and stays at $h_x=0$.

\section{Results: the cavity phase structure}
\label{sec:results}

This section feeds the bare-matter energies $\emat$ of \cref{sec:models} through the machinery of
\cref{sec:general}. We begin with the one case where every step is exact---the one-dimensional chain
at $\varepsilon=0$ (\cref{sec:results_guide})---and then read all the models against the general onset
classification (\cref{sec:results_map}). The bulk of the section then works out the central model, the
Dicke--Ising model in general dimension (\cref{sec:results_DI}); the remaining cavity-coupled magnets
follow in \cref{sec:results_beyond}.

Most boundaries are analytic, and the analytic work is three handles on one object---the collective
$\hat S_z^2$ problem of \cref{eq:Sz2}: a low-field series about the $m=0$ vacuum (a
linked-cluster series in Takahashi's formalism~\cite{Kato1949,Takahashi1977}, carried to sixth
order (\cref{app:takahashi}), the order at which the tricritical character is fixed; the onsets and
the tricritical points, \cref{sec:results_tri}), a high-field series about the saturated state that
$\lambda\to\infty$ prepares (the triangular lattice, \cref{sec:results_triangular}), and a $1/d$
expansion about the self-consistent mean field in between (\cref{sec:results_1d}). The perturbative
route to the cavity problem was first set up in the precursor work~\cite{LeibigBSc}; the
structureless collective limit (Lipkin--Meshkov--Glick, where $1/N$ is the small
parameter~\cite{Dusuel2004,Dusuel2005}) is no precedent for it---there is no dimension and there are
no bonds. These expansions deliver every \emph{onset}---including the rays at the quadruple point
itself, whose PN--PS and AN--AS angles follow from the same $a_2$ machinery
(\cref{sec:results_qp})---and the $1/d$ expansion delivers the AS--PS line too, in closed form at
large $d$ (\cref{sec:results_1d}). What no expansion reaches is that line in the \emph{physical}
dimensions $d=1,2,3$: its exact location at $d=1$, and its order, which turns on a susceptibility divergence
that is non-analytic in $1/d$ (\cref{sec:lp}). There the Dicke--Ising chain---non-integrable even
in one dimension---is solved numerically, by infinite-system density-matrix renormalisation group
(iDMRG/VUMPS, variational uniform matrix product states~\cite{White1992,ZaunerStauber2018}, in the MPSKit.jl library~\cite{MPSKit}), away from
the corner and at it. The full derivations are
collected in \cref{app:a4,app:1d,app:lp,app:idmrg}.

\subsection{A guided example: the one-dimensional chain at \texorpdfstring{$\varepsilon=0$}{eps=0}}
\label{sec:results_guide}

It is worth seeing the whole machinery on the simplest case first, where every step is exact: the
one-dimensional Dicke--Ising chain at zero longitudinal field ($\varepsilon=0$, $d=1$). The bare
matter \cref{eq:H_DI_mat} is then the transverse-field Ising chain---the cavity field $h=\lambda m$
\emph{is} the transverse field, and $m=\langle\sigma^z\rangle$ is the superradiant order parameter,
solved exactly by Jordan--Wigner. Its uniform magnetisation $\mumat(h)=\langle\sigma^z\rangle$ is
known in closed form, vanishing in the Ising-ordered state at $h=0$ and rising toward saturation as
the field grows, with a quantum critical point at $h_c=J$, a continuous transition to the polarised phase; the uniform
susceptibility $\xmat=-\emat''$ \emph{diverges logarithmically} there (the $2$D-Ising universality of
the $(1{+}1)$-dimensional critical point).

\begin{figure*}[tbp]
\centering
\includegraphics[width=\linewidth]{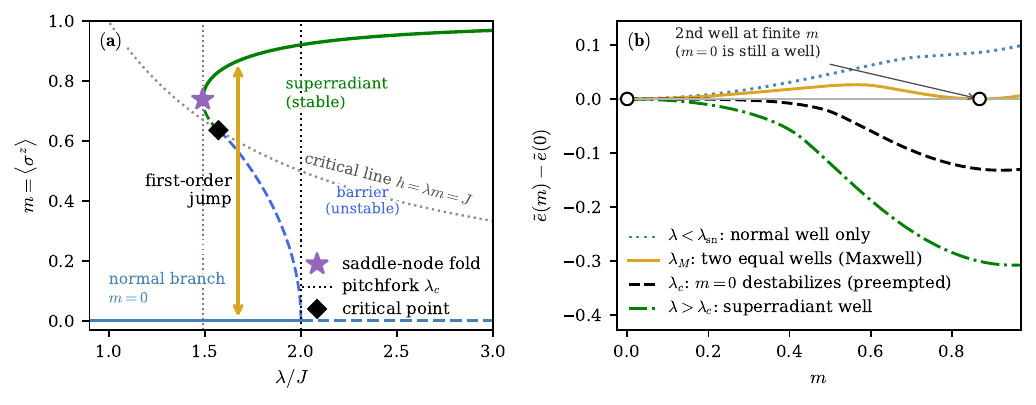}
\caption{The guided example: the one-dimensional Dicke--Ising chain at $\varepsilon=0$, bare matter the
transverse-field Ising chain (exact, Jordan--Wigner). \textbf{(a)} The single connected stationarity
curve $\lambda(m)=\mumat^{-1}(m)/m$ in the coupling--magnetisation plane; on the curve the abscissa
equals the field per unit magnetisation, $\lambda=h/m$, since the self-consistent field is
$h=\lambda m$. From the pitchfork at $m=0$
($\lambda_c=2|J|$) the unstable \emph{barrier} branch (royalblue, dashed) bends back toward smaller coupling to a
\emph{saddle-node fold} at $(\lambda_{\mathrm{sn}},m)\approx(1.49,0.74)$ and turns into the stable
\emph{superradiant} branch (green); it crosses the bare critical line $h=\lambda m=J$ (grey dotted) at the
Ising critical point $(\pi/2,2/\pi)$, where the logarithmically divergent susceptibility forbids a
minimum, so the curve passes as a maximum. Grey labels mark the Ising-ordered ($h<J$) and polarised
($h>J$) sides of the bare critical line. The onset is \emph{first order}: the normal state ($m=0$) jumps to the superradiant
branch at the Maxwell coupling $\lambda_M$ (arrow), preempting the pitchfork.
\textbf{(b)} The landscape $\evar(m)$ at representative couplings---a single normal well below the
fold, degenerate wells at $\lambda_M$, the destabilised (preempted) landscape at the bare pitchfork $\lambda_c$, and the superradiant well beyond.}
\label{fig:tfim}
\end{figure*}

Feeding this $\emat$ through \cref{sec:general} reproduces every element of the general analysis in
one picture (\cref{fig:tfim}). The photon penalty holds a normal phase up to $\lambda_c=1/\xmat(0)=2|J|$,
where $m=0$ loses stability; but the response about $m=0$ is \emph{subcritical} ($a_4<0$), so the stationarity curve
$\lambda(m)=\mumat^{-1}(m)/m$ leaves the axis toward \emph{smaller} coupling as an unstable barrier
branch and bends back at a saddle-node fold into the stable superradiant branch (\cref{fig:tfim}a).
The bare critical point $(m_c,\lambda_*)=(2/\pi,\pi/2)$ sits on that unstable branch, where the
divergent susceptibility \emph{forbids a minimum}---the curve can only pass as a \emph{maximum} (the
Larkin--Pikin mechanism of \cref{sec:lp}). The onset is therefore \emph{first order}: the normal state
jumps to the superradiant branch at the Maxwell coupling $\lambda_M\approx1.673\,J$ (between
the saddle-node fold at $1.49\,J$, where $\lambda\xmat(\lambda m)=1$, and the pitchfork at $\lambda_c=2|J|$), by a finite $\Delta m\approx0.87$,
preempting the pitchfork (\cref{fig:tfim}b). The same functional gathers this into one object---the
stationarity curve folds back at the saddle node, and the Larkin--Pikin maximum stands \emph{above} the
normal-state ($m=0$) energy, the barrier separating the two wells (\cref{fig:tfim}b). This
first-order superradiant onset is consistent with the
matter-interaction mechanism of Lee and Johnson~\cite{lee2004}, the displaced-frame analysis of Rohn
\emph{et al.}~\cite{Rohn2020}, the wormhole quantum Monte Carlo phase diagram of Langheld
\emph{et al.}~\cite{langheld}, and the independent exact coherent-state solution of Rao
\emph{et al.}~\cite{Rao2025unilateral}, whose perpendicular-field model reduces at $B_y=0$ to
this chain; it is the divergence-forced fold of \cref{sec:lp} in its most
transparent, exactly solvable form.

That the magnetically ordered ($m=0$) and superradiant ($m\neq0$) phases break \emph{different}
$\mathbb Z_2$ symmetries already rules out a \emph{direct} second-order transition between them. True,
but it misses what lies between: the ordered-superradiant state is present on the connected curve as
its \emph{unstable} barrier branch, and the first order is the fold across that branch (the
Larkin--Pikin maximum just traced). The intermediate is never absent, only here unstable; the
symmetry argument sees the two endpoints without the curve that joins them.

Its spectral footprint has been computed as well: the polariton
spectrum obtained for this very chain by Rom\'an-Roche
\emph{et al.}~\cite{roman2025linear,roman2025bound} shows the lower polariton
\emph{harden} rather than soften across the transition---the first-order jump preempts the
softening, and the spinodal where the mode would go soft (\cref{sec:spectral}) is never reached by
the equilibrium branch. Their transition point, $\lambda\approx1.674\,J$ in our units, matches the
Maxwell coupling $\lambda_M\approx1.673\,J$ above. The softening the equilibrium branch
escapes is not gone, only displaced onto the \emph{metastable} continuations. There the soft-mode
reading of \cref{sec:spectral} still holds, since the bare matter stays gapped along them. The normal
branch goes soft at its spinodal $\lambda_c=1/\xmat(0)=2|J|$, with onset exponent
$\Omega\propto|\lambda-\lambda_c|^{1/2}$; the superradiant branch goes soft at its saddle-node fold
$\lambda_{\mathrm{sn}}\approx1.49\,J$, with $\Omega\propto|\lambda-\lambda_{\mathrm{sn}}|^{1/4}$. This
quarter-power is a property of the fold: $\Omega^2=Z\evar''$ (\cref{sec:spectral}), and at the saddle
node $\evar''\to0$ linearly in $m-m_{\mathrm{sn}}$ while the magnetisation approaches the fold as a
square root, $m-m_{\mathrm{sn}}\propto|\lambda-\lambda_{\mathrm{sn}}|^{1/2}$; hence
$\evar''\propto|\lambda-\lambda_{\mathrm{sn}}|^{1/2}$ and
$\Omega=\sqrt{Z\evar''}\propto|\lambda-\lambda_{\mathrm{sn}}|^{1/4}$. It is not
the order-parameter quarter-power of a tricritical onset (\cref{sec:results_tri}), which the mode does not inherit. The two spinodals bracket the Maxwell coupling, the fold sitting just past the
bare critical field where the divergence of \cref{sec:lp} forces it. Ramping the coupling across the
onset on the hysteretic branch then reveals the soft mode before the branch turns over---a
measurable precursor of the first-order switch, completing the hardening of the equilibrium branch
rather than contradicting it.

We start here because the $\varepsilon=0$ chain isolates the divergence-driven fold in its simplest
form. The $\varepsilon=0$ curve and the antiferromagnetic curve at
$\varepsilon>\varepsilon_{\mathrm{tri}}^{\mathrm{AF}}$ in one dimension are both single connected
curves whose superradiant--superradiant transition is forced first order by the same divergent
$(d{+}1)$-Ising susceptibility; they differ in how the curve leaves the onset. At $\varepsilon=0$
($a_4<0$) it leaves the axis as the unstable barrier and bends back at a \emph{single} saddle-node
fold, the onset itself first order. Once $\varepsilon>\varepsilon_{\mathrm{tri}}^{\mathrm{AF}}$
($a_4>0$) it leaves as a clean second-order pitchfork (the AN--AS onset), and the divergence
reappears downstream as an \emph{S-shaped} AS--PS transition with \emph{two} saddle-node folds
(\cref{sec:results_tri}). The curve of
\cref{fig:tfim} is the reference picture for the antiferromagnetic analysis below: the $1/d$
expansion (\cref{sec:results_1d}) tracks how its fold moves with dimension, and
\cref{sec:results_qp} examines the corner of the phase diagram---the quadruple point---where the
AS--PS line ends. With this concrete picture in hand we turn to the
general classification.

\subsection{The response-class map}
\label{sec:results_map}

The guided example traced one model's \emph{entire} curve. We now step back and classify just the
\emph{onset}---how each model first leaves the $m=0$ axis---across all of them, before returning to the
full curves model by model. Read against the leading-singularity classification of \cref{sec:landau},
each model of \cref{sec:models} declares its onset at a glance:
\begin{center}
\begin{tabular}{llll}
\hline
model & $\emat(h)-e_0$ as $h\to0$ & $\xmat(0)$ & onset \\
\hline
Dicke--Ising (gapped) & $-\tfrac12\xmat(0) h^2$ & finite & threshold $\lambda_c=1/\xmat(0)$ \\
triangular AF; quadruple point & $-c\,|h|$ (cusp) & divergent & immediate, $\lambda_c=0$ \\
compass, symmetric ($\Delta=0$) & $-\tfrac12 A h^2\ln\tfrac1{|h|}$ & log-divergent & BKT, $m\sim e^{-b/\lambda}$ \\
compass, gapped ($\Delta>0$) & $-\tfrac12\xmat(0) h^2$ & finite & threshold $\lambda_c$ \\
Heisenberg, XX & $-\tfrac12\xmat(0) h^2$ & finite & threshold $\lambda_c$ \\
\hline
\end{tabular}
\end{center}
The mechanism behind each row is the way the matter responds to the field. A \emph{gapped} chain
rearranges locally, so $\xmat(0)$ is finite and given by local perturbation theory---the regular row.
An extensively \emph{degenerate} ground manifold instead responds collectively, with $\xmat(0)$ fixed
not by a local calculation but by degenerate perturbation theory on the manifold; an \emph{odd}
(linear) response there makes $\xmat(0)$ diverge---the cusp row, onset at vanishing coupling---while an
\emph{even} ($\propto m^2$) response diverges only logarithmically (the marginal log row) or stays
finite when a gap opens (the regular row). Degeneracy by itself fixes neither the row nor the phase
selected, and the two axes are independent. The triangular antiferromagnet has the odd, cusp response
\emph{and} the field selects an ordered (clock) state---\emph{order} by disorder
(\cref{sec:model_tri}); the quadruple point has the same odd cusp but the field selects the
\emph{disordered} paramagnet---\emph{disorder} by disorder (\cref{sec:model_QP}); the compass manifold
has the even, marginal response and---like the quadruple point, not the triangular lattice---selects
a disordered state, disorder by disorder again, but
in the log/regular row rather than the cusp (\cref{sec:model_compass}). Three degenerate manifolds,
three fates.

Beyond the onset, what the cavity does to each \emph{matter} transition is governed by \cref{sec:lp}:
a divergent matter susceptibility forces the superradiant--superradiant transition first order; a
finite one decides by size instead---the transition stays continuous only if
$\xmat(h_c)<1/\lambda_*$, the original (classical) Larkin--Pikin inequality of \cref{sec:lp}, and
turns first order otherwise. The two parts
that follow work the consequences out---first the central Dicke--Ising model (\cref{sec:results_DI}:
tricriticality, the $1/d$ expansion and the antiferromagnetic phase, and the quadruple point), then the
other cavity-coupled magnets (\cref{sec:results_beyond})---and in every case the first-order jumps are
the folds of the single connected curve of \cref{sec:folds}, fixed by a Maxwell construction.

\subsection{The Dicke--Ising model}
\label{sec:results_DI}

\begin{figure}[tbp]
\centering
\setlength\abovecaptionskip{2pt}
\includegraphics{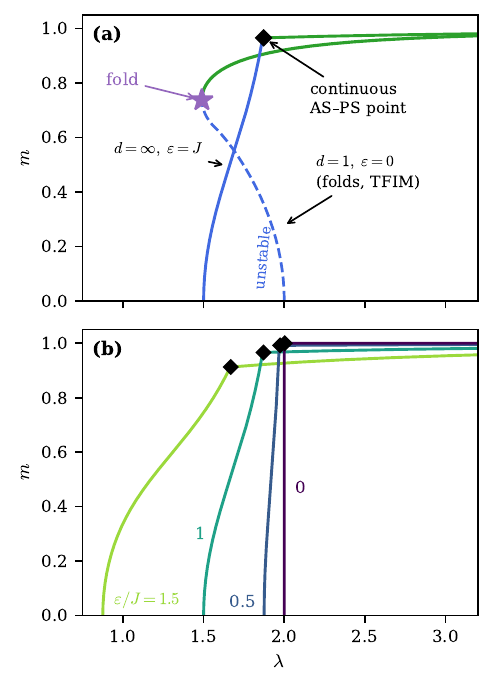}
\caption{The Dicke--Ising stationarity curve $\lambda(m)=\mumat^{-1}(m)/m$ with and without the
fold ($J=\omega_c=1$; diamonds mark continuous critical points, stars saddle-node folds).
\textbf{(a)} The same model in two limits, overlaid. At $d=1$, $\varepsilon=0$ (exact
transverse-field Ising matter) the subcritical response ($a_4<0$) sends the curve off the axis toward
smaller coupling as an unstable barrier branch (dashed); it crosses the bare critical line $h=\lambda m=J$,
where the logarithmically divergent susceptibility forbids a minimum and the curve passes as a maximum
(Larkin--Pikin), then folds back \emph{beyond it} at a saddle node (star) into the stable superradiant
branch, a first-order normal-to-superradiant onset (the fully annotated guided example, with the
critical line and Maxwell jump, is \cref{fig:tfim}). At $d=\infty$,
$\varepsilon=J$ (two-sublattice mean field, solid) the matter response is finite: the curve does not
fold back---each coupling supports a single magnetisation---and the AS--PS transition (diamond, where
the staggered order vanishes) is continuous. The fold is the finite-$d$, fluctuation-induced effect of \cref{sec:lp},
invisible to the $d=\infty$ mean field. \textbf{(b)} The $d=\infty$ antiferromagnetic family for
$\varepsilon/J=0,0.5,1,1.5$. The curve passes through each AS--PS critical point (diamond) with
finite positive slope (\cref{eq:Sinfty})---$\lambda(m)$ has no turning point, hence no fold---so the
AS--PS transition is second order, matching Zhang \textit{et al.}~\cite{Zhang2014}. The $\varepsilon=0$ curve is
the degenerate flat ridge $\lambda=2|J|$.}
\label{fig:dinf}
\end{figure}

We now treat the central model in depth, in general dimension $d$. Its phase structure has three distinct origins, and conflating them obscures the physics. (i) The
onsets out of the normal vacuum, ferromagnetic and antiferromagnetic alike, are continuous where the
next Landau coefficient permits and turn first order through a \emph{shape fold} of the stationarity
curve where it does not; the ferromagnetic tricritical point ($a_4=0$ at $m=0$, with $a_6>0$) is a
genuine, ordinary onset feature, while the antiferromagnetic $a_4=0$ locus ($a_6<0$) is instead the
first-order endpoint at which the second-order AN--AS line terminates (\cref{sec:results_tri}). (ii)
The antiferromagnetic AS--PS transition is the Larkin--Pikin case: a divergent matter susceptibility at
the bare AN--AS critical line folds the curve and drives it first order at finite $d$, invisible to the
$d=\infty$ mean field (\cref{sec:lp}). (iii) The quadruple point, where all four phases meet, is not a
fold at all but a line-crossing on the extensively degenerate matter manifold at $\varepsilon=2|J|$
(\cref{sec:results_qp}). We take these in turn, with \cref{fig:dinf}(a) as the organising picture: the same
stationarity curve, folded by the divergent susceptibility in one dimension, unfolded at $d=\infty$
where the mean-field response stays finite. Everything below fills in the picture between these two
limits.

\subsubsection{Ferromagnetic and antiferromagnetic tricriticality}
\label{sec:results_tri}

Away from the quadruple point the Dicke--Ising chain sits in the regular row: the bare matter is
gapped, $\xmat(0)$ is finite, and the superradiant onset is a continuous second-order line at
$\lambda_c=1/\xmat(0)$ (\cref{sec:landau}). Where that line turns first order is fixed by the next
Landau coefficient. Around a normal vacuum ($m=0$, $\langle\sigma^z\rangle=0$ exactly by symmetry)
the fourth-order coefficient is the competition of two cluster processes---a single-flip
contribution $W_1>0$ and an adjacent bound-pair contribution $W_{\mathrm{bond}}<0$,
\begin{equation}
a_4 = \lambda^4\big(W_1 + d\,W_{\mathrm{bond}}\big),
\end{equation}
each a closed-form function of the gaps $\Delta E_s,\Delta E_b$ above the vacuum and of the vacuum
type (\cref{app:a4}). The sign of $a_4$ decides the order through \cref{sec:landau}, and the
construction is vacuum-type-generic---the same mechanism governs the ferromagnet, the
antiferromagnet, and (in \cref{sec:results_beyond}) non-Ising matter.

The two clusters carry the entire microscopic content of the onset, and it is worth telling once in
plain terms. Here the normal vacuum is an \emph{uncorrelated product state}, so the picture of
\cref{sec:spectral} is literal: single flips on it are the matter mode of an effective Dicke model,
and $W_1>0$ (four flips on one site) is the single flip, renormalised---the part a structureless
collective model (Dicke, Lipkin--Meshkov--Glick) also has; by itself it would make every onset
continuous. The bond term is what one collective mode cannot see: two flips on \emph{adjacent}
sites share a bond, the pair costs $\Delta E_b<2\Delta E_s$---they bind---and the lattice supplies
$d$ such bonds per site, hence the $d\,W_{\mathrm{bond}}$ with
$W_{\mathrm{bond}}\propto(\Delta E_b-2\Delta E_s)<0$. The density bookkeeping is elementary: a flip density $\sim m^2$ makes single-flip
energetics the $m^2$ term and flip-\emph{pair} energetics the $m^4$ term, which is where the
Landau quartic lives. The competition is then transparent---one on-site repulsion against $d$ bond
attractions---and where the attractions win ($a_4<0$), this is the misprediction anticipated in
\cref{sec:spectral}, now with its mechanism in hand. One can of course absorb the binding into the
quartic of the collective mode---numerically that is what $a_4$ is---but its sign, and its linear
growth with $d$, come from spatial two-flip correlations that no single-mode truncation of the
matter supplies.

\paragraph{Ferromagnet.} For the polarised (PN) vacuum, in the rescaled units $J\to J/d$ under
which mean field is exact at $d=\infty$ (\cref{app:1d}), the single-flip gap is
$\Delta E_s=2\varepsilon+4|J|$ and the bound-pair gap $\Delta E_b=4\varepsilon+4(2-\tfrac1d)|J|$,
its $1/d$ correction the seed of what follows; the on-site flip energetics set the $m^2$ term, the
$d$ bond-binding terms the $m^4$, and
\begin{equation}
a_4^{\mathrm{ferro}} = \frac{4\lambda^4(\varepsilon-|J|/d)}{(\Delta E_s)^3\,\Delta E_b}
\;\Longrightarrow\;
\boxed{\;\varepsilon_{\mathrm{tri}}^{\mathrm{ferro}}=|J|/d\;}
\label{eq:eps_tri_ferro}
\end{equation}
($=|J|$ at $d=1$, $\to0$ as $d\to\infty$). In the second-order regime
$\varepsilon>\varepsilon_{\mathrm{tri}}^{\mathrm{ferro}}=|J|/d$ the onset line
$\lambda_c=\varepsilon+2|J|$ is exact; below it the PN--PS transition is first order. The
first-order window has width $|J|/d$ and \emph{closes as $d\to\infty$}: at $d=\infty$ the onset is
second order for every $\varepsilon>0$, the tricritical point sliding to $\varepsilon=0$. The
window is thus a genuine finite-$d$ effect, opened by the bound-pair (double-flip) channel of
\cref{app:boson} that the $1/d$ correction to $\Delta E_b$ first exposes. The value \cref{eq:eps_tri_ferro} for the tricritical field agrees, as a
cross-method check within the same project, with the numerical linked-cluster expansion plus DMRG (NLCE+DMRG) determination in $d=1$ to within $6\times10^{-4}$ in units of $|J|$~\cite{leibig2025}. The next coefficient settles its character:
at $a_2=a_4=0$ the sextic $a_6^{\mathrm{ferro}}>0$ in every dimension [\cref{eq:c6_tri}], so the
superradiant branch leaves $m=0$ as a minimum and \cref{eq:eps_tri_ferro} is a genuine,
\emph{ordinary} tricritical point. There the order parameter grows with the mean-field tricritical
exponent, $m\sim(\lambda-\lambda_c)^{1/4}$ (\cref{sec:landau})---but the soft polariton of
\cref{sec:spectral} does not inherit the $1/4$: on the emerging branch
$\evar''(m_*)=-8a_2\propto(\lambda-\lambda_c)$ in place of the $-4a_2$ of an ordinary pitchfork, so
the mode gap reopens with the ordinary exponent $\tfrac12$, only with a $\sqrt2$-enhanced amplitude.

\begin{figure}[tbp]
\centering
\includegraphics[width=\linewidth]{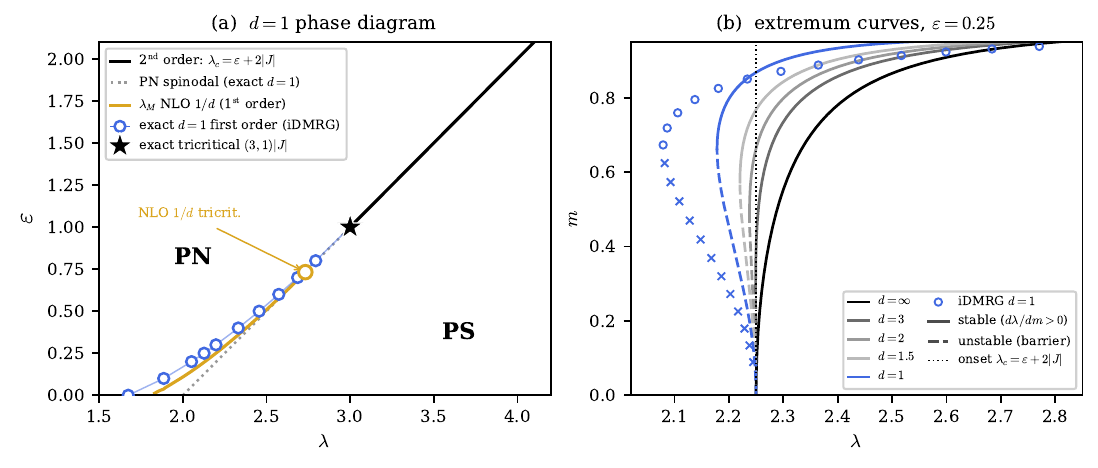}
\caption{Ferromagnetic phase diagram in the rescaled $d{=}1$ units of the text
($|J|=\omega_c=1$; $\lambda=g^2/2\omega_c$). \textbf{(a)} \emph{Solid line:} the exact second-order
PN--PS onset $\lambda_c=\varepsilon+2|J|$, valid above the tricritical point (star) at
$(\lambda,\varepsilon)=(3,1)\,|J|$ [\cref{eq:eps_tri_ferro}]. \emph{Dotted line:} its continuation
below the tricritical point as the PN spinodal (metastability limit). \emph{Open circles:} the numerically exact
$d=1$ first-order line from infinite-system DMRG (Maxwell construction), from
$\lambda_M=1.673\,|J|$ at $\varepsilon=0$ to the star, the jump shrinking toward the
tricritical point. \emph{Goldenrod line:} the next-to-leading-order $1/d$ first-order line (Maxwell on the
NLO functional), ending at its own NLO tricritical point (small circle) at
$(\sqrt3+1,\sqrt3-1)\,|J|\approx(2.73,0.73)$; its offset from the exact tricritical point is the
truncation artifact of \cref{sec:results_1d}, not a second transition. \textbf{(b)} Extremum curves
$\lambda(m)=h/m$ at fixed $\varepsilon=0.25$ for $d=\infty,3,2,1.5,1$: monotone (second order) at
$d=\infty$, folding for $d<|J|/\varepsilon=4$, the $d=1$ iDMRG curve (circles) folding most
strongly---the analytic picture of how the first-order window of width $|J|/d$ opens as $d$
decreases.}
\label{fig:ferro}
\end{figure}

\begin{figure}[tbp]
\centering
\includegraphics{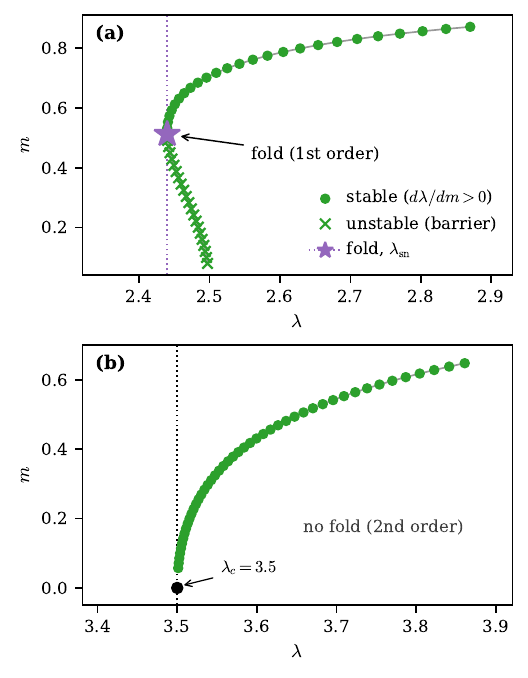}
\caption{Ferromagnetic stationarity curves $\lambda(m)=h/m$ from $1$D iDMRG ($J=-1$).
\textbf{(a)} For $\varepsilon=0.5<\varepsilon_{\mathrm{tri}}^{\mathrm{ferro}}=|J|$
the curve back-bends at a saddle-node fold at $(\lambda,m)\approx(2.44,\,0.51)$ (purple star)---a
shape fold, \cref{sec:lp} (first order); circles mark stable points ($d\lambda/dm>0$), crosses
the unstable barrier branch. \textbf{(b)} For
$\varepsilon=1.5>\varepsilon_{\mathrm{tri}}^{\mathrm{ferro}}$
the curve is monotone (second order, no fold), leaving the axis at $\lambda_c=3.5$ (black dot).
Because the bare ferromagnet at any $h>0$ has a single phase and no critical line, the
back-bend in (a) is a pure \emph{shape} fold (\cref{sec:lp})---not the Larkin--Pikin fold of a
diverging susceptibility. This is the iDMRG
realisation of the $a_4=0$ ferromagnetic tricritical point \cref{eq:eps_tri_ferro}.}
\label{fig:b5ferro}
\end{figure}

\Cref{fig:ferro} collects the resulting $d=1$ phase diagram. Panel (a) shows the exact second-order
onset $\lambda_c=\varepsilon+2|J|$ (valid for $\varepsilon\ge|J|$), the numerically exact $d=1$ first-order line
traced point-by-point by infinite-system DMRG (open circles, from the free-fermion value
$\lambda_M=1.673\,|J|$ at $\varepsilon=0$ to the tricritical point at
$(\lambda,\varepsilon)=(3,1)\,|J|$), and the $1/d$ next-to-leading-order (NLO) Maxwell line, which bends left of the
spinodal and ends at its own NLO tricritical point---displaced from the exact one by the
truncation, a visual reading of the caveat of \cref{sec:results_1d}. Panel (b) follows the extremum
curve as $d$ grows: at fixed $\varepsilon$ it folds for $d<|J|/\varepsilon$ and unbends to the
monotone $d=\infty$ curve, the analytic picture of the closing first-order window. Infinite-system
DMRG realises the change of order directly in one dimension (\cref{fig:b5ferro}): below
$\varepsilon_{\mathrm{tri}}^{\mathrm{ferro}}$ the stationarity curve back-bends at a saddle-node
fold and the transition is first order; above it the curve is monotone and leaves the axis at
$\lambda_c=\varepsilon+2|J|$, as the exact second-order line requires. In the taxonomy of \cref{sec:lp} this is a \emph{shape} fold (\cref{fig:folds}): the
longitudinal field breaks the matter $\mathbb{Z}_2$ explicitly, so no critical line lies beneath
the curve and the fold comes from the subcritical shape of the response alone---in contrast to the
Larkin--Pikin folds of the antiferromagnetic problem below (\cref{sec:results_1d}), which sit on a
genuine critical line.

The same ferromagnetic value and the underlying correlated nearest-neighbour double-spin-flip
mechanism were also obtained, from a Landau analysis of the same functional, by
Koziol~\cite{Koziol2026}. As established in \cref{sec:landau}, both the second-order line
$\lambda_c=1/\xmat(0)$ and the tricritical point \cref{eq:eps_tri_ferro} are exact
\emph{locations}---here fixed by the single coefficients $a_2$ and $a_4=\lambda^4(W_1+d\,W_{\mathrm{bond}})$ of
the $1/N$-free functional, independent of every higher-order term.\footnote{These results---the closed-form ferromagnetic and antiferromagnetic tricritical criteria, the
analysis of the antiferromagnetic superradiant phase, and the quadruple point below---were obtained by
the author in November~2025 during the preparation of Ref.~\cite{leibig2025} in the group of
K.~P.~Schmidt, recorded in the collaboration's shared Overleaf project where they remain in the
version history to this day, and withdrawn by the author from that contribution on 16~November~2025.
The finite-size scaling of the tricritical point above the upper critical dimension is the distinct
contribution of Ref.~\cite{Koziol2026}, complementary to the transition locations and order
established here.}

\begin{figure}[tbp]
\centering
\includegraphics[width=0.50\linewidth]{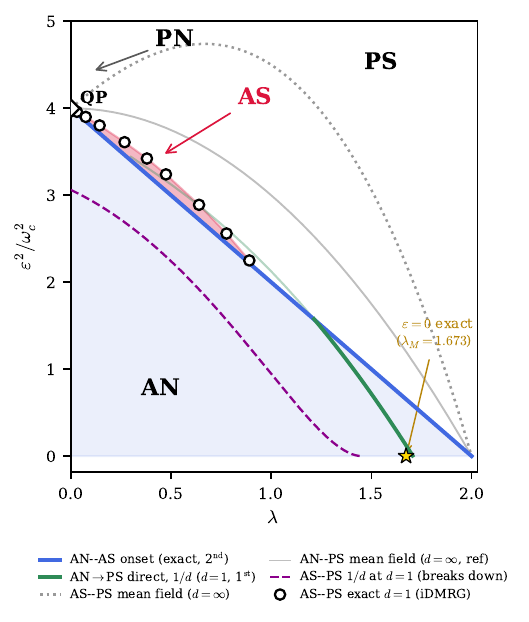}
\caption{Antiferromagnetic Dicke--Ising phase diagram in the rescaled units $|J|=\omega_c=1$,
$\lambda=g^2/2\omega_c$, in the $(\lambda,\varepsilon^2)$ plane.
\emph{Blue:} the exact second-order AN--AS onset $\varepsilon^2=2|J|(2|J|-\lambda)=4-2\lambda$,
straight and $d$-independent. \emph{Open circles:} the AS--PS transition from infinite-system DMRG,
numerically exact in $d=1$---saddle-node folds at $\varepsilon=1.5$--$1.8$ (\cref{fig:extremum_af})
and near-corner Maxwell crossings at $\varepsilon=1.85$--$1.99$ (\cref{fig:qpjumps})---bounding a
thin realised AS sliver (shaded) just above the AN--AS line. \emph{Green:} the first-order
AN$\to$PS line, where the polarised energy crosses the exact N\'eel energy $e_{\rm AN}=-|J|$,
at next-to-leading order in $1/d$ at $d=1$; because $e_{\rm AN}$ is exact this crossing is
clean (no tangency) and reaches $\varepsilon=0$ at $\lambda\approx1.70$, beside the
free-fermion Maxwell value $\lambda_M=1.673$ (gold star). The line is solid where realised ($\varepsilon<\varepsilon^*$) and
faint where it is the metastable AN--PS crossing ($\varepsilon>\varepsilon^*$, where AS wins): the
realised onset kinks from the AN--AS to the AN$\to$PS line at $\varepsilon^*$
($\approx1.26$ here, $\approx1.32$ from iDMRG). \emph{Grey dotted:} the $d=\infty$ mean-field AS--PS line, which
\emph{overestimates} the AS region; \emph{thin grey:} its AN--PS counterpart
$\varepsilon^2=4-\lambda^2$. \emph{Magenta dashed:} the AS--PS boundary from the $1/d$ shift formula---a controlled estimate of its location
and of how the AS region narrows toward $d=1$; it overshoots near $d=1$, where the location is read instead from
the iDMRG and the \emph{order} from the binding curvature and the iDMRG, not from this curve. The open diamond is the quadruple
point ($\lambda=0$, $\varepsilon=2|J|$); the opposite end
($\lambda=2|J|$, $\varepsilon=0$) is the zero-field pitchfork $\lambda_c=2|J|$
(\cref{sec:results_guide}).}
\label{fig:trikrit}
\end{figure}

\paragraph{Antiferromagnet.} For the antiferromagnetic (AN) vacuum at weak field $\varepsilon<2|J|$,
the two sublattice single-flip gaps $\Delta E_A=4|J|-2\varepsilon$, $\Delta E_B=4|J|+2\varepsilon$
are field-dependent but $d$-independent (they fix the $d$-independent onset
$\lambda_c=(4J^2-\varepsilon^2)/2|J|$), while sublattice cancellation makes the bound-pair gap
\emph{field-independent}, $\Delta E_b^{\mathrm{AF}}=4(2-\tfrac1d)|J|$, and
\begin{equation}
a_4^{\mathrm{AF}}=\frac{64\lambda^4 J^2\big[(8d-3)\varepsilon^2-4J^2\big]}{d\,(\Delta E_A\Delta E_B)^3\,\Delta E_b^{\mathrm{AF}}}
\;\Longrightarrow\;
\boxed{\;\varepsilon_{\mathrm{tri}}^{\mathrm{AF}}=\frac{2|J|}{\sqrt{8d-3}}\;},
\label{eq:eps_tri_AF}
\end{equation}
which retains $d$-dependence---a direct structural consequence of the field-independent
$\Delta E_b^{\mathrm{AF}}$, in contrast to the ferromagnet, where the single-flip and bound-pair
scalings combine into $\varepsilon_{\mathrm{tri}}^{\mathrm{ferro}}=|J|/d$. Both windows close as
$d\to\infty$, but at different rates---$|J|/d$ against $2|J|/\sqrt{8d-3}\sim1/\sqrt d$, the
antiferromagnetic one parametrically the wider---a contrast the bare units obscure.\footnote{Both
$a_4$ follow from the same cluster expansion of the quartic coefficient (\cref{eq:a4_master}): the ferromagnet from the polarised vacuum, the antiferromagnet from the N\'eel vacuum with sublattice-resolved gaps.} Here, by contrast, the sextic reverses
sign: at the antiferromagnetic $a_4=0$ point $a_6^{\mathrm{AF}}<0$ in every dimension [\cref{eq:c6_af}], so the
branch leaves $m=0$ as a \emph{maximum}, $m=0$ is locally unstable, and \cref{eq:eps_tri_AF} is not
an ordinary tricritical point but the first-order point at which the second-order AN--AS line
terminates. At strong field $\varepsilon>2|J|$ the vacuum is polarised and
$a_4^{\mathrm{AF,strong}}=4\lambda^4(\varepsilon+|J|/d)/[(\Delta E_s)^3\Delta E_b]>0$ throughout
(with the strong-field gaps $\Delta E_s=2\varepsilon-4|J|$, $\Delta E_b=4\varepsilon-4(2-\tfrac1d)|J|$), so the PN--PS
transition is second order with no tricritical point. The boundary $\varepsilon=2|J|$ is the quadruple
point (\cref{sec:results_qp}). \Cref{fig:trikrit} collects the resulting antiferromagnetic diagram,
whose AN--AS arc and AS--PS line are mapped out in \cref{sec:results_1d}.

\subsubsection{The \texorpdfstring{$1/d$}{1/d} expansion and the antiferromagnetic phase}
\label{sec:results_1d}

Two structural questions motivate a controlled expansion in the inverse coordination $1/d$
(rescaling $J\to J/d$, for which matter mean field is exact at $d=\infty$). First, is the
$a_4=0$ point of \cref{eq:eps_tri_AF}---where the AN--AS onset would turn locally unstable---ever
reached, in \emph{any} dimension, or does the first-order AN$\to$PS jump always preempt it? Second,
does the second-order AS--PS transition of the $d=\infty$ mean field survive at finite $d$? Both
get sharp answers below: the $a_4=0$ point is preempted in \emph{every} dimension
[$J_{\mathrm{cross}}>J_{\mathrm{tri}}$, \cref{eq:Jcross}], and the AS--PS line stays second order
in its interior at large $d$, bounded by two tricritical points near each end [$\mathcal{S}_{\mathrm{AS}}^{(\infty)}>0$, \cref{eq:Sinfty}]. What the expansion delivers is the
\emph{geometry} in the $(\lambda,m)$ plane---the four dressed boundaries and their $O(1/d)$
shifts---together with that large-$d$ statement about the order; what it cannot do is locate the
changeover to the physical dimensions, which turns on the onset of the susceptibility
\emph{divergence}, non-analytic in $1/d$ and invisible to the series at any order (the order is
settled numerically in one dimension, away from the corner, \cref{fig:extremum_af}, and at it,
\cref{sec:results_qp}; the NLO branch crossing is a valid $1/d$ estimate of the boundary's location
and shift, not an order indicator, \cref{app:1d}). Tracking how the four boundaries shift with $d$ is
a by-product, not the aim.

The $1/d$ crossings deserve a word, because they recur. A finite truncation of the series estimates
where each boundary sits and how it shifts with the dimension---for the AS--PS line, the
antiferromagnetic region narrowing toward small $d$---exactly as the same expansion shifts any
transition (the hypercubic transverse-field Ising critical field, for one). What it does \emph{not}
fix is the \emph{order} of a superradiant--superradiant transition: at $d=\infty$ the AS--PS
bifurcation is a second-order tangency, and the first order that appears in the physical dimensions is
forced by the susceptibility divergence, non-analytic in $1/d$ and invisible to the series at any
order. The order is therefore read off three handles, none of them a bare branch crossing: the exact
Landau coefficients where the onset is from $m=0$ (\cref{sec:results_tri}); the binding curvature
$\mathcal{S}_{\mathrm{AS}}$ where the series is analytic, at large $d$ (\cref{eq:Sinfty}); and the
numerics in the physical dimensions (the iDMRG folds of \cref{fig:extremum_af}, the Larkin--Pikin
criterion of \cref{sec:lp}). The one place a crossing genuinely misleads is the ferromagnetic NLO
line of \cref{fig:ferro}: there the exact analysis fixes the tricritical point at
$\varepsilon_{\mathrm{tri}}^{\mathrm{ferro}}=|J|/d$ ($a_6>0$), so its NLO-displaced tricritical point
is a truncation artifact, not a second transition.

What it \emph{can} do, it does with a single method, the same for every phase: rotate each site
onto its self-consistent mean-field axis; the rotated Ising bond then carries, besides a single-site piece
absorbed into the reference (which is precisely the mean-field condition), a pure two-site
fluctuation: the interaction now \emph{sits on the bonds}, and only bond terms expand cleanly in
$1/d$. A single bond at second order on top of the reference is the entire $O(1/d)$
(\cref{app:1d}), and because the reference is stationary, the terms linear in the fluctuation
cancel (the envelope theorem). All that distinguishes the phases is \emph{which} reference state
one rotates to---and for the AS phase that choice is the entire subtlety, taken up with the AS
phase below.

The polarised side is obtained the same way---the longitudinal field cants the polarised reference
too, so the rotation is already at work there---matter mean field plus single-bond second-order
perturbation theory (\cref{app:1d}); at $\varepsilon=0$ the (ferro/AF-identical) benchmark is
\begin{equation}
|J|_c^{\mathrm{AN\text{-}PS}}(0,d)=\tfrac14+\tfrac1{32d}+O(d^{-2})
\end{equation}
(in the scale-fixed units of \cref{app:1d}, $\lambda=\tfrac12$; invariantly
$J_c/\lambda=\tfrac12+\tfrac1{16d}$). Continued, uncontrolled, down to $d=1$, the benchmark reads
$\lambda_M=2|J|(1-\tfrac1{8d})=1.75\,|J|$ for the zero-field first-order
transition---within $5\%$ of the exact Maxwell value $1.673\,|J|$ of the chain
(\cref{sec:results_guide}), where leading order alone is $20\%$ off: the $O(1/d)$ term removes
three quarters of the mean-field error even where the expansion is least trusted.
The figure's $\lambda\approx1.70$ and this series value $1.75\,|J|$ are the same NLO-in-$1/d$ crossing read off two ways: the former solves $\evar^*_{\mathrm{PS}}(\lambda)=e_{\rm AN}$ numerically with the exact $e_{\rm AN}=-|J|$, while the latter truncates the root's $1/d$ series $\lambda_M=2|J|(1-\tfrac1{8d})$. They share identical energies and differ only at $O(d^{-2})$, the numerically solved $1.70$ being the more accurate.
The ferromagnetic PN--PS first-order line below $\varepsilon_{\mathrm{tri}}^{\mathrm{ferro}}$
follows from the same energy match; its phase diagram and its direct iDMRG realisation were shown
with the tricritical analysis (\cref{fig:ferro,fig:b5ferro}, \cref{sec:results_tri}).

\paragraph{Two routes, one functional.} The normal-phase Landau coefficients of
\cref{sec:results_tri} and the $1/d$ superradiant expansion of this section are two expansions of
the \emph{same} functional, transposed in their double series in $(m,1/d)$: the former exact in $d$
but truncated in $m$ (the flip-cluster $a_2,a_4,\dots$), the latter full in $m$ but truncated in
$1/d$ (mean field plus single-bond fluctuations). Where they overlap they must agree term by term,
and they do: expanding the superradiant functional about $m=0$ reproduces the closed-form $a_2$ and
$a_4$ of the ferromagnetic and antiferromagnetic PS onsets to the digit, including the $O(1/d)$
slopes. The antiferromagnetic $a_4$ comes out negative below
$\varepsilon_{\mathrm{tri}}^{\mathrm{AF}}=2|J|/\sqrt{8d-3}$, the counterpart of the ferromagnetic
$|J|/d$---wider, by the $1/\sqrt d$ scaling, and reinforced by $a_6<0$ into the first-order endpoint
of \cref{eq:eps_tri_AF}. The AS sheet taken up next is the only onset with no free limit to expand around, so no series of its
own fixes its finite-$d$ coefficient; requiring the two routes---the normal-phase Landau coefficients
and the $1/d$ superradiant expansion---to agree there determines it, a stringent check on both.

\paragraph{The AS phase (closed form).}
Here is the reference question the method statement deferred, and it is two difficulties at once:
what to expand \emph{around}---the self-consistent AS state is anchored to no free limit---and what
to expand \emph{in}---at fixed $d$ no physical coupling stays small. The bare ordered phase by
itself is unproblematic---its flip expansions start from the staggered
product vacuum, as in \cref{sec:results_tri}. The \emph{self-consistent} AS state at finite $m$ is
different: the field it sits in is the superradiant order it must itself sustain, so it is anchored
to no free limit at all---unlike the polarised ($m=0$) and saturated ($m=1$) ends it has no
perturbative reference state, it is bounded by \emph{two} critical edges (the matter AN--AS line
and the Dicke AS--PS line), and its magnetisation rises as a square root,
$m\sim\sqrt{\varepsilon-\varepsilon_c^{\mathrm{AN\text{-}AS}}}$, out of the AN--AS edge. The
envelope theorem supplies the way out---it already underwrites the bare-matter $1/d$ step itself,
stationarity of the canted reference cancelling the linear terms, with or without a cavity---and it
extends to the self-consistent field: the linear fluctuation terms cancel about \emph{any}
stationary reference, perturbatively reachable or not, so we expand about the exact $d=\infty$
solution at finite $m$ itself. The cost is that the small parameter is $1/d$ and nothing else:
unlike the polarised side, no expansion in a physical coupling exists at fixed $d$, so the result
is controlled only at large $d$, far from $d=1$.
Writing $\eta=\sqrt{1-\lambda/(2J)}$ and $h_*\equiv\lambda m$ for the self-consistent cavity
\emph{field}, the
bipartite mean field plus single-bond PT (in the rotated frame of \cref{app:1d}, where the cavity
magnetisation is $\langle\sigma^x\rangle$ and the staggered Ising order $\langle\sigma^z\rangle$;
the result below is in the frame-independent $\varepsilon,\lambda$) gives the AS energy density to
$O(1/d)$,
\begin{equation}
\evar_{\mathrm{AS}} = -J-\frac{(\varepsilon-2J\eta)^2}{2\lambda}-\frac{\lambda(\varepsilon-2J\eta)^3}{32\,d\,J\eta^3 h_*^2}+O(1/d^2),
\quad h_*^2=\frac{\varepsilon(4J-\lambda)}{\eta}-4J^2-\varepsilon^2,
\label{eq:AS_1d}
\end{equation}
with the two boundaries of the AS phase
\begin{equation}
\varepsilon_c^{\mathrm{AN\text{-}AS}}=2J\eta,
\qquad
\varepsilon_c^{\mathrm{AS\text{-}PS}}=(2J+\lambda)\,\eta .
\label{eq:AS_lines}
\end{equation}
The $O(1/d)$ energy needs only the leading self-consistent cavity \emph{field}
$h_*\equiv\lambda m$, since the shift
$\delta m$ cancels by the envelope theorem (\cref{app:1d}). At $d\to\infty$ \cref{eq:AS_1d,eq:AS_lines}
match the mean-field result of Zhang~\textit{et al.}~\cite{Zhang2014} exactly under
$g_Z=g/2$ (the $\tfrac12$ in $\hat S_z$). The AS branch is the global minimum between the two
lines by a standard Landau bifurcation, so at $d\to\infty$ both AS boundaries are
\emph{second order} (\cref{fig:dinf}).

\paragraph{The first question: preemption of the $a_4=0$ point.}
On the antiferromagnetic side the AN--AS $a_4=0$ point of \cref{eq:eps_tri_AF} is \emph{preempted}
by the first-order AN--PS line in every dimension. The test is an energy comparison at that point
itself (scale-fixed units of \cref{app:1d} throughout this paragraph): \cref{eq:eps_tri_AF}
expands as $J_{\mathrm{tri}}(d)=\tfrac14+\tfrac1{32d}+O(d^{-2})$---coincidentally identical, at
this order, to the $\varepsilon=0$ benchmark above---and
$\varepsilon_{\mathrm{tri}}^2(d)=\tfrac1{32d}+O(d^{-2})$, and evaluating the polarised-branch
energy $\evar^*_{\mathrm{PS}}$ [\cref{eq:PS_1d}] there against the field-independent antiferromagnetic energy
$e_{\mathrm{AN}}=-J$ per site (the $J\to J/d$ rescaling of this $1/d$ analysis; $-dJ$ un-rescaled) gives
\begin{equation}
\Delta(d)\equiv\big[\evar^*_{\mathrm{PS}}-e_{\mathrm{AN}}\big]_{\mathrm{tri}}=-\frac{1}{64\,d}+O(d^{-2})<0 .
\end{equation}
The polarised superradiant state thus already lies below the antiferromagnetic state at the $a_4=0$
point. Extending the comparison \emph{along} the AN--AS line---the $d$-independent curve
$\varepsilon^2=J(4J-1)$---the same energy difference is, at leading order,
$\evar^*_{\mathrm{PS}}-e_{\mathrm{AN}}=(J-\tfrac14)/(4J+1)$ minus a positive $O(1/d)$ term; near the
zero-field end of the line the $O(1/d)$ term wins and keeps the polarised state lower (preempted), and the
difference turns positive (antiferromagnetic lower, so AS can appear) only beyond
\begin{equation}
\boxed{\;J_{\mathrm{cross}}(d)=\tfrac14+\tfrac1{16d}+O(d^{-2})\;>\;J_{\mathrm{tri}}(d)=\tfrac14+\tfrac1{32d}+O(d^{-2})\;.}
\label{eq:Jcross}
\end{equation}
The entire initial arc of the AN--AS line up to $J_{\mathrm{cross}}$---the $a_4=0$ point
included---is therefore preempted by the first-order AN--PS line (\cref{fig:trikrit}). Only \emph{near} the quadruple point does the AS phase
reappear, as the quadruple-point analysis shows (\cref{sec:results_qp}). Two consequences follow
together. The AN--AS onset is \emph{second order wherever it is realised}: where its quartic would
turn it first order ($a_4<0$), the polarised state already lies lower and the system jumps AN$\to$PS
instead, so a first-order AN--AS onset is never reached. That this exhausts the competing
phases---leaving no fourth, hidden branch---is the completeness of \cref{sec:completeness}.

\begin{figure}[tbp]
\centering
\includegraphics{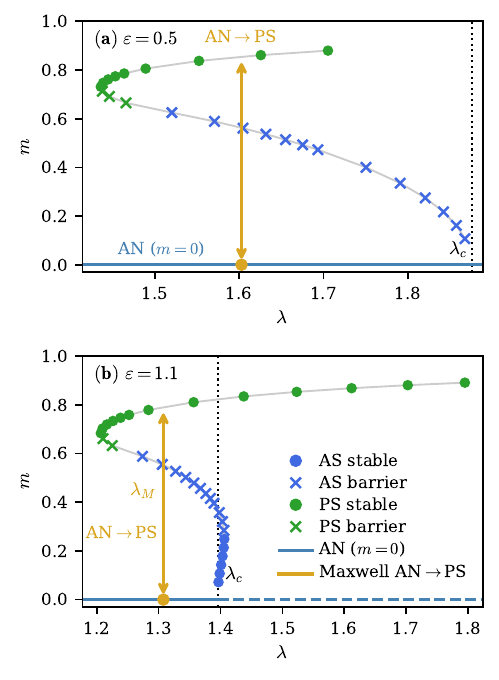}
\caption{Preemption of the antiferromagnetic phase in one dimension (iDMRG), and the role of the $a_4$
sign. Stationarity curves $\lambda(m)=h/m$; stable stationary points (ascending)
are filled circles, the unstable barrier (descending) crosses; antiferromagnetic ($|m_{\mathrm{stag}}|>0$)
blue, polarised superradiant green; the arrow is the global Maxwell jump from the antiferromagnetic
normal state ($m=0$). \textbf{(a)} $\varepsilon=0.5<\varepsilon_{\mathrm{tri}}^{\mathrm{AF}}=2/\sqrt5$
($a_4<0$): the curve \emph{descends} from $m=0$, so the entire AS region lies on the unstable barrier
and is never a minimum---AN$\to$PS directly. \textbf{(b)} $\varepsilon=1.1>\varepsilon_{\mathrm{tri}}^{\mathrm{AF}}$
($a_4>0$): the curve \emph{ascends} from $m=0$, so AS is a genuine local minimum (second-order onset at
$\lambda_c$, dotted line); but the AN$\to$PS jump occurs at $\lambda_M<\lambda_c$ (both marked),
\emph{before} the AS branch
appears, so AS is metastable and again bypassed. Either way, away from the QP the antiferromagnetic phase
is preempted; it is realised only nearer the QP (\cref{fig:extremum_af}).}
\label{fig:c2preempt}
\end{figure}

\begin{figure*}[tbp]
\centering
\includegraphics[width=\linewidth]{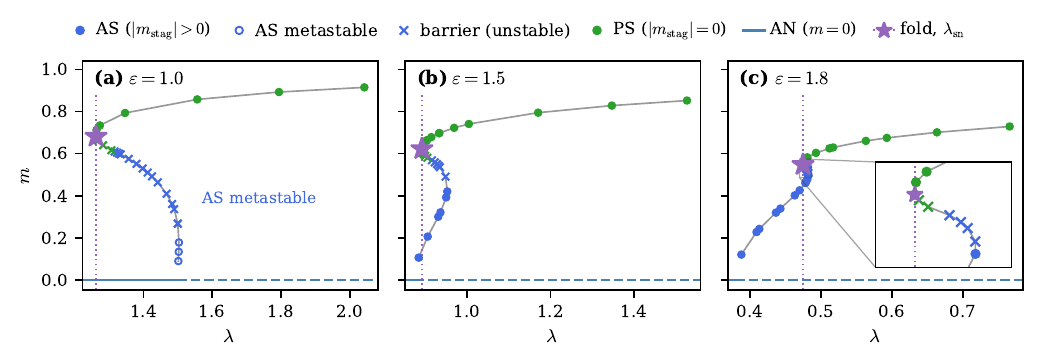}
\caption{The Larkin--Pikin fold in one dimension, from infinite-system DMRG. Stationarity curves
$\lambda(m)=h/m$ versus the superradiant order parameter $m$ for the bare antiferromagnetic chain at
$\varepsilon=1.0,1.5,1.8$ ($J=1$).
Antiferromagnetic points (staggered $|m_{\mathrm{stag}}|>0$, blue) and polarised superradiant points
($|m_{\mathrm{stag}}|\approx0$, green) are distinguished, with the marker convention of
\cref{fig:c2preempt}: ascending, locally stable points are filled circles, the descending barrier
segment between the local maximum and the fold (unstable) crosses, and the steelblue line is the
antiferromagnetic normal state ($m=0$); each curve back-bends at a saddle-node fold
(purple star; the purple dotted vertical line marks its coupling $\lambda_{\mathrm{sn}}$), the signature of the
first-order AS--PS onset (\cref{sec:lp}). As $\varepsilon\to2J$ the fold
tightens and moves to weaker coupling ($\lambda_{\mathrm{sn}}\approx1.26,\,0.89,\,0.47$ at
$m\approx0.68,\,0.62,\,0.55$), the curves marching toward the quadruple point; the inset of
panel (c) magnifies the fold region, where the back-bend is too tight to read off the main panel's
axes; the corner jump itself is resolved directly in \cref{sec:results_qp}. At $\varepsilon=1.0$ the
antiferromagnetic branch is metastable (open markers in panel (a))---the realised transition is the
direct AN$\to$PS jump
(\cref{sec:results_1d})---so the fold shown there is the locally stable AS--PS extremum, not a realised
onset. This is the direct $d=1$ counterpart of the finite-$d$ fold absent from the $d=\infty$ mean field
of \cref{fig:dinf}.}
\label{fig:extremum_af}
\end{figure*}

Infinite-system DMRG confirms the $1/d$ verdict directly in $d=1$: $J_{\mathrm{cross}}>J_{\mathrm{tri}}$
(\cref{eq:Jcross}) already puts the preempted arc beyond the $a_4=0$ point, into the $a_4>0$ regime, and
the polarised state is found to undercut the AS branch there. Even where the AN--AS onset is locally \emph{second order}
($a_4>0$, i.e.\ $\varepsilon>\varepsilon_{\mathrm{tri}}^{\mathrm{AF}}=2/\sqrt5\approx0.894$), the
first-order AN$\to$PS jump can still preempt it: the Maxwell coupling $\lambda_M$ falls \emph{below}
the antiferromagnetic onset $\lambda_c$, so the locally stable AS branch never becomes the global minimum
and the transition runs AN$\to$PS directly, with AS metastable (\cref{fig:c2preempt}). The data
bracket the crossover between $\varepsilon=1.1$ and $1.5$; only closer to the quadruple point does
$\lambda_M$ exceed $\lambda_c$, making AS the realised ground state (\cref{fig:extremum_af}). In one
dimension, then, the antiferromagnetic phase occupies only a narrow wedge just below the quadruple point,
in line with the wormhole quantum Monte Carlo map~\cite{langheld}.

\paragraph{The second question: the order of the AS--PS line at large dimension.}
Whether the AS--PS transition stays second order away from $d=\infty$ is the Larkin--Pikin question
of \cref{sec:lp}: it turns on whether the bare $(d{+}1)$-Ising susceptibility diverges at the
critical field. It does in every physical dimension $d\le3$, forcing the onset first order; for
$d\ge4$ the matter sits above its upper critical dimension, $\xmat$ stays finite, and second order
becomes \emph{possible}---though not automatic. Above $d_{\mathrm{uc}}$ the order is fixed by the sign of the \emph{binding}
(antiferromagnetic-branch) curvature
$\mathcal{S}_{\mathrm{AS}}=\evar''(m_c)/\lambda_*=1-\lambda_*\xmat^R(0)$, the full self-consistent
response in which the staggered tilt relaxes with the field, not the regular polarised-side
$\chi_{\mathrm{reg}}$ (the case list, dimension by dimension, is \cref{app:lp}). The $1/d$ expansion
settles its sign at large dimension. At $d=\infty$ the closed forms of \cref{app:1d} give, along the
whole line,
\begin{equation}
\boxed{\;\mathcal{S}_{\mathrm{AS}}^{(\infty)}=\frac{2\lambda(2|J|-\lambda)}{|J|\,(2|J|+\lambda)}\;>\;0\quad(0<\lambda<2|J|)\;,}
\label{eq:Sinfty}
\end{equation}
positive throughout the interior but vanishing at \emph{both} ends, the quadruple point
($\lambda\to0$) and the zero-field pitchfork ($\lambda=2|J|$), where $\lambda_*\xmat^R(0)\to1$. (The
regular polarised-side curvature $1-\lambda_*\chi_{\mathrm{reg}}=\tfrac12+\lambda/4|J|\in(\tfrac12,1)$
is larger and never vanishes, but it is the non-binding response and overstates the stability, by the
gap $(3\lambda-2|J|)^2/[4|J|(2|J|+\lambda)]$.) The leading $1/d$ correction is negative throughout, so
$\mathcal{S}_{\mathrm{AS}}(d)=\mathcal{S}_{\mathrm{AS}}^{(\infty)}+\mathcal{S}^{(1)}/d$ with
$\mathcal{S}^{(1)}<0$ drives the marginal ends negative and opens a first-order sliver near each:
\emph{two} tricritical points, a tricritical line in the $(d,\varepsilon)$ plane that closes onto the
corner and the zero-field end as $d\to\infty$ ($\lambda_{\mathrm{tri}}^{+}\approx2|J|-0.75|J|/d$ on the
clean zero-field side; the quadruple-point side is truncation-sensitive, and the corner is
independently first order, \cref{sec:results_qp}). The interior second order is thus robust at large
$d$ but bounded by these two tricritical points, a structure distinct from the $d\le3$ first order,
which is forced non-perturbatively by the $\xmat$ divergence (\cref{sec:lp}) and is invisible to the
$1/d$ series at any order. This is the model-specific side of the ``quantum annealed criticality'' of
Chandra \emph{et al.}~\cite{Chandra2018QAC,Chandra2020QAC}: above the upper critical dimension the
fluctuation-induced \emph{forcing} of first order is removed (\cref{sec:lp}), and the residual,
quantitative question---the sign of the binding curvature---comes out positive in the interior at
large $d$.

\paragraph{The fold in one dimension.}
The fold the $d=\infty$ curves of \cref{fig:dinf} lack---the finite-$d$ back-bend that
\cref{sec:lp} predicts---is exhibited directly in $d=1$, where infinite-system DMRG solves the bare
antiferromagnetic chain directly (\cref{app:idmrg}). \Cref{fig:extremum_af} plots the resulting
stationarity curves $\lambda(m)=h/m$ at $\varepsilon=1.0,1.5,1.8$: each \emph{back-bends} at a
Larkin--Pikin fold, the logarithmically divergent $2$D-Ising susceptibility of the bare matter
forcing the AS--PS onset first order---on the realised ground state at $\varepsilon=1.5,1.8$, on
the metastable AS branch at $\varepsilon=1.0$ (the preemption above). The fold tightens and shifts
to weaker coupling as $\varepsilon\to2J$---these folds are the open circles of the $d=1$
AS--PS line of \cref{fig:trikrit}, running into the quadruple point. The two questions of this
subsection are thus answered: the $a_4=0$ point is never realised, and in one dimension the AS phase survives only as
the wedge near the corner; the interior second order at large $d$, bounded by two tricritical points near each end, yields in the
physical dimensions ($d\le3$) to the divergence-forced fold. The corner itself is examined next.

\subsubsection{The quadruple point}
\label{sec:results_qp}

At $\varepsilon=2|J|$, $g=0$ the four phases meet: the quadruple point (QP) is the corner of the
$2\times2$ of \cref{sec:model_DI}, where the matter's antiferromagnetic-ordering boundary and the superradiant
onset cross and all four cells---PN, PS, AN, AS---touch. There the matter reorganises onto the
independent-set (Rydberg-blockade) manifold of \cref{sec:model_QP}, whose extensive golden-ratio
degeneracy ($\sim\varphi^N$ in one dimension) makes this the one corner immune to perturbation theory
($\lambda\to0$). The existence of the AS phase in one
dimension has been questioned~\cite{mendoncca2025} and affirmed by the numerical
maps~\cite{langheld,leibig2025}, but the corner itself ($\lambda\to0$) was not examined directly
there, and they never solved the corner. The corner is where everything in
this paper meets---the matter ordering line, the superradiant onset, the contested AS phase, and the
completeness count---and it brings structure of its own: the functional minimised here is not that
of the bare chain but the \emph{sector} functional built on the independent-set chain \cref{eq:IS}
(the Fibonacci Hilbert space of a detuned Rydberg-blockade chain, \cref{sec:model_QP}), with the
blockade constraint enforced exactly within each cell and by a penalty between cells
(\cref{app:idmrg}); and the corner is scale-invariant, so its physics lives on rays. The frustration that makes it hard is also what makes it rich: on the
degenerate manifold the antiferromagnetic and the polarised order lie energetically close, the
functional alternating between favouring one and the other---the coexistence that the blockade sets
up and that the rest of this section resolves.

\paragraph{Ray structure.} Near the QP the effective model on the independent-set manifold is
homogeneous in $(\lambda,\delta)$ with $\delta=\varepsilon-2|J|$: the radius
$\rho=\sqrt{\lambda^2+\delta^2}$ enters only as a linear energy prefactor, $E=\rho\,\mathcal F(\phi)$,
so the phase selected depends only on the ray, which we take as $r=\lambda/(2|\delta|)$ (the factor
$2$ is the independent-set blockade: creating one up-spin costs $2\delta$). The two-parameter
neighbourhood thus collapses to a one-dimensional circle of rays, and two of the three boundaries follow
from the closed-form one-dimensional onset lines through
$\lambda_c=1/\xmat(0)$. On the polarised side a single flip costs $2\delta$, so
$\xmat(0)=1/\delta$ and $\lambda_c=\delta$, placing PN--PS at $r=\tfrac12$. On the antiferromagnetic side
the bipartite response is $\xmat(0)\simeq1/(2|\delta|)$ and $\lambda_c=2|\delta|$---twice the
polarised slope, the ``blockade factor two''---placing AN--AS at $r_{c1}=1$. The upper AS--PS
boundary, by contrast, is fixed by iDMRG (\cref{app:idmrg}) at $r_{c2}\approx1.5$. Here $\xmat(0)$ \emph{diverges}
as the gap closes ($\sim1/|\delta|$) while $\lambda_c=1/\xmat(0)$ vanishes at the same rate, so these
finite ray ratios are set by the linearly closing blockade gap rather than by a finite
susceptibility; the onset is nonetheless second order ($a_4>0$) at any $\delta\neq0$.

Along the perpendicular ray $\delta=0$---the $90^\circ$ ray straight out of the QP, the detuning
measured from the polarised ($\delta>0$) axis---the detuned chain \cref{eq:IS} reduces to the pure
PXP Hamiltonian, all of whose off-diagonal matrix elements in the occupation basis are
non-positive; by Perron--Frobenius its ground state is therefore unique and nodeless, with no
spontaneous sublattice imbalance---exact at all $g$. (Being single-scale, $-h_z\sum_i\tilde X_i$ moreover has the \emph{same} ground state at
every coupling, with gap $\propto h_z$.) That this paramagnet is gapped and \emph{disordered} is a
fact of the bare phase diagram, not of Perron--Frobenius: the period-2 ($\mathbb{Z}_2$) ordering
transition occurs only at finite detuning---the Fendley \textit{et al.}\ hard-boson point
$(U/w)_c\simeq-1.308$~\cite{fendley2004competing}, i.e.\ $\phi_c\simeq123^\circ\neq90^\circ$---so
$\delta=0$ lies a finite distance inside the disordered phase, as the strict-blockade iDMRG of
\cref{app:idmrg} confirms (gap $\approx0.97\,h_z$, saturating in $N$). Perron--Frobenius alone would
not settle it: in an ordered phase the unique nodeless state is the symmetric combination of the
two N\'eel patterns, with vanishing local order $\langle n_i-n_{i+1}\rangle$ yet long-range
staggered correlations---so nodelessness is compatible with order, and it is the disordered-phase
input, not Perron--Frobenius, that excludes staggered order here. With it, the disorder-by-disorder
selection (\cref{sec:results_map}) picks the PS state directly along $\delta=0$, with no intervening
antiferromagnetic order, so the QP itself lies on the PS side. Off that ray the detuning lifts the manifold degeneracy and the
selection is an ordinary gapped onset: for $\delta>0$ (rays below $90^\circ$, polarised side) the
cavity drives the continuous PN$\to$PS onset at $r=\tfrac12$; for $\delta<0$ (above $90^\circ$,
antiferromagnetic side) the matter orders antiferromagnetically and increasing the coupling carries it
AN$\to$AS$\to$PS across the analytically known AN--AS ray $r_{c1}=1$ and the first-order AS--PS ray
$r_{c2}$ (\cref{app:idmrg}). Only \emph{away} from the QP is the AN--AS $a_4=0$ point preempted by the
first-order AN--PS line (\cref{sec:results_1d}); near the QP the AS phase is always entered.

\paragraph{Finite extent of the AS phase.} That the AS phase nonetheless occupies a finite wedge is established directly by the strict-blockade
iDMRG (\cref{app:idmrg}), which resolves a stable AS branch between the analytically known AN--AS ray
$r_{c1}=1$ and the first-order AS--PS ray $r_{c2}\approx1.5$.
The AS phase therefore has finite extent in every neighbourhood of the QP, a narrow wedge
$r\in(1,r_{c2})$ (radial width $\Delta r\approx0.5$), consistent
with the numerical phase diagrams~\cite{langheld,leibig2025} (and contrary to
the claimed absence of AS in $1$D~\cite{mendoncca2025}).

\paragraph{Order of the AS--PS transition.} The QP is the least accessible point of the phase
diagram to perturbation theory: along every ray into the corner $\lambda\to0$, so the expansion is invalid there. We therefore
settle the order of the AS--PS transition \emph{at} the QP directly, solving the sector functional
numerically---infinite-system DMRG as the solver (\cref{app:idmrg}), with the metastable-branch
hysteresis protocol of Ref.~\cite{leibig2025}, whose DMRG+NLCE maps covered the full chain away
from the corner but not the corner itself. The outcome:
\begin{quote}
\emph{the AS--PS transition is first order along its entire length, up to the quadruple point.}
\end{quote}
The same calculation establishes, free of any mean-field input, that the AS phase occupies the
finite wedge above. The QP makes the result the cleanest \emph{verification} of the cavity Larkin--Pikin mechanism in the
whole phase diagram: there the regular quartic is stabilising ($a_4>0$ at every
$\delta\neq0$), so the Landau quartic drives nothing and the
transition is forced first order \emph{purely} by the divergent critical line, with no help from the
quartic. The \emph{bare} matter \cref{eq:IS} undergoes
a single \emph{second-order} transition in the $2$D-Ising class. Three independent
studies agree on this class: the exact hard-boson solution of Fendley
\textit{et al.}~\cite{fendley2004competing} fixes the universality class, our iDMRG gives the
order-parameter exponent $\beta=1/8$ (and $\nu=1$ from the correlation length), and the DMRG of
Chepiga and Mila~\cite{chepiga2019dmrg} the central charge $c\approx1/2$ (\cref{app:idmrg}). The \emph{dressed} (self-consistent) transition is
nonetheless first order, and for the reason of \cref{sec:lp}: the bare susceptibility diverges (here
$d=1$, the $2$D-Ising logarithm), so the critical magnetisation $m_c$ is a local \emph{maximum} of
$\evar$, while the completeness result of \cref{sec:completeness} guarantees the resulting min--max--min
landscape hides no further branch. A provably second-order matter quantum critical point (QCP) is thus rendered first order
purely by the cavity back-action, exactly as \cref{sec:lp} requires. Here the marginality is
explicit, and it is a $d=1$ statement: the \emph{regular} part of the landscape carries positive
curvature, $K_2=\lambda_*(1-\lambda_*\chi_{\mathrm{reg}})>0$, so on its own it would leave the
transition second order, and it is solely the logarithmically divergent critical line (the
$2$D-Ising Case~B of \cref{app:lp}) that bends the stationarity curve into a fold---robust, not
knife-edge, as the corner limit below makes precise.

\begin{figure}[tbp]
\centering
\includegraphics{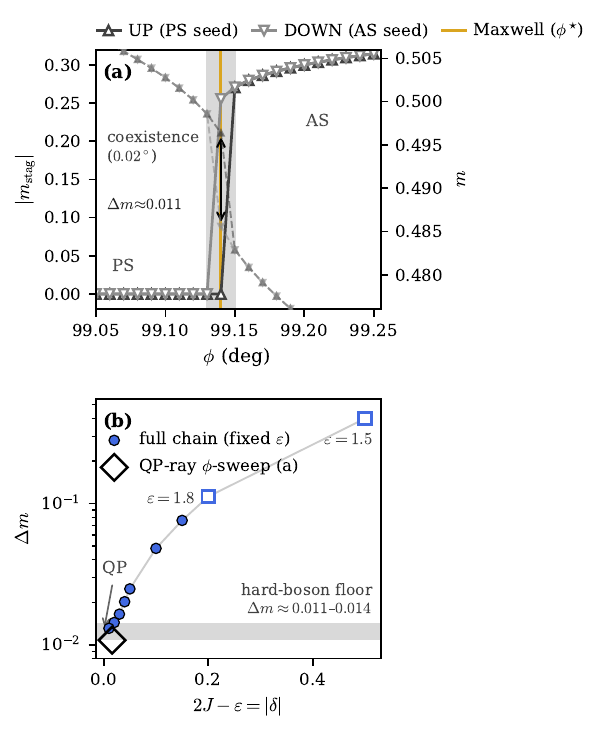}
\caption{First-order AS--PS jumps near the quadruple point (QP), from strict-blockade iDMRG.
\textbf{(a)} Hysteresis across the transition in a narrow window: sweeping the QP-ray angle
$\phi$ up (seeded from the polarised superradiant branch) and down (from the antiferromagnetic
superradiant branch) gives curves that coincide except inside a $\sim\!0.02^\circ$-wide coexistence
window, where the antiferromagnetic order $|m_{\mathrm{stag}}|$ and the cavity order parameter $m$
jump together---a first-order loop. \textbf{(b)} The full-chain AS--PS jump $\Delta m$ (log scale)
versus distance from the corner $2J-\varepsilon=|\delta|$
($\delta=\varepsilon-2|J|$ as in the text): a dense fixed-$\varepsilon$ scan ($\varepsilon=1.85$--$1.99$,
filled) shows it shrink toward the QP and \emph{flatten to a finite floor} $\approx0.011$--$0.014$---the
scale-invariant hard-boson QP jump. The
$\varepsilon=1.5,1.8$ points (open) lie on fixed-$\varepsilon$ rays farther out and are compared only
loosely; they too descend toward the same floor. Validated by the bond-dimension- and
method-independent agreement of \cref{app:idmrg}.}
\label{fig:qpjumps}
\end{figure}

\paragraph{Shrinking but finite jumps.}
The first-order jump is therefore the generic outcome at every point of the AS--PS line, and the
iDMRG resolves it directly (\cref{app:idmrg}). The same first-order superradiant--superradiant
transitions are seen numerically in a driven-Rydberg variant by stochastic-series-expansion quantum
Monte Carlo~\cite{Dong2025anisotropic}, where they are reported without a mechanism; the
divergent-susceptibility Larkin--Pikin argument here supplies it. At coarse angular resolution the AS--PS transition
looks continuous; resolving the ray angle to $\sim0.01^\circ$ while continuing each metastable branch
to its self-consistent fixed point (\cref{app:idmrg}) instead uncovers a clear first-order
hysteresis loop (\cref{fig:qpjumps}a). The jump grows steadily as one
moves away from the QP and stays finite as one approaches it (\cref{fig:qpjumps}b):
$\Delta m\approx0.40$ at $\varepsilon=1.5$ and $0.11$ at $\varepsilon=1.8$ along the full
antiferromagnetic chain, and a dense fixed-$\varepsilon$ scan resolves the near-corner approach
directly: the jump shrinks smoothly toward the QP and flattens to a finite floor. At the QP itself the matter
reduces to the scale-invariant hard-boson chain, which gives the corner jump directly:
$\Delta m\approx0.011$ ($0.0105(5)$, \cref{app:idmrg}), consistent with the $\approx0.014$ obtained by
extrapolating the near-corner fixed-$\varepsilon$ trend to $\rho\to0$. This corner jump is
\emph{independent} of the distance from the corner: that effective Hamiltonian is linear in the
detuning $\delta$ and the field $h_z$, so its ground state---and the jump---depend only on the ray
$r=\lambda/(2|\delta|)$, not on the radius $\rho$. Consistently, the near-corner fixed-$\varepsilon$
crossings all lie on a \emph{single} AS--PS ray ($\phi\approx99.1^\circ$ in the $(\delta,g^2)$ plane),
the same ray the $\phi$-sweep of \cref{fig:qpjumps}a cuts---and since the jump is the order-parameter
discontinuity at the boundary, independent of the crossing direction, both trace one curve $\Delta
m(\rho)$ into the same finite floor (\cref{fig:qpjumps}b; convergence ladder in \cref{app:idmrg}). The
jump still varies along this ray because the exact scale invariance is only approached as
$\rho\to0$: at finite radius the matter is the full antiferromagnetic Ising chain, not yet the strict
hard-boson (PXP) limit, so $\Delta m(\rho)$ relaxes smoothly onto its $\rho$-independent floor only at
the corner. (The $\varepsilon=1.5,1.8$ points sit farther
out, where the boundary has curved away from this ray, so they are compared only loosely.) The log-marginal Larkin--Pikin case (\cref{app:lp}) forces the transition first order, and the jump
does not vanish into the corner: $\Delta m$ is the global Maxwell discontinuity, here the finite
scale-invariant floor $\Delta m\approx0.011$. The exponential factor $\exp(-K_2/2B\lambda_*^2)$ of
\cref{app:lp} is \emph{not} this jump; it is the half-width of the local barrier in the order
parameter $\delta=m-m_c$, and it too stays finite at the corner: there $\xmat\sim1/\lambda_*$
diverges, so $K_2=\lambda_*(1-\lambda_*\chi_{\mathrm{reg}})\sim\lambda_*$ and the log amplitude
$B\sim1/\lambda_*$ scale together and the exponent stays $O(1)$. The first order is thus \emph{robust}
at the quadruple point, not marginal.
The QP is thus a
four-phase multicritical point where AN, AS, PN, and PS meet; its underlying \emph{bare} matter is in
the $2$D-Ising class ($\beta=1/8$, $c=1/2$, \emph{not} the tricritical-Ising $\beta=1/24$),
while the \emph{dressed} superradiant transitions there are first order.

\paragraph{A tetracritical point.}
The two $\mathbb Z_2$ symmetries of \cref{sec:model_DI}---with $\psi$ the staggered
(antiferromagnetic) order parameter and $m$ odd under photon
parity---allow as lowest coupling the biquadratic $m^2\psi^2$, and the four phases are its
coexistence table: the symmetry content of a tetracritical point, though not its textbook
realisation, since one of the meeting lines (AS--PS) arrives \emph{first order}. What differs from
the textbook scenario~\cite{KosterlitzNelsonFisher1976} is that $m$ is not a co-equal fluctuating
order parameter but a single \emph{slaved}, non-critical cavity mode, fixed by $m=\mumat(\lambda m)$.
The superradiant--superradiant first order is therefore not the bicritical kind, where two competing
orders exclude one another, but the Larkin--Pikin kind: the divergent matter susceptibility forbids a
minimum at the bare critical line, so the connected curve of \cref{sec:folds} must pass it as a
maximum, and a maximum between the stable phases is what forces the fold.

The role of the coupling geometry is made explicit by a recent counterpart that shares exactly the
ingredient just demonstrated, a divergent susceptibility at a quantum-critical point. Sur \emph{et
al.}~\cite{Sur2026} couple a cavity \emph{longitudinally}, along the quantum-critical order parameter
itself; the diverging susceptibility then simply amplifies the cavity response, and the superradiant
onset is \emph{continuous}, with a critical coupling that vanishes at the matter QCP ($g_c\to0$). Here
the coupling is \emph{orthogonal}: the cavity sees $\sigma^z$ while the antiferromagnetic order lies
along $\sigma^x$, and the two no longer commute. Order and superradiance now compete rather than
reinforce, the competition Sur \emph{et al.}\ themselves identify in their non-commuting geometry, and
that competition is what renders the superradiant--superradiant transition first order.

\subsection{Other cavity-coupled magnets}
\label{sec:results_beyond}

The same leading-singularity reading classifies the other cavity-coupled magnets of
\cref{sec:model_tri,sec:model_compass,sec:model_iso} and fills the rows of \cref{sec:results_map}
with worked numbers, keeping the paper-wide normalisation (\cref{tab:conventions}); a saturation field $h_{\mathrm{sat}}$ sets the lower coupling
$\lambda_{\mathrm{low}}=h_{\mathrm{sat}}/\mu_{\mathrm{sat}}$ and a finite zero-field susceptibility
$\xmat(0)$ the upper one $\lambda_{\mathrm{up}}=1/\xmat(0)$, between which the saturation onset can turn the
transition first order.

\subsubsection{Frustrated antiferromagnets: the triangular lattice (cusp row)}
\label{sec:results_triangular}

The geometrically frustrated triangular Ising
antiferromagnet at $\varepsilon=0$ (\cref{sec:model_tri}) realises the cusp row: its extensively
degenerate classical manifold responds linearly to the cavity field, $\emat(h)-e_0\sim-c|h|$, so
$\xmat(0)$ diverges, $\lambda_{\mathrm{up}}=1/\xmat(0)\to0$, and the onset is immediate---the normal phase
collapses to the $g=0$ axis (\cref{eq:lambdac_general}). Which superradiant phase appears is the
order-by-disorder question of \cref{sec:model_tri}: on the triangular lattice the field selects the
clock-ordered state, so the immediate phase is AS, and the clock order melts in a further AS$\to$PS
transition at $x_c^{\mathrm{AF}}\equiv J/\Gamma_c\approx0.61$
($\Gamma_c/J=1.65$~\cite{IsakovMoessner2003,Humeniuk2016}).

\begin{figure}[!htbp]
\centering
\includegraphics{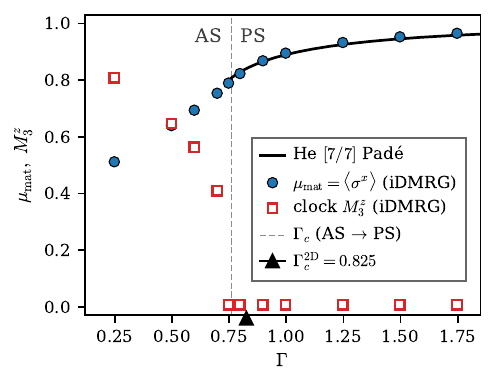}
\caption{Triangular-lattice transverse-field Ising antiferromagnet---the $\varepsilon=0$
cavity matter of \cref{sec:model_tri}---in the convention
$\hat H=\sum_{\langle ij\rangle}S^z_iS^z_j+\Gamma\sum_i S^x_i$ with $S=\tfrac12$, $J=1$.
Markers: published $L_y=6$ iDMRG of Saadatmand \textit{et al.}~\cite{Saadatmand2018}---the
field-axis (uniform) magnetisation $\mumat=\langle\sigma^x\rangle$ that the cavity
couples to (circles) and the clock order parameter $M^z_3$ (squares). Solid curve: the
He--Hamer--Oitmaa high-field $[7/7]$ Pad\'e for $\mumat$, which reproduces the
iDMRG to $<1\%$ throughout the polarised (PS) phase. The clock parameter vanishes near
$\Gamma_c\approx0.75$ at $L_y=6$, consistent with the continuous melting at the AS$\to$PS
transition established in \cref{sec:results_triangular}, just below the two-dimensional thermodynamic value
$\Gamma_c=0.825$ in these $S=\tfrac12$ units, the offset being the expected finite-$L_y$ shift.
The main text quotes this same point in the Pauli convention as $\Gamma_c/J=1.65$, equivalently
$x_c^{\mathrm{AF}}\equiv J/\Gamma_c=0.61$~\cite{IsakovMoessner2003}.}
\label{fig:tri_mu}
\end{figure}

\begin{figure}[!htbp]
\centering
\setlength\abovecaptionskip{5pt}
\includegraphics{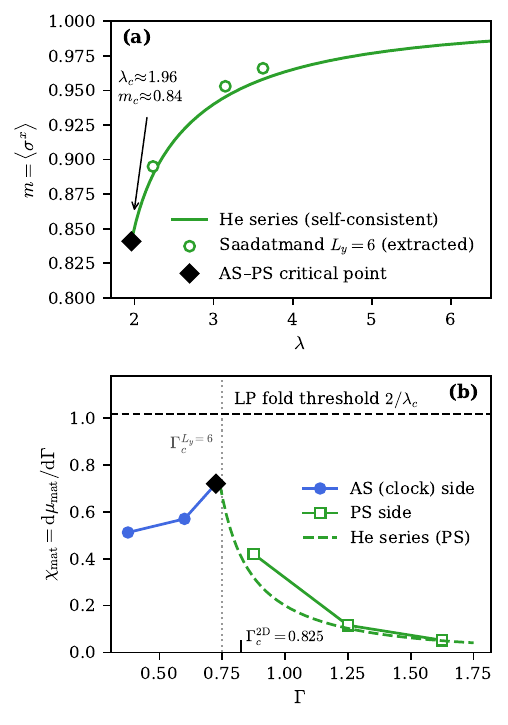}
\caption{Self-consistent solution and second-order test for the triangular antiferromagnet
($\varepsilon=0$); as in \cref{fig:tri_mu}, $m=\langle\sigma^x\rangle=\mumat$ is the rotated-frame
magnetisation the cavity couples to, not the clock order parameter. \textbf{(a)} The polarised
(PS) $m(\lambda)$ solved self-consistently
from the He--Hamer--Oitmaa $[7/7]$ Pad\'e (solid); open circles: the
published $L_y=6$ iDMRG of Saadatmand \textit{et al.}~\cite{Saadatmand2018},
reproduced to $<1\%$. Here $\lambda=h/\mumat$ ($\lambda=g^2/2$, $J=\omega_c=1$); the branch ends at the AS--PS critical point (black diamond)
$(\lambda_{\mathrm{AS\text{-}PS}},m_c)\approx(1.96,0.84)$, at the thermodynamic $\Gamma_c=0.825$~\cite{IsakovMoessner2003}. \textbf{(b)} The matter
susceptibility $\xmat=\mathrm{d}\mumat/\mathrm{d}\Gamma$ ($h=2\Gamma$, so the slope is twice the per-$h$ susceptibility)
from the same $L_y=6$ iDMRG (filled circles, AS side; open squares, PS side;
coarse finite differences) traces a \emph{finite} peak at the critical point (black diamond), not a
divergence---excluding a Larkin--Pikin fold; the He high-field series (dashed) covers only the
PS branch above $\Gamma_c$, where it agrees. The horizontal dashed line is the fold threshold
$2/\lambda_{\mathrm{AS\text{-}PS}}\approx1.02$: the peak stays clearly below it. Dotted vertical (grey): the $L_y=6$
clock-melting field $\Gamma_c\approx0.75$~\cite{Saadatmand2018}; black tick: the
two-dimensional thermodynamic value $\Gamma_c^{\rm 2D}=0.825(25)$~\cite{IsakovMoessner2003} (the
$\sim10\%$ offset is the finite-$L_y$ shift). Together with $\evar''(m_c)\approx0.6>0$ this fixes
the AS--PS transition as second order at all resolved scales.}
\label{fig:tri_sc}
\end{figure}

\emph{The self-consistent magnetisation.} The polarised (PS) branch follows explicitly from the
matter response. The He--Hamer--Oitmaa high-field series~\cite{HeHamerOitmaa1990} gives the bare
magnetisation $\mumat(h)=-\emat'(h)$ as a $[7/7]$ Pad\'e in $x=J/h$ (the antiferromagnet from the
ferromagnetic series by $x\to-x$). The cavity magnetisation is the self-consistent solution of
$m=\mumat(\lambda m)$, built up order by order in the self-consistent $m$; inverting with $h=\lambda
m$ gives $\lambda=h/\mumat(h)$, so the PS branch is the parametric curve
$\big(\lambda,m\big)=\big(h/\mumat(h),\,\mumat(h)\big)$, running from full polarisation $m\to1$ at
large $\lambda$ down to $m_c\approx0.84$ at $\lambda_{\mathrm{AS\text{-}PS}}\approx1.96$, where the clock order sets in
(\cref{fig:tri_sc}a). The construction deserves its name spelled out: it is the saturation end of
the $\hat S_z^2$ problem [\cref{eq:Sz2}] in practice---the high-field series expands the
interacting matter about the saturated state that the $\lambda\to\infty$ limit prepares, and the
self-consistency turns that bare series, order by order, into a perturbatively exact series for the
\emph{cavity} problem, the route of Ref.~\cite{LeibigBSc}. No fold interrupts it: the PS branch of
the stationarity curve is monotone all the way down to the critical point, as shown next. The
construction is validated against the published $L_y=6$ iDMRG of
Saadatmand \textit{et al.}~\cite{Saadatmand2018} (read off their Fig.~2), whose uniform
magnetisation the $[7/7]$ Pad\'e reproduces to $<1\%$ across the polarised phase (\cref{fig:tri_mu}).

\emph{The AS--PS transition is second order.} Two independent readings of the matter \emph{curvature}
fix this. First, the self-consistent PS minimum never folds into a maximum before the clock order
appears: the stationarity curve $\lambda(h)=h/\mumat(h)$ is monotonic along the whole PS branch (no
turning point, no fold), and at the critical point the curvature is comfortably positive,
\begin{equation}
\evar''(m_c)\propto1-\lambda_{\mathrm{AS\text{-}PS}}\,\xmat(h_c)=1-1.96\times0.20\approx0.60>0,
\end{equation}
with $\xmat(h_c)\approx0.20$ from the series. The margin is wide: the curvature flips to a maximum
(first order) only if $\lambda_{\mathrm{AS\text{-}PS}}\,\xmat(h_c)\ge1$, i.e.\ $\xmat(h_c)\ge1/\lambda_{\mathrm{AS\text{-}PS}}\approx0.51$---a
factor $2.5$ above the series value, equivalently a factor-$2.5$ error in $\lambda_{\mathrm{AS\text{-}PS}}$. The Pad\'e
$\xmat$ lies in the polarised phase where the high-field series converges (it reproduces the iDMRG
magnetisation to $<1\%$), and the coarse finite differences of the published $L_y=6$ data probe the
response independently from the antiferromagnetic side, staying clearly below the threshold
$1/\lambda_{\mathrm{AS\text{-}PS}}$ even at their peak (in the per-$\Gamma$ units of \cref{fig:tri_sc}b the threshold
is $2/\lambda_{\mathrm{AS\text{-}PS}}\approx1.02$, drawn as the dashed line); even a generous $50\%$ error on
$\xmat(h_c)$ together with a $20\%$ error on $\lambda_{\mathrm{AS\text{-}PS}}$ still leaves $1-\lambda_{\mathrm{AS\text{-}PS}}\xmat\approx0.3>0$.
Within everything the series and the $L_y=6$ data resolve, the minimum is therefore
robust---though percent-level agreement of the magnetisation does not by itself bound the
derivative at the endpoint, which is why the universality argument below carries the remaining
weight. Second, the matter susceptibility itself stays
finite and merely \emph{peaks} at the transition. Extracting
$\xmat=\mathrm{d}\mumat/\mathrm{d}h$ by finite differences from the published $L_y=6$ iDMRG of
Saadatmand \textit{et al.}~\cite{Saadatmand2018}, it rises monotonically through the AS (clock) phase
to a finite maximum at $\Gamma_c$ and falls away again in the PS phase (\cref{fig:tri_sc}b)---the
signature of a continuous transition with a finite response, not the divergence that would force a
Larkin--Pikin fold. Both readings place the AS--PS line in the $\alpha<0$, finite-$\xmat$ escape of
\cref{sec:lp}: the bare clock ordering is in the $3$D-XY class, whose negative specific-heat exponent
makes the \emph{singular} part of $\xmat$ vanish at criticality, so no divergence is available to
bend the curve---\emph{no fold} is forced, and the verdict reduces to the quantitative
criterion $\lambda_{\mathrm{AS\text{-}PS}}\xmat(h_c)<1$, met with the margins above. One caveat is worth stating:
$\alpha\approx-0.015$ is small, so the singular cusp of $\xmat$ is, within our resolution, numerically
indistinguishable from a logarithm, and its non-universal amplitude at $h_c$ is resolved by
neither the $[7/7]$ Pad\'e nor the $L_y=6$ differences; were it anomalously large, a fold would
re-enter---confined exponentially close to the transition, with a jump far below the present
resolution. The transition field is itself uncertain at the $\sim10$--$15\%$ level---$\Gamma_c\approx0.75$--$0.85$ in $S=\tfrac12$ units across the recent $L_y=6$ iDMRG~\cite{Saadatmand2018} and two-dimensional quantum Monte Carlo~\cite{IsakovMoessner2003} estimates---but the verdict rests on the $\alpha<0$ universality, which is independent of where $\Gamma_c$ sits, and the quantitative margin $1-\lambda_{\mathrm{AS\text{-}PS}}\xmat$ stays positive across this spread. At all resolved scales the AS--PS line is second order; what the $3$D-XY
assignment excludes outright is the divergent alternative. Because the cavity
couples only to the non-critical \emph{uniform} magnetisation, the AS--PS line then simply \emph{is}
the matter clock transition under a smooth self-consistent reparametrisation of the control field
($h=\lambda m$, regular precisely because $\xmat$ is finite); a smooth reparametrisation preserves
critical exponents, so the line inherits the matter's $3$D-XY universality with no exponent
measurement of its own.

Spectrally (\cref{sec:spectral}), this is the interacting counterpart of the exactly solvable
$d=\infty$ AS--PS line (\cref{app:dinf_dicke}): the lower polariton's photon weight would again
collapse, now with a $3$D-XY exponent ($Z\sim|t|^{2\nu z+\alpha}$, against the mean-field $|t|$ at
$d=\infty$), but here the soft mode overlaps the gapless matter continuum, so whether it persists as
a sharp polariton is left open (\cref{sec:conclusions}).

\emph{Continuous symmetry from a discrete model.} The $\alpha<0$ escape deserves a remark, since
$\alpha<0$ is the generic situation for \emph{continuous}-symmetry breaking and for more exotic
critical points (\cref{sec:lp})---and a cavity coupling to
$\sigma^z$ acts as a field, under which genuine continuous-symmetry breaking is normally unavailable
(the bare Dicke transition breaks only the discrete photon parity). Here it arrives through an
\emph{emergent} continuous symmetry: at its ordering transition the three-sublattice clock order
parameter sits at the $(2{+}1)$-dimensional XY fixed point, the six-fold ($\mathbb{Z}_6$) clock
anisotropy being dangerously irrelevant, so the discrete clock symmetry is enlarged to an emergent
continuous $U(1)$ and the criticality is $3$D-XY. The triangular antiferromagnet is thus an unusual cavity setting: a purely Ising interaction whose
frustration generates an emergent continuous ($U(1)$) symmetry, hence $\alpha<0$, so the
superradiant--superradiant transition escapes the Larkin--Pikin fold and stays second order.

On the kagome lattice, where the field instead selects a
\emph{disordered} paramagnet (\cref{sec:model_tri}), the cavity
drives a single PS phase, with no AS wedge---just as on the $\delta=0$ ray out of the quadruple
point. What the present framework adds is the
organising fact: within the functional the cavity acts on the degenerate manifold as nothing but
the uniform field $h=\lambda m$, so the selection \emph{is} the bare matter's own order or
disorder by disorder, read through the cusp of $\emat$ and the landscape of
\cref{sec:landscape}---what required a dedicated analysis there is, here, a corollary.

\subsubsection{The compass chain (marginal and regular rows)}
\label{sec:results_compass}

\begin{figure}[!htbp]
\centering
\includegraphics{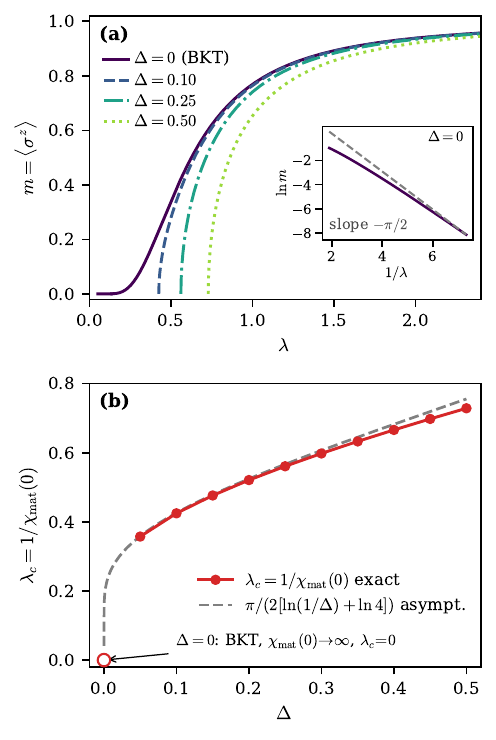}
\caption{Cavity-coupled compass chain at the symmetric compass angle ($J_1=1+\Delta$,
$J_2=1-\Delta$), bare chain solved exactly by Jordan--Wigner. \textbf{(a)} Self-consistent
superradiant magnetisation $m(\lambda)$ ($\lambda=g^2/2\omega_c$): the
onset is smooth and threshold-free at the symmetric point $\Delta=0$
(Berezinskii--Kosterlitz--Thouless, $m\sim e^{-\pi/(2\lambda)}$; inset: $\ln m$ versus
$1/\lambda$ at $\Delta=0$ approaches the straight line of slope $-\pi/2$, dashed guide, the slow
bend away from it reflecting the subleading $1/\lambda$ prefactor) and develops an increasingly sharp
threshold as $\Delta$ grows. \textbf{(b)} The onset threshold $\lambda_c(\Delta)=1/\xmat(0)$
for $\Delta>0$ (gapped in the cavity-coupled channel, finite $\xmat(0)$): it opens continuously from the $\Delta=0$ point
(open circle), where $\xmat(0)$ is logarithmically divergent and $\lambda_c=0$---only an apparent
clash with the finite normal phase: the threshold closes logarithmically slowly,
$\lambda_c=\pi/(2K(1-\Delta^2))$ exactly, asymptotically
$\pi/(2[\ln(1/\Delta)+\ln4])$ (grey dashed; \cref{app:compass}). Degeneracy with an
\emph{even} ($m^2$) response thus leaves a finite normal phase, unlike the linear-cusp
quadruple point.}
\label{fig:compass}
\end{figure}

The bond-alternating compass chain
(\cref{sec:model_compass}) is a second degenerate manifold and a revealing counterpoint to the
quadruple point. Its $\emat(h)$ is \emph{even} in $h$ (no longitudinal cusp), so the cavity energy
gain near $m=0$ is quadratic, $\propto m^2$, not linear in $|m|$. The cusp is absent because the
cavity field does not couple \emph{within} the degenerate manifold: unlike the triangular Ising case,
where a single flip lowers the energy already at first order in $h$ (the linear cusp), here the
manifold degeneracy is lifted only at \emph{second} order in the field, so the leading response is
$\propto m^2$. The symmetric point $\Delta=0$ is the gapless, maximally frustrated point of the bare
matter---a critical point of the bare chain; its macroscopic ground-state degeneracy is in fact
present at every $\Delta$, but inert in the cavity-coupled channel---and
it is the gapless free-fermion modes there---not a finite local rearrangement---that make the uniform
susceptibility diverge, but only \emph{logarithmically}: the onset out of the normal phase is then a marginal,
Berezinskii--Kosterlitz--Thouless-type essential singularity $m\sim\exp(-\pi/(2\lambda))$ with no sharp
$\lambda_c$ (the marginal row). Read as perturbation theory: at $\Delta=0$ the reference is gapless
and degenerate, so no convergent series in the field exists and no finite $\lambda_c$
emerges---only the essential singularity; any $\Delta>0$ restores convergence, and with it the
threshold. A bond asymmetry $\Delta>0$ gaps the chain, $\xmat(0)$ becomes
finite, and the onset hardens into a \emph{finite} threshold $\lambda_c(\Delta)=1/\xmat(0)$ that
opens continuously from the $\Delta=0$ point (the regular row). Both onset quantities are fully
\emph{analytic} in $\Delta$ from the free-fermion solution: as the gap closes linearly the zero-field
susceptibility diverges as $\xmat(0)\simeq(2/\pi)\ln(1/\Delta)$, so the threshold closes only
logarithmically slowly, $\lambda_c(\Delta)\simeq\pi/(2\ln(1/\Delta))\to0$ as $\Delta\to0$---a
multiplicative log factor that makes the approach to the marginal point gentle rather than abrupt
(the same coefficient that sets the $\Delta=0$ onset $m\sim e^{-\pi/(2\lambda)}$). The same logarithm governs finite-size numerics: near the symmetric (gapless) point the finite-size
scaling of the polariton threshold is likely difficult, possibly converging only logarithmically in
$N$---a slowness the exact solution sidesteps, and which any finite-size
simulation of the full light--matter problem (quantum Monte Carlo
included~\cite{langheld}) would inherit. Both regimes are visible
in the exact free-fermion solution (\cref{fig:compass}). Because $\emat$ is even and concave with no
matter first-order transition, the stationarity curve is a \emph{single} connected curve throughout
(\cref{sec:folds})---for $\Delta>0$ leaving the axis at a finite $\lambda_c$, at $\Delta=0$ at
$\lambda_c=0$. This completes the trichotomy of \cref{sec:results_map}: compass and quadruple
point are both disorder-by-disorder, the compass differing only by its even, marginal row (a
normal phase of finite extent instead of one collapsing to the $g=0$ axis).

The Landau machinery of \cref{sec:landau} runs unchanged at $\Delta>0$ and, with the
free-fermion bands in hand, lands in closed form (\cref{app:compass}):
$\xmat(0)=(2/\pi)\,K(1-\Delta^2)$ exactly, $K$ the complete elliptic
integral---identifying the additive constant of the logarithm as $\ln4$---and the quartic
$a_4=\lambda^4E(1-\Delta^2)/(4\pi\Delta^2)>0$ at every bond asymmetry: the onset is continuous
along the entire $\Delta>0$ line. More is true: $\mumat(h)$ is strictly concave for every
$\Delta>0$, so the stationarity curve rises monotonically from the axis---fold-free, first order
nowhere---while the extensively degenerate zero modes, present at every $\Delta$, stay inert:
they carry no weight in the uniform $\sigma^z$ channel, and neither $\xmat(0)$ nor any higher
response the curve reads acquires a contribution from them. Degeneracy alone, like gaplessness
alone, drives none of the mechanisms of \cref{sec:general}.

\subsubsection{Isotropic chains: Heisenberg and XX, and the Dicke--Heisenberg diagram}
\label{sec:results_iso}

\begin{figure}[tbp]
\centering
\setlength\abovecaptionskip{5pt}
\includegraphics{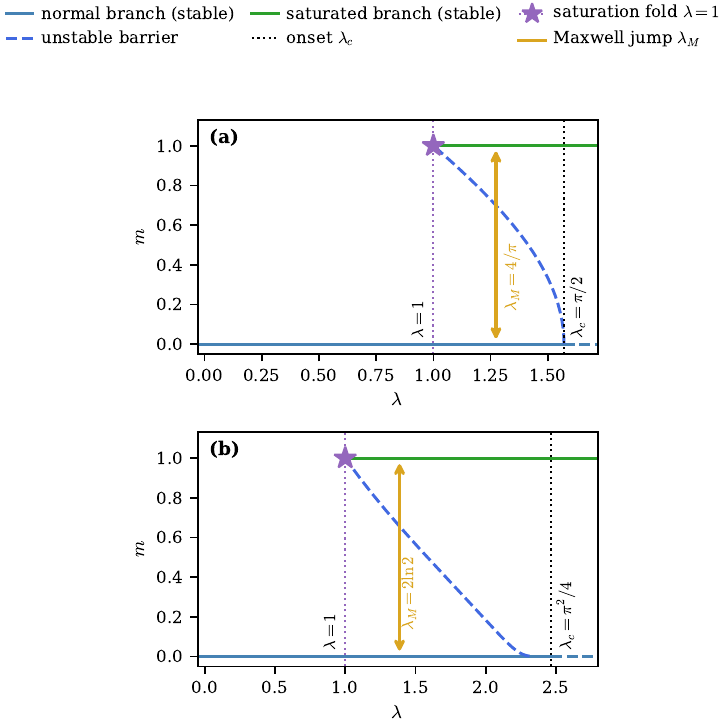}
\caption{Stationarity curves $\lambda(m)=\mumat^{-1}(m)/m$ for the isotropic chains, both first order. \textbf{(a)} XX chain, the exact closed form $\lambda(m)=\sin(\pi m/2)/m$
(\cref{eq:xx_curve}): a monotone unstable barrier from the pitchfork $\lambda_c=\pi/2$ down to the
saturation spinodal $\lambda=1$, so the onset jumps from $m=0$ to saturation at the Maxwell coupling
$\lambda_M=4/\pi$. \textbf{(b)} Heisenberg chain, the same structure with $\mumat(h)$ from
the Bethe ansatz, $\lambda_c=\pi^2/4$ and $\lambda_M=2\ln2$. The XX branch leaves with the ordinary vertical square-root tangent,
while the marginal logarithmic corrections in the Heisenberg susceptibility make its branch lift
off flatter than any power, $m\sim e^{-b/(\lambda_c-\lambda)}$. The normal branch ($m=0$,
solid) is drawn up to $\lambda_c$; at $\lambda_M$ (arrow) the ground state jumps to the saturated
branch ($m=1$, solid, metastable for $\lambda<\lambda_M$), the dashed curve the unstable barrier between. In both, $\xmat$ diverges at the
$z=2$ saturation edge, and the Larkin--Pikin mechanism (\cref{sec:lp}) makes the whole branch a barrier. Markers: the onset $\lambda_c$ (black dotted) carries no
polariton softening (the cavity couples through a \emph{conserved} operator---spectrally silent), and
the saturation fold at $\lambda=1$ (purple dotted, purple star) is non-differentiable,
unlike the smooth, spectrally-active folds of the other figures.}
\label{fig:isocurves}
\end{figure}

The antiferromagnetic
Heisenberg and XX chains are gapless yet have \emph{finite} uniform susceptibilities
($\xmat(0)=4/\pi^2$ and $2/\pi$; the mechanism---gaplessness at the antiferromagnetic point, not in the
$q=0$ channel the cavity couples to---is recalled in \cref{sec:model_iso}). They sit in the regular row. Both chains also conserve
the very operator the cavity couples to, $[\hat H,\sum_i\sigma^z_i]=0$, with a consequence for the
spectral reading of \cref{sec:spectral}: the \emph{dynamic} uniform susceptibility vanishes identically
(the conserved $\sum_i\sigma^z_i$ has no matrix elements at nonzero frequency), so the photon never
hybridises with the matter---there is no polariton precursor---and the superradiant onset proceeds
by ground-state level crossings between magnetisation sectors; the static functional analysis is
untouched, resting only on the thermodynamic response. In fact the statement is exact to all
orders, not only in linear response: since the conserved $\sum_i\sigma^z_i$ is a number in each
magnetisation sector, the full light--matter model solves sector by sector as a displaced
oscillator---the photon line sits exactly at $\omega_c$ in every sector, and no composite
photon--matter excitation can descend. What replaces the soft mode is the $O(1/N)$ fan of sector
ground states, whose stiffness changes sign at $\lambda_c$. This conserved case is the boundary of
the observation that (almost) every correlated light--matter system maps, in its normal phase, onto
an effective Dicke model~\cite{Schellenberger2024}: that mapping couples the photon to a matter
operator that \emph{creates} excitations, while a conserved operator creates none, so the effective
Dicke model collapses to a bare photon at $\omega_c$. These chains are the \emph{almost}. (Couple the same Heisenberg magnet
\emph{transverse} to a polarising field instead, so that the coupled operator is no longer
conserved, and the zero-momentum magnon does hybridise into polaritons~\cite{roman2025linear};
the conserved geometry here is the complementary, spectrally silent case.) The XX chain even gives the
stationarity curve in \emph{closed form}: from the exact field-axis magnetisation
$m=\mumat(h)=\langle\sigma^z\rangle=\tfrac2\pi\arcsin(h)$ ($m\in[0,1]$, $\mu_{\mathrm{sat}}=1$ at $h=1$) one
inverts $h=\sin(\pi m/2)$, so
\begin{equation}
\lambda(m)=\frac{\mumat^{-1}(m)}{m}=\frac{\sin(\pi m/2)}{m},
\label{eq:xx_curve}
\end{equation}
an exact stationarity curve that falls monotonically from $\lambda_c=\pi/2$ at $m\to0$ to $\lambda=1$ at
saturation (\cref{fig:isocurves}a). Leaving the axis toward \emph{smaller} coupling ($a_4<0$), the
whole branch is the unstable barrier between the normal and the saturated state, so the onset is
\emph{first order}---a Maxwell jump from $m=0$ straight to saturation, with spinodals $\lambda=\pi/2$
and $\lambda=1$, and $\lambda_c=\pi/2=1/\xmat(0)$
recovering $\xmat(0)=2/\pi$. The first order here is the Larkin--Pikin mechanism at the
\emph{saturation} (band-edge) transition rather than at an interior critical point: as $h\to
h_{\mathrm{sat}}=1$ the response is a dynamical-exponent $z=2$ critical point, the susceptibility
$\xmat=\mathrm{d}\mumat/\mathrm{d}h=2/(\pi\sqrt{1-h^2})\sim1/\sqrt{1-h}$ diverging from below (and
dropping to zero in the gapped saturated phase above)---a single critical peak that, by \cref{sec:lp},
forbids a continuous onset and makes the saturated end of the curve a maximum. The Maxwell coupling
itself is analytic ($\lambda_M$ below). The Heisenberg chain shows the verdict is unchanged by genuine interactions. Now $\mumat(h)$ comes
from the Bethe ansatz---the zero-temperature magnetisation curve first computed from the Bethe
integral equations by Griffiths~\cite{Griffiths1964}---rather than from the elementary form
(\cref{fig:isocurves}b), and the uniform susceptibility carries marginal logarithmic corrections near
zero field. Nonetheless $\xmat(0)=4/\pi^2$ stays finite and the first order is again fixed at the
saturation edge, so the interactions and the marginal log leave the classification untouched; the
Landau coefficients $a_4<0$, $a_6>0$ of \cref{eq:landau}
confirm the subcritical, first-order character, giving
$\lambda_{\mathrm{low}}<\lambda_M<\lambda_{\mathrm{up}}$:
\begin{center}
\begin{tabular}{lcccc}
\hline
 & $\xmat(0)$ & $\lambda_{\mathrm{low}}$ & $\lambda_M$ & $\lambda_{\mathrm{up}}$ \\
\hline
Heisenberg & $4/\pi^2$ & $1$ & $2\ln2\approx1.39$ & $\pi^2/4\approx2.47$ \\
XX & $2/\pi$ & $1$ & $4/\pi\approx1.27$ & $\pi/2\approx1.57$ \\
\hline
\end{tabular}
\end{center}

\subsubsection{The Dicke--Heisenberg diagram: an exact tricritical point at the saturation corner}
\label{sec:results_dheis}
\begin{figure}[tbp]
\centering
\includegraphics[width=\linewidth]{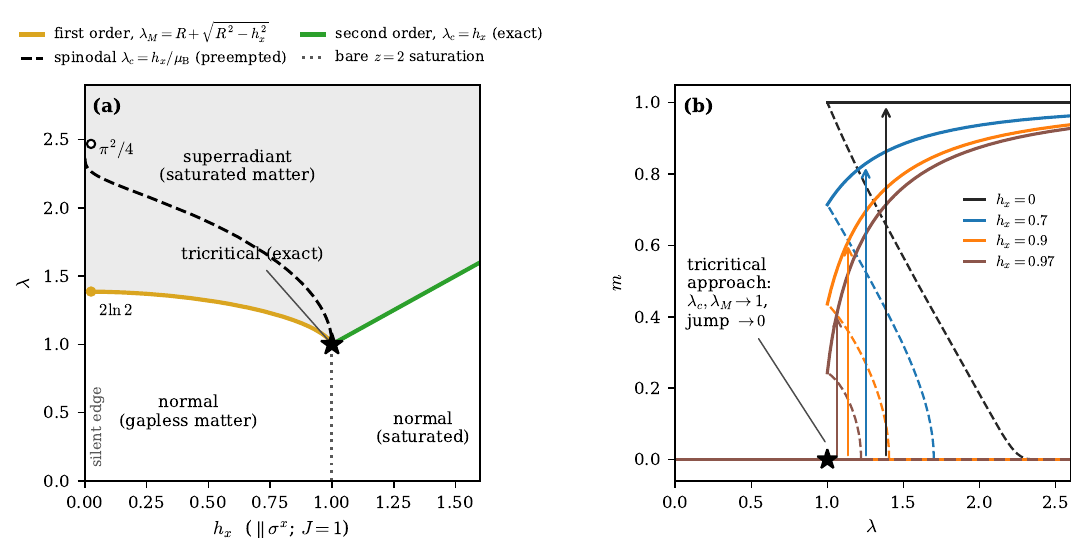}
\caption{The Dicke--Heisenberg diagram. \textbf{(a)} The Heisenberg chain of
\cref{fig:isocurves}b with a
uniform field $h_x$ along $\sigma^x$, every line in closed form from the Bethe magnetisation
curve [\cref{eq:dheis,eq:dheis_tr}; $J=1$]. For $h_x<1$ the transition is first order (gold), the
jump shrinking continuously to zero at the saturation corner $(1,1)$ (star), where the
second-order line $\lambda_c=h_x$ (green, exact) takes over: an exactly solved tricritical point
on the bare $z=2$ saturation line (dotted). The dashed spinodal
$\lambda_c=h_x/\mu_{\mathrm{B}}$ is preempted throughout (no stable gapless superradiant phase).
The conserved, spectrally silent case is the $h_x=0$ edge; for $h_x>0$ a sharp Larmor polariton
rides on the gapless matter (see text). \textbf{(b)} How the
fold dies: stationarity curves at fixed $h_x$, from the same Bethe data---normal branch ($m=0$),
gapless interior branch (dashed, the unstable barrier), saturated branch (solid),
with the Maxwell jump as an arrow. As $h_x\to1$ the spinodal and Maxwell coupling close
on the corner ($\lambda_c,\lambda_M\to1$, star) and the jump vanishes---the
tricritical point of panel (a) in curve space.}
\label{fig:dheis}
\end{figure}

The silent, conserved case is in fact the edge of a larger diagram that remains exactly
solvable. Switch on, for the Heisenberg chain, a uniform field $h_x$ along $\sigma^x$
(\cref{sec:model_iso}): the conservation is broken, and because the exchange is SU(2)-symmetric
the external and the self-consistent field combine into one tilted field,
\begin{equation}
\evar(m)=\tfrac{\lambda}{2}m^2+e_{\mathrm{B}}\!\Big(\sqrt{h_x^2+\lambda^2m^2}\Big),
\label{eq:dheis}
\end{equation}
with $e_{\mathrm{B}}$ the same Bethe energy as above---every statement below reduces to a closed-form expression in
the magnetisation curve $\mu_{\mathrm{B}}$ (\cref{fig:dheis})---itself closed form for XX, and obtained from the Bethe integral equations for the Heisenberg chain. The transverse response at
$m=0$ is exact by a rotation argument, $\xmat(0)=\mu_{\mathrm{B}}(h_x)/h_x$, so the would-be
onset sits at $\lambda_c(h_x)=h_x/\mu_{\mathrm{B}}(h_x)$. But $\mu_{\mathrm{B}}/h_x$ \emph{rises}
monotonically across the whole gapless phase (Bethe data; in closed form for XX), so $a_4<0$
along the entire line: the gapless branch is barrier everywhere---\emph{a stable gapless
superradiant phase exists nowhere in the diagram}, the same preemption that strips the
antiferromagnetic wedge in \cref{sec:results_1d}---and the realised transition is the Maxwell
jump to the saturated branch,
\begin{equation}
\lambda_M(h_x)=R+\sqrt{R^2-h_x^2},\qquad R(h_x)=1+e_{\mathrm{B}}(1)-e_{\mathrm{B}}(h_x),
\label{eq:dheis_tr}
\end{equation}
recovering $2\ln2$ at $h_x=0$. The jump $\Delta m=\sqrt{\lambda_M^2-h_x^2}/\lambda_M$
shrinks from $1$ and vanishes \emph{continuously} at the saturation corner $(h_x,\lambda)=(1,1)$;
beyond it the matter is gapped and the onset $\lambda_c=h_x$ is exact and second order ($a_4>0$).
The corner is therefore an exactly solved \emph{tricritical point}, sitting on the bare $z=2$
saturation line. In the language of \cref{sec:lp} it is a single statement: the Larkin--Pikin
anchor---the point on the curve where the bare matter is critical, here the saturation edge
$\sqrt{h_x^2+\lambda^2m^2}=1$, i.e.\ $\lambda m=\sqrt{1-h_x^2}$---slides to the axis as
$h_x\to1^-$ and \emph{merges with the superradiant onset} at the corner, having left the physical
curve altogether for $h_x>1$, where the matter is saturated throughout. What singles the
saturation edge out is that it is \emph{one-sided}: $\xmat$ diverges only from the gapless side,
so the sign of $h_x-1$ flips the Larkin--Pikin verdict, and the corner is the flip point. It is
in this sense a tricritical point of an unusual kind: the onset quartic does not soften but
\emph{diverges}, $a_4\to-\infty$ with the saturation divergence of $\xmat$, while the realised
jump instead dies by confinement, squeezed against the saturation cap---a divergent local
coefficient coexisting with a vanishing jump, the order and size of the transition fixed by the
global curve, not by the local Landau coefficients. And it
makes a sharp pair with the quadruple point of \cref{sec:results_qp}: both are corners where a
bare matter-critical line meets the superradiant boundary, but there the first-order jump stays
finite and scale-invariant into the corner, while here it dies continuously. The spectral
silence lifts the moment $h_x>0$, and in the sharpest possible way:
$[\hat H_m,S^\pm_{x}]=\pm 2h_xS^\pm_{x}$ exactly (the ladder operators about the field axis;
the factor $2$ is the $\sigma$-convention's Zeeman splitting), so
the \emph{entire} uniform weight of the cavity channel sits in a single Larmor line at
$\omega=2h_x$---a sharp polariton riding on gapless matter, its weight
$\propto\mu_{\mathrm{B}}(h_x)$ vanishing into the silent edge as $h_x\to0$. With the channel
gapped and the coupling non-conserved, the spectral reading of \cref{sec:spectral} applies
verbatim: the lower polariton closes \emph{exactly} where the stationarity curve leaves the
axis, at the spinodal $\lambda_c(h_x)=1/\xmat(0)$ (on the metastable normal branch, the realised
jump having occurred at $\lambda_M<\lambda_c$). At $h_x=0$, by contrast, the same
onset has \emph{no spectral precursor at all} (\cref{sec:results_iso}): one thermodynamic
onset, spectrally visible at any $h_x>0$ and spectrally invisible at the conserved edge.

The lesson is that gaplessness alone does not produce the marginal exponential onset; only a \emph{divergent}
cavity-coupled susceptibility does---realised by the compass point above. In particular, nothing in
the framework \emph{requires} a gapped normal phase: the functional and the curve see only the
uniform response, and these gapless chains sit in the regular row with a perfectly ordinary
threshold. This populates the
classification of \cref{sec:landau} with concrete worked cases, and the closed form
\cref{eq:xx_curve} is the simplest exactly-solvable instance of a first-order superradiant onset read
as a fold of the single stationarity curve---with the Dicke--Heisenberg diagram as its
two-parameter completion: an exactly solved tricritical point on a bare critical line, and the
silent case resolved into the edge of a spectrum that elsewhere carries a sharp,
symmetry-protected polariton.

\section{Conclusions and outlook}
\label{sec:conclusions}

The thermodynamic Dicke--Ising problem reduces to a single self-consistent matter
functional; how its minima appear and split (its bifurcations) sets the entire phase diagram.
From it we obtained the exact second-order boundaries, the closed-form $a_4=0$ points
$\varepsilon_{\mathrm{tri}}^{\mathrm{ferro}}=|J|/d$ (a tricritical point) and
$\varepsilon_{\mathrm{tri}}^{\mathrm{AF}}=2|J|/\sqrt{8d-3}$ (the first-order endpoint of the
second-order line---itself shown to be preempted by the direct AN$\to$PS transition in every
dimension), both in the rescaled units $J\to J/d$ and both vanishing as $d\to\infty$. A $1/d$ expansion then
locates where all four phase boundaries sit, including the AS phase; the AS--PS boundary is first order
for $d\le3$, forced non-perturbatively by Larkin--Pikin, with two tricritical points in the
$(d,\varepsilon)$ plane above. At the quadruple point the matter reduces to a detuned
Rydberg-blockade chain with a finite AS wedge and a first-order AS--PS line along its full extent. A completeness statement ties the global stationary structure to the number of
bare-matter phases, so the local Larkin--Pikin verdict is never overturned by a hidden branch.

The thread tying these results together---the exact second-order boundaries, the $a_4=0$ points, the
$1/d$ boundary locations, the quadruple point, and completeness---is the \emph{continuity} of the bare
matter transitions. Because the
matter orders continuously, the stationarity curve is one \emph{connected} equation of state: the
Dicke onset and the superradiant--superradiant transitions are arms of the same curve, and the
superradiant first orders are jumps across its folds rather than level crossings of unrelated
sheets. A fold can
straighten where a level crossing of two disconnected sheets---the picture a first-order \emph{bare}
matter transition forces, with $m$ slaved to it---cannot straighten by cavity tuning alone; so the
superradiant--superradiant line can turn continuous, its order parameter vanishing at a tricritical
point above the matter's upper critical dimension ($d_{\mathrm{uc}}=3$ for the $\mathbb{Z}_2$
transition), where the AS--PS interior is second order between two tricritical points---and the same
connectedness is what lets one functional fix the onset, the order, and the
superradiant--superradiant lines at once. Had the matter transitions been first order, the curve would
have broken into disconnected pieces and far less could be said.

The core mechanism is not special to Ising matter: the decoupling holds for arbitrary $\hat H_m$, the
fourth-order mechanism is vacuum-type generic, and the onset classification applies to any
collectively coupled matter---as the compass, Heisenberg, XX, and frustrated triangular examples make
concrete.

Several directions remain open. They fall into four groups, taken in turn: how far the equilibrium
criterion reaches onto more complex matter---intermediate phases, disorder, and higher dimension; the
cavity-dressed \emph{spectrum} built on the superradiant and critical vacua; the coupling and
entanglement questions \emph{specific} to the cavity; and the finite-$N$ and non-equilibrium regimes
where the irreducibly quantum cavity effects enter. A natural direction \emph{within} the present equilibrium framework is to push
the criterion onto matter with an \emph{intermediate} phase: the same triangular antiferromagnet
(\cref{sec:results_triangular}) in quasi-one dimension---a triangular ladder, equivalently an axial
next-nearest-neighbour Ising chain in a transverse field~\cite{Selke1988}---develops a \emph{floating
phase}, a gapless, critical phase of incommensurate, algebraically decaying
order~\cite{BeccariaCampostriniFeo2007} bounded by \emph{two} transitions of opposite character, on
which the criterion makes opposite predictions. The ordered-side boundary is a
commensurate--incommensurate (Pokrovsky--Talapov) transition with a one-sided divergent susceptibility
$\xmat\sim|\Delta|^{-1/2}$~\cite{PokrovskyTalapov1979}---already enough (\cref{sec:lp}) to fold the
curve into a first-order transition and preempt the floating phase---while the disordered-side
boundary is a Berezinskii--Kosterlitz--Thouless
transition~\cite{Berezinskii1971,KosterlitzThouless1973} whose essential singularity leaves $\xmat$
finite, so the curve need not fold and a continuous, non-mean-field superradiant line can survive;
which of the two wins is tunable by the lattice
anisotropy~\cite{IglovikovScalettarSingh2013,ChepigaMila2022}. (Our preliminary stationarity-curve
calculations on the zigzag strip, not shown, indeed find the Pokrovsky--Talapov side preempting---the
BKT side, by contrast, leaves $\xmat$ finite through its essential singularity, so the curve there is
locally stable against folding---and earlier numerics on cavity-coupled frustrated chains likewise
did not find the floating phase realised~\cite{Schiller2021thesis}; whether
fine-tuning can instead realise a BKT-terminated cavity transition we leave open.)

A natural counterpart is to look
deliberately for matter that \emph{violates} the single-hump susceptibility assumption of
\cref{sec:completeness}: any extra superradiant minimum off the single hump would generically appear
by a fold, i.e.\ as a first-order transition, and the global stationary structure could then depart
from the local verdict. The sharper version asks not merely for an extra minimum but for a
\emph{continuous} one---a second minimum splitting off the curve at $m>0$ by a critical bifurcation
rather than a fold. This would need a fine-tuned degeneracy ($\evar''=\evar'''=0$) there and does not
occur generically: a \emph{continuous} off-axis bifurcation needs both conditions to coincide,
whereas a generic $q=0$ hump in the curve at $m>0$---critical or not---delivers only a first-order
fold.

Quenched randomness pushes the criterion from a third side. Weak randomness rounds first-order
transitions in low dimension~\cite{AizenmanWehr1989}, and at disordered fixed points the
correlation-length bound $\nu\ge2/d$~\cite{Harris1974,CCFS1986} translates, \emph{where
hyperscaling holds}, into a non-positive specific-heat exponent. A non-positive $\alpha$ leaves the
singular part of the field-conjugate response finite at criticality, and it is precisely the
\emph{divergence} of that response that Larkin--Pikin requires to fold the curve; a random fixed point
therefore offers exactly the finite-response, $\alpha\le0$ criticality on which a \emph{continuous}
superradiant--superradiant transition can survive (activated scaling included). Quenched randomness is thus a candidate route to
\emph{continuous} superradiant--superradiant transitions in the physical dimensions---does the
fold survive disorder?

A cleaner knob on the same fold is the range of the matter interaction. At zero longitudinal field
the ferromagnet is itself $\mathbb{Z}_2\times\mathbb{Z}_2$, the matter Ising symmetry alongside the
photon parity, and the Landau reading of a single first-order FN$\to$PS line, both breaking together,
misses what \cref{sec:results_guide} made explicit: an ordered-superradiant state, a ferromagnetic
superradiant (FS) phase that is the counterpart of the AS phase, lies on the connected curve, present
but unstable, folded away by the divergent $(d{+}1)$-Ising susceptibility through Larkin--Pikin. With
sufficiently long-ranged interactions ($1/r^{d+\sigma}$, $\sigma<2d/3$) the matter criticality turns
long-range Gaussian; because the cavity-conjugate response inherits the matter's specific-heat
singularity (\cref{sec:lp}), it stays finite ($\alpha<0$, with decay-dependent $\nu=1/\sigma$,
$z=\sigma/2$)~\cite{FeySchmidt2016,Koziol2021}, so the divergence that \emph{forces} the fold is gone.
What this removes is the forcing, not the verdict. With $\chi$ now finite, the FS phase is governed by
two bare-matter responses (as for the AS phase, \cref{sec:results_1d}, though here, with no second
sublattice and so no staggered order to relax, the near-critical response is simply $\chi_{\mathrm{reg}}$): the onset coefficient
$a_4$, a low-order response (\cref{app:a4}) whose sign the long range could potentially change from
its short-range $a_4<0$, and the curvature $1-\lambda_*\chi_{\mathrm{reg}}$ where the curve meets the
matter critical line, the near-critical response taken from the matter at $h_c$. Their signs give
four cases. (i)~$a_4>0$, $\lambda_*\chi_{\mathrm{reg}}<1$: FN$\to$FS and FS$\to$PS both second order,
a genuine FS phase between two continuous lines (the shape the large-$d$ antiferromagnet already
shows). (ii)~$a_4>0$, $\lambda_*\chi_{\mathrm{reg}}>1$: FN$\to$FS second order, FS$\to$PS first order.
(iii)~$a_4<0$, $\lambda_*\chi_{\mathrm{reg}}<1$: FN$\to$FS first order, FS$\to$PS second order, a case
the present model does not realise. (iv)~$a_4<0$, $\lambda_*\chi_{\mathrm{reg}}>1$: both first order.
Across all four, the global Maxwell construction can still pre-empt FS with a direct first-order
FN$\to$PS jump, leaving the FS branch metastable, as it does for the short-range $\varepsilon=0$ chain
(\cref{fig:tfim}) and further away from the quadruple point in the one-dimensional antiferromagnet
(\cref{fig:c2preempt}). Trapped-ion chains, which tune the decay across $\sigma=2/3$ ($p<5/3$ in one
dimension), make the concrete model a potentially realistic test case for experimental platforms.

A fourth direction is dimensional: in higher dimensions the
Dicke--Heisenberg diagram of \cref{sec:results_iso} sits exactly on the tricritical margin
(classically $a_4\equiv0$), and the sign of the quantum correction decides whether a stable,
Goldstone-carrying gapless superradiant phase exists. This question is already contained in the
magnetisation curve $m(h)$ of the square-lattice antiferromagnet, since $a_4$ is fixed by that curve's
low-field expansion (\cref{app:a4}); the curve is accessible to sign-free quantum
Monte Carlo and, analytically, to the $1/S$ expansion.

And since the decoupling and the functional hold at \emph{any} temperature, the
onset/order/fold analysis is expected to carry over to thermal phase boundaries and tricritical lines---an
axis we have not worked out, but one with a first exactly solvable case: Otake and
Bamba~\cite{otake2025} couple a \emph{classical} Ising chain to the cavity along its ordering
axis and solve the self-consistency in closed form, the photon-mediated field producing a
finite-temperature transition the bare chain cannot have on its own. A natural \emph{quantum} counterpart is the symmetric compass chain ($\Delta=0$), whose free-fermion solution extends to finite temperature: a Lee--Gammelmark-style analysis there could track how its zero-temperature, Berezinskii--Kosterlitz--Thouless-type onset (\cref{sec:results_compass}) fares at finite $T$.

A different equilibrium direction is the \emph{spectrum} rather than the ground state.
\Cref{sec:spectral} gives its first word---stable branches carry a true $q=0$ polariton, and a fold
is exactly where it softens. The known effective-Dicke mappings for correlated matter are confined to
the \emph{normal} phase~\cite{Schellenberger2024}; the corresponding effective Dicke model \emph{on a
superradiant vacuum} we construct exactly at $d=\infty$ (\cref{app:dinf_dicke}), where the lower
polariton on the antiferromagnetic-superradiant vacuum softens at the AS--PS transition with its
photon weight vanishing as $|t|$. Doing the same at finite $d$, where this soft mode overlaps the
matter continuum (below) rather than a sharp pole, is open. The \emph{matter} side of the spectrum is equally open. Three questions stand out: how the
matter gaps close along the transition lines (in ordinary form, as one would expect on the
continuous $3$D-XY line of the triangular antiferromagnet?); what the corresponding
\emph{dynamic} correlation functions would show; and what becomes of multi-particle excitations, in particular whether cavity-induced bound states form in
our spectrum, as the bound polaritons Ref.~\cite{roman2025bound} finds in the related
\emph{transverse}-field Dicke--Ising chain. All three are unexplored. Because the matter vacuum
is itself correlated---and, along this $3$D-XY line
(\cref{sec:results_triangular}), critical---this is richer than the bare Dicke case: the collective
mode lives on a matter continuum it can hybridise with rather than in a gap, and the displacement
decoupling underlying our analysis is controlled only at the level of the \emph{extensive}
ground-state energy, leaving its fate at finite, $O(1)$ excitation energy open. Settling this
superradiant--superradiant transition quantitatively, beyond the present resolution, calls for dedicated
large-scale quantum Monte Carlo: it means pinning both the critical field $\Gamma_c$---currently known
only to $\sim10$--$15\%$ across iDMRG and quantum Monte Carlo estimates---and the matter
susceptibility at it. The data in hand favour the continuous, $3$D-XY scenario. A further reach
of the same spectrum is upward in energy: the unstable stationary points our construction
crosses---the barriers---are not eigenstates but loci of excited-state quantum phase transitions,
where the many-body density of states is singular along the classical
separatrix~\cite{Caprio2008,Emary2003}; their fate in the full excited spectrum, and in driven,
non-equilibrium settings, is left open here.

The remaining directions are specific to the cavity. Still within equilibrium: for which
light--matter couplings does the decoupling survive? Photon self-interactions (a Kerr or other
anharmonic term) and couplings nonlinear in the matter operator preserve the normal
(non-superextensive) extensivity of the ground-state energy yet are no longer removed by a pure
displacement; whether an exact matter functional still emerges defines the reach of the framework.
The entanglement question sharpens the same point: what does \emph{genuine} light--matter
entanglement in the thermodynamic limit require---is it zero for \emph{any} finite number of
collective modes, and does generating it take extensively many? To anchor the question: in a
symmetry-broken superradiant phase the displacement-decoupled ground state is a pure product of a
coherent photon and the matter state---zero light--matter entanglement at leading order---and the
only finite piece, the $\ln 2$ of the $\mathbb Z_2$ cat doublet, is subextensive, leaving the
energy density and hence the entire phase diagram untouched. What the question targets is the \emph{genuine}, $O(1/N)$ entanglement
beyond it. With $M$ modes the present
construction generalises to $M$ coupled self-consistency conditions---a landscape
$\evar(m_1,\dots,m_M;\lambda)$ whose geometry (here: one curve and its folds) is itself an open
structural question. The simplest instance is already at hand: a $U(1)$-symmetric cavity coupling
gives the two-component isotropic case $\evar(m_x,m_y)$, where the broken symmetry selects only
the modulus, Goldstone fluctuations decouple from the transition, and the amplitude carries the
whole bifurcation structure. And the displacement-decoupled regime treated here is not the only
one: global cavity \emph{fluctuations} acting in a symmetric sector can themselves stabilise
strongly correlated matter~\cite{Mann2025}, a direction complementary to everything above. For
that repulsive---necessarily drive-engineered---sign of the collective term the functional's
verdict is in fact immediate: the displacement vanishes and the energy density stays bare, so
all such physics is subextensive state selection, beneath the density-level resolution of the
present theory.

At finite $N$ the question becomes structural. The strict $N\to\infty$ result is a minimisation
over a single global variable; can the $1/N$ corrections be captured by \emph{local} effective
Hamiltonians with additional \emph{global} self-consistent parameters---in which case the
framework is systematically improvable---or does genuine spatial non-locality enter, making
$N=\infty$ a singular limit? The question has practical bite: variational treatments that
truncate the light--matter correlations to a few global parameters---such as the single
polaritonic dressing of Ref.~\cite{mendoncca2025}, whose ground states do not show the narrow AS
wedge established here, while the truncation-free wormhole quantum Monte Carlo of
Ref.~\cite{langheld} does see it---are precisely such ans\"atze; whether the wedge is lost to the
ansatz or simply to resolution we cannot tell from outside, and knowing which global parameters
suffice would tell where such truncations can be trusted. Computing the first $1/N$ correction for a
tractable case---with the quadruple point and the frustrated magnets as natural testing
grounds---would decide; the $1/N$ graph expansions of light--matter systems now being
developed~\cite{Schellenberger2026} are one but not the only route to address that problem. Beyond equilibrium lie quench and driven dynamics, with the \emph{folds} as
natural targets (a soft collective mode meeting a first-order jump is the equilibrium shadow
of switching dynamics). Sharper still is whether \emph{transient} light--matter
entanglement appears at large $N$ during relaxation, which asks whether the decoupling is a property of the
thermodynamic limit itself or only of its equilibrium states. These axes---finite $N$ and beyond
equilibrium---are where the irreducibly quantum cavity effects might enter: at finite $N$, the
$O(1/N)$ entanglement and photon squeezing established in \cref{sec:decoupling}; beyond equilibrium,
whatever transient light--matter correlations may survive relaxation. The equilibrium thermodynamics treated
here introduces no phase types beyond the matter phases plus self-consistency, a consequence of
the exact thermodynamic-limit decoupling rather than an assumption.

\begin{acknowledgments}
I thank Sumeet for reviewing the figures. I gratefully acknowledge Anthropic's Claude
(Claude Code, primarily the Claude Opus~4.8 and Claude Fable~5 models): we discussed the
established literature back and forth, with Claude cross-checking it and pointing to connections with
known mechanisms (notably the Larkin--Pikin mechanism); it also helped to carry out and check
the calculations, produce the figures, and write the manuscript. The direction of the research
and responsibility for its content are mine.
\end{acknowledgments}

\noindent\textit{Code and data availability.}---The scripts and data that reproduce the figures
and numerical results of this paper---the free-fermion and $1/d$ analyses, the strict-blockade
infinite-system DMRG at the quadruple point, and the self-consistent He--Hamer--Oitmaa Pad\'e
construction for the triangular antiferromagnet (with the digitised $L_y=6$ iDMRG data of
Saadatmand \textit{et al.}~\cite{Saadatmand2018} used to validate it)---are openly available on
Zenodo (DOI: \href{https://doi.org/10.5281/zenodo.20746670}{10.5281/zenodo.20746670}).

\appendix
\addtocontents{toc}{\protect\appendixtocfix}

\section{The low-field expansion: linked clusters and the tricritical coefficients}
\label{app:a4}

\subsection{Vacuum, broken symmetry, and the background field}
\label{app:a4_setup}

The Landau coefficients of \cref{eq:landau} are low-field Taylor coefficients of the bare-matter
energy: writing the per-site ground-state energy in the field $h$ conjugate to $\sigma^z$ as
\begin{equation}
\emat(h)=e_0-\tfrac12\,\xmat(0)\,h^2+\frac{a_4}{\lambda^4}\,h^4+\frac{a_6}{\lambda^6}\,h^6+O(h^8),
\label{eq:lowfield}
\end{equation}
the substitution $h=\lambda m$ in \cref{eq:functional} returns exactly the Landau series
\cref{eq:landau} [the $h^4$ coefficient is $-c_4/24$ in the notation $c_4=\mumat'''(0)$ used
there]. Three ground rules fix the calculation. First, the expansion starts from an \emph{eigenstate} of the bare matter at $h=0$: the
$\sigma^x$-polarised product state for the polarised vacuum, and \emph{one} of the two N\'eel
products for the antiferromagnet. The bare antiferromagnetic ground state is doubly degenerate, so
one chooses a symmetry-broken member. The choice is immaterial in the thermodynamic limit: the two
N\'eel products differ by $N$ flips, so as $N\to\infty$ no finite-order process connects them, and the
broken-symmetry series is therefore the thermodynamic-limit series. Second, the vacuum carries no
$\sigma^z$, so odd orders vanish and $\emat$ is even in $h$---the regular row of
\cref{sec:landau}. Third, the standard pitfall: a flip cluster is never detached from the
lattice. Every cluster site keeps its full coordination $2d$, and the bonds leaving the cluster
attach to the frozen vacuum background, so cluster site $i$ sees the total longitudinal field
$\varepsilon+s_i\,(2d-n_{\mathrm{int},i})\,|J|$ ($n_{\mathrm{int},i}$ its
bonds inside the cluster, $s_i$ the sublattice sign of the background). Each embedded process
therefore carries its \emph{full-lattice} gap---an isolated flip costs the same $\Delta E_s$
wherever it sits---and dropping these boundary bonds is the quickest route to a wrong
coefficient.

\subsection{One transformation, many names}
\label{app:takahashi}

We organise the series in Takahashi's linked-cluster form~\cite{Takahashi1977}, built on Kato's
expansion of the perturbed projector $\bar P$~\cite{Kato1949}. The projected frame
$\{\bar P|i\rangle\}$ is not orthonormal---$\bar P P$ is not an isometry---and the canonical
repair is L\"owdin's symmetric orthonormalisation~\cite{Lowdin1950},
\begin{equation}
T=\bar P P\,(P\bar P P)^{-1/2},
\label{eq:takahashi_T}
\end{equation}
the isometry from the model space onto the perturbed subspace. The resulting Hermitian effective
Hamiltonian $T^\dagger\hat HT$ is the canonical block diagonalisation---the unique one closest to
the identity---that appears in the literature under several names, the exact two-block Schrieffer--Wolff
transformation among them; see Ref.~\cite{HoermannSchmidt2023} for the dictionary and a
cluster-additive generalisation. Nothing below hangs on the choice: the quantities in
\cref{eq:lowfield} are ground-state \emph{energy} coefficients, and the eigenvalue series is the
same in every perturbation scheme; we work in Takahashi's form.

\subsection{Clusters, denominators, and the quartic coefficient}
\label{app:a4_master}

\begin{figure}[tbp]
\centering
\includegraphics{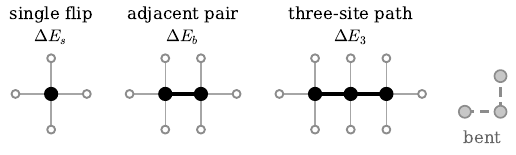}
\caption{The three connected clusters of the low-field expansion through sixth order, drawn for
$d=2$. Filled sites are the flipped (cluster) sites, thick lines their internal bonds; the grey
bonds leave the cluster and attach to the frozen vacuum background (open sites), giving cluster
site $i$ the longitudinal background field $\varepsilon+s_i(2d-n_{\mathrm{int},i})|J|$ of
\cref{app:a4_setup}. Bent (L-shaped) three-site paths (dashed, far right) embed with the same
gaps---$\mathbb{Z}^d$
has no triangles---and count toward the same $d(2d-1)$ paths per site.}
\label{fig:clusters}
\end{figure}

At order $h^{2k}$ each insertion of $V=-h\sum_i\sigma^z_i$ flips one spin and the $2k$ insertions
must return to the vacuum, so every site is flipped an even number of times and a connected
cluster spans at most $k$ sites: at fourth order a single site or an adjacent pair; at sixth
order additionally the three-site path (the bipartite hypercubic lattice has no
triangles)---\cref{fig:clusters}. The fourth-order term of the expansion reads
\begin{equation}
E^{(4)}=\langle0|VSVSVSV|0\rangle-\langle0|VS^2V|0\rangle\,\langle0|VSV|0\rangle,
\qquad S=\frac{Q}{E_0-\hat H_0},\quad Q=1-|0\rangle\langle0|,
\end{equation}
its second term removing the disconnected piece (extensivity). For nearest-neighbour matter on
the hypercubic lattice this assembles, per site, into
\begin{equation}
\frac{a_4}{\lambda^4}=W_1+d\,W_{\mathrm{bond}},\qquad
W_1=\frac{1}{(\Delta E_s)^3}>0,\qquad
W_{\mathrm{bond}}=\frac{2\,(\Delta E_b-2\Delta E_s)}{(\Delta E_s)^3\,\Delta E_b},
\label{eq:a4_split}
\end{equation}
where $\Delta E_s$ is the single-flip and $\Delta E_b$ the adjacent-pair gap above the vacuum.
The bound-pair channel is attractive ($W_{\mathrm{bond}}<0$) whenever $\Delta E_b<2\Delta E_s$,
so the order of the onset is the competition of one unfrustrated single-flip repulsion against
$d$ correlated bound-pair attractions. Collecting,
\begin{equation}
a_4=\lambda^4\,\frac{(1+2d)\,\Delta E_b-4d\,\Delta E_s}{(\Delta E_s)^3\,\Delta E_b},
\label{eq:a4_master}
\end{equation}
written here for a vacuum with a single flip gap; for the N\'eel vacuum the same assembly runs
with the sublattice-resolved denominators below and yields \cref{eq:coeff_af_24}.

The denominators are the gaps of the virtual flip configurations, and they are worth listing.
\emph{Polarised vacuum:} a single flip costs $\Delta E_s=2\varepsilon+4d|J|$ (field plus all $2d$
bonds); an adjacent pair $\Delta E_b=4\varepsilon+4(2d-1)|J|$ (the shared bond is not counted
twice); at sixth order also the separated pair $2\Delta E_s$ (the two ends of the
three-site path) and the path triple flip $\Delta E_3=6\varepsilon+4(3d-2)|J|$. \emph{N\'eel vacuum} ($J>0$,
$0\le\varepsilon<2|J|$): the two sublattices give distinct single-flip gaps
$\Delta E_A=4d|J|-2\varepsilon$ and $\Delta E_B=4d|J|+2\varepsilon$, but sublattice
cancellation---an adjacent pair flips one site on each sublattice---makes
the bound-pair gap \emph{field-independent}, $\Delta E_b^{\mathrm{AF}}=4(2d-1)|J|$. The
contrast between the field-dependent single-flip gaps and the field-independent bound-pair gap is
the structural origin of the antiferromagnetic tricritical scaling: after the rescaling $J\to J/d$,
the locus \cref{eq:eps_tri_AF} goes as $1/\sqrt d$ against the ferromagnet's
$|J|/d$ \cref{eq:eps_tri_ferro}, both vanishing as $d\to\infty$. The two path types contribute
$\Delta E_3^{ABA}=4(3d-2)|J|-2\varepsilon$ and $\Delta E_3^{BAB}=4(3d-2)|J|+2\varepsilon$,
and the separated pairs $2\Delta E_A$, $2\Delta E_B$. The remaining work at sixth order is bookkeeping---the
orderings of the six insertions over the cluster, each with its weight from the sixth-order term
of the expansion of $T^\dagger\hat HT$; the exhaustive enumeration is reproduced by the
exact-rational script in the reproducibility package.

\subsection{The coefficients through sixth order}
\label{app:a4_list}

All coefficients and gaps below are written in \emph{bare} units ($|J|$ literal). The rescaling
$J\to J/d$ of \cref{app:1d}---under which matter mean field is exact at $d=\infty$---is applied
only when converting the bare $a_4=0$ roots and onsets to the tricritical loci quoted in the main
text; it is the positive substitution $|J|\to|J|/d$, which moves the loci but cannot change the
\emph{sign} of any coefficient, so the ordinary-tricritical versus first-order-endpoint
classification is convention-invariant.

For the polarised vacuum (any $\varepsilon\ge0$, every $d$; at $\varepsilon=0$ from one
broken-symmetry member, rule one above),
\begin{align}
\xmat(0)&=\frac{2}{\Delta E_s},\qquad
\frac{a_4}{\lambda^4}=\frac{4\,(\varepsilon-|J|)}{(\Delta E_s)^3\,\Delta E_b},
\label{eq:coeff_ferro_24}\\
\frac{a_6}{\lambda^6}&=\frac{64\,\big[-3\varepsilon^3+10\varepsilon^2|J|
+(12d^2+10d-11)\,\varepsilon|J|^2-4(2d-1)|J|^3\big]}{(\Delta E_s)^5\,(\Delta E_b)^2\,\Delta E_3}.
\label{eq:c6}
\end{align}
The susceptibility puts the second-order onset at $\lambda_c=1/\xmat(0)=\varepsilon+2d|J|$ (the
rescaled $\varepsilon+2|J|$ of the main text). The
quartic vanishes at the bare root $\varepsilon=|J|$
($\varepsilon_{\mathrm{tri}}^{\mathrm{ferro}}=|J|/d$ after rescaling, \cref{eq:eps_tri_ferro}),
and there the sextic collapses to
\begin{equation}
\frac{a_6}{\lambda^6}\bigg|_{\varepsilon=|J|}=\frac{6d+1}{D_d}>0,\qquad
D_d=32\,d\,(6d-1)\,(2d+1)^5\,|J|^5,
\label{eq:c6_tri}
\end{equation}
positive for all $d$: the superradiant branch leaves $m=0$ as a \emph{minimum}, and the
ferromagnetic $a_4=0$ point is a genuine, \emph{ordinary} tricritical point in every dimension.
At $\varepsilon=0$ the sextic reduces to $-1/\big[256\,d^5(2d-1)(3d-2)\,|J|^5\big]$, at $d=1$ the
exact free-fermion value $-1/256$ of the transverse-field Ising chain---one of the anchors below.

For the N\'eel vacuum ($0\le\varepsilon<2|J|$),
\begin{align}
\xmat(0)&=\frac{1}{\Delta E_A}+\frac{1}{\Delta E_B}=\frac{8d|J|}{\Delta E_A\,\Delta E_B},\qquad
\frac{a_4}{\lambda^4}=\frac{64\,d^2J^2\,\big[(8d-3)\,\varepsilon^2-4d^2J^2\big]}
{d\,(\Delta E_A\,\Delta E_B)^3\,\Delta E_b^{\mathrm{AF}}},
\label{eq:coeff_af_24}\\
\frac{a_6}{\lambda^6}&=-\frac{4096\,d\,\mathcal{B}(\varepsilon,J,d)}
{(\Delta E_A)^5(\Delta E_B)^5\,\Delta E_b^{\mathrm{AF}}\,\Delta E_3^{ABA}\,\Delta E_3^{BAB}},
\label{eq:c6_af}\\
\mathcal{B}&=-(8d-3)\,\varepsilon^6J^2+4(2d-1)(32d^2-43d+10)\,\varepsilon^4J^4\nonumber\\
&\quad+16d^2(72d^3-141d^2+94d-20)\,\varepsilon^2J^6+64d^4(3d-2)\,J^8.\nonumber
\end{align}
The susceptibility gives the antiferromagnetic onset in closed form,
$\lambda_c=1/\xmat(0)=(4d^2J^2-\varepsilon^2)/(2d|J|)$ (rescaling to the main-text
$(4J^2-\varepsilon^2)/(2|J|)$, the AN--AS line linear in the
$(\lambda,\varepsilon^2)$ plane of \cref{fig:trikrit}). The quartic vanishes at the bare root
$\varepsilon=2d|J|/\sqrt{8d-3}$ ($\varepsilon_{\mathrm{tri}}^{\mathrm{AF}}=2|J|/\sqrt{8d-3}$ after
rescaling, \cref{eq:eps_tri_AF}), and there
$a_6<0$ in \emph{every} dimension---manifestly, since on that locus the polynomial reduces to
$\mathcal{B}=1024\,d^4(2d-1)^2(9d^2-8d+2)\,J^8/(8d-3)^2$, positive for all
$d\ge1$ (the quadratic $9d^2-8d+2$ has negative discriminant), while every gap in the denominator stays positive: the
branch leaves $m=0$ as a \emph{maximum}, so $m=0$ is locally unstable and the antiferromagnetic
$a_4=0$ locus marks a first-order point rather than an ordinary tricritical one
(\cref{sec:results_tri}).

All coefficients are evaluated in exact rational arithmetic on the three clusters and
cross-checked four ways: against chain exact diagonalisation ($d=1$, $L\le16$) and torus exact
diagonalisation ($d=2$) at and away from the tricritical points; at $\varepsilon=0$ against the
free-fermion chain; and through the bipartite identity that the polarised and N\'eel series must
coincide at $\varepsilon=0$---they do, order by order, with $h^2,h^4,h^6$ coefficients
$-1/(4d)$, $-1/\big[64d^3(2d-1)\big]$, $-1/\big[256d^5(2d-1)(3d-2)\big]$ (units $|J|=1$).

\subsection{Embedding in \texorpdfstring{$d$}{d}: bonds per site, not coordination}
\label{app:lce}

The per-site assembly on the hypercubic lattice $\mathbb{Z}^d$ is
\begin{equation}
\frac{a_{2k}}{\lambda^{2k}}=W_1+d\,W_2+d(2d-1)\,W_3+\dots,
\label{eq:lce_embed}
\end{equation}
with $W_b$ the reduced contribution of the $b$-site cluster and the prefactors the embedding
numbers of $\mathbb{Z}^d$: $d$ bonds per site and $\binom{2d}{2}=d(2d-1)$ three-site
paths per site (for the N\'eel vacuum the paths split evenly into $ABA$ and $BAB$ types). Two
remarks. The variable that organises the series is $d$---the number of \emph{bonds per site},
$z/2$---not the coordination number $z$: the linear-in-$d$ growth of the bound-pair attraction in
\cref{eq:a4_split} counts bonds. And the clean one-parameter family in $d$ is a \emph{hypercubic}
statement: on other lattices the embedding numbers differ and, from sixth order on, triangle and
loop clusters enter (the embedding machinery is the standard linked-cluster
one~\cite{OitmaaBook}); ``in every dimension'' in this paper always means the hypercubic lattice
$\mathbb{Z}^d$.

\subsection{The hardcore-boson reading}
\label{app:boson}

On the product vacuum, flips are hardcore bosons $b_i^\dagger$, and the cavity couples only to
their uniform superposition $B=N^{-1/2}\sum_ib_i$---the $k=0$ mode. At the quadratic level the
coupled problem is exactly the effective Dicke model of \cref{sec:spectral}---gap $\Delta E_s$,
coupling from the matrix elements of $\xmat(0)$---and this level is rigorous: the onset is its
lower polariton softening at $\lambda_c=1/\xmat(0)$. The quartic
level reads as boson--boson interactions: $W_1>0$ is the on-site (hardcore) repulsion---the
only piece a structureless collective model keeps---and
$d\,W_{\mathrm{bond}}<0$ (when $\Delta E_b<2\Delta E_s$) is a nearest-neighbour pair binding on
the $d$ bonds per site. This reading is an \emph{interpretation} of the fourth-order
coefficients---their structure and the factor $d$ are exact---not an independent boson
derivation; what it makes plain is that the $k=0$ truncation by itself retains only $W_1$ and so
always predicts a continuous onset: when the binding dominates, the single-mode picture
mispredicts the order (\cref{sec:spectral,sec:results_tri}).

\section{The \texorpdfstring{$1/d$}{1/d} expansion of the superradiant phases}
\label{app:1d}

Under the standard rescaling $J\to J/d$, matter mean field becomes exact at $d=\infty$, and the
phases are obtained by minimising $\evar(m)=\tfrac\lambda2 m^2+\emat(\lambda m)$
[\cref{eq:functional}] jointly over the
matter product state and the cavity mean field $m$. The $O(1/d)$ correction to $\emat$ is a
single-bond second-order perturbation theory about the mean-field reference---the one method of
\cref{sec:results_1d}, run identically for the polarised and the ordered phases.

For the algebra of this appendix we relabel the axes so the Ising interaction is diagonal---a
global $\sigma^x\!\leftrightarrow\!\sigma^z$ rotation of \cref{eq:H} that changes nothing physical
(the convention used in the numerics, where the figures are labelled directly in $m$): the bond is
then $(J/d)\sigma^z_i\sigma^z_j$, the longitudinal field $\varepsilon$ couples to $\sigma^z$, the
cavity (superradiant) magnetisation is $m=\langle\sigma^x\rangle$, and the staggered Ising order is
$\langle\sigma^z\rangle$. \emph{Throughout the paper $m$ denotes the cavity (superradiant)
magnetisation regardless of the axis label}---$\langle\sigma^z\rangle$ in the main text,
$\langle\sigma^x\rangle$ here, the same physical quantity in both.

\emph{The counting that everything rests on.} The bond coupling is $J/d$ and each site has $d$
bonds, so there are only two classes of terms, a power of $d$ apart: a term contributing at
\emph{first} order in the bond coupling enters at $d\cdot(J/d)=O(1)$ per site---no perturbation at
all---while one that starts at \emph{second} order enters at $d\cdot(J/d)^2=O(1/d)$. (A coupling
acting on single sites would sit at $O(1)$ outright; only the \emph{bond} interaction, its $1/d$
coupling set against the $d$-fold bond count, can be organised this way.) The expansion exists
because the rotation onto the self-consistent mean-field axis sorts the Hamiltonian \emph{exactly}
into these two classes. Rotating a bond $(J/d)\sigma^z_i\sigma^z_j$ onto the canted single-site
quantisation axis produces, besides the genuine two-flip fluctuation, a \emph{single-site} piece;
summed over the $2d$ bond ends at a site it is an $O(1)$ field---first class, not a
perturbation---so it must be absorbed into $\hat H_0$, and demanding the absorbed field be
stationary is precisely the mean-field canting condition. After the absorption the bookkeeping
closes exactly: the subtracted fluctuation $\delta\sigma^z$ has no diagonal matrix element in the
rotated ground state---it acts as a pure single flip---so each bond's first-order contribution
vanishes \emph{identically}, and its leading contribution is second order,
$d\cdot(J/d)^2=J^2/d$: the entire $O(1/d)$ term, with no $O(1)$ remainder. The counting also
closes upward---one-flip cross-terms between adjacent (L-shaped) bonds vanish, and two-bond
clusters first enter at $(J/d)^4 d^2=J^4/d^2$: after the mean-field rotation the bond interaction
truncates \emph{exactly} in $1/d$.

Explicitly, for our model: let the single-site reference $|0_i\rangle$ point along its mean-field
axis, with longitudinal and transverse weights $s_i=\langle\sigma^z_i\rangle$ and $t_i$
($s_i^2+t_i^2=1$; the two sublattice angles of the AS paragraph below give
$s_{A,B}=\sin(\beta\pm\alpha)$, $t_{A,B}=\cos(\beta\pm\alpha)$). Then
\begin{equation}
\frac{J}{d}\,\sigma^z_i\sigma^z_j
=\frac{J}{d}\,s_is_j
+\frac{J}{d}\big(s_j\,\delta\sigma^z_i+s_i\,\delta\sigma^z_j\big)
+\frac{J}{d}\,\delta\sigma^z_i\,\delta\sigma^z_j,
\qquad
\delta\sigma^z_i\,|0\rangle=t_i\,\ket{\mathord\uparrow_i},
\label{eq:bond_split}
\end{equation}
with $\ket{\mathord\uparrow_i}$ the flipped site. The first term is a constant; the middle,
single-site terms---summed over the $2d$ bond ends at a site---are the $O(1)$ field absorbed into
$\hat H_0$; and the last term has exactly \emph{one} nonzero matrix element from the reference,
$(J/d)\,t_it_j$ onto the strict two-flip state. Its first-order expectation therefore vanishes,
and second order gives, per site,
\begin{equation}
\frac{e^{(1)}_{\mathrm{mat}}}{d}
=-\frac{J^2\,(t_it_j)^2}{d\,\Delta E_{2\mathrm{flip}}},
\label{eq:bond_nlo}
\end{equation}
with $\Delta E_{2\mathrm{flip}}$ the gap of the doubly flipped state above the reference (its
$O(1/d)$ bond part is beyond this order). Evaluated at each phase's reference,
\cref{eq:bond_nlo} \emph{is} the entire $O(1/d)$ term: in the polarised phase $t=\cos\beta$ and
$\Delta E_{2\mathrm{flip}}=4\Delta_0$ ($\Delta_0$ the polarised single-site mean-field gap
parameter, $\Delta_0^2=h^2+(\varepsilon-2J\sin\beta)^2$, derived at the end of this appendix; a
single flip costs $2\Delta_0$) give
$-J^2\cos^4\beta/(4d\,\Delta_0)$, and at the canted
antiferromagnetic saddle $t_At_B=\cos^2\alpha+\cos^2\beta-1$ reproduces the
$e^{(1)}_{\mathrm{mat}}$ of the antiferromagnetic paragraph below.
We do not pursue the expansion beyond next-to-leading order. Two
structural facts then make the $O(1/d)$ term explicit.

\emph{(i) Envelope theorem.} The mean-field reference is chosen stationary, so terms linear in
the fluctuation $\delta m$ vanish and the $O(1/d)$ energy is evaluated at the leading
self-consistent magnetisation alone---the shift $\delta m$ does not contribute (Hellmann--Feynman).
\emph{(ii) Self-consistent collapse.} In the polarised phase the leading-order cavity
self-consistency $m=\mumat(\lambda m)$ together with the matter mean field fixes no number by
itself---it leaves a scale freedom, which we use to choose energy units with $\lambda=\tfrac12$
at leading order, independently of $\varepsilon,J$ (the scale-fixed units
of this appendix; invariant statements are ratios such as $J_c/\lambda$ in
\cref{sec:results_1d}), collapsing the
mean-field quartic and giving the closed polarised energy
\begin{equation}
\evar^{*}_{\mathrm{PS}}(\varepsilon,J,d)=-\tfrac14-\frac{\varepsilon^2}{4J+1}
-\frac{8J^2}{d}\Big[\tfrac14-\frac{\varepsilon^2}{(4J+1)^2}\Big]^2+O(1/d^2).
\label{eq:PS_1d}
\end{equation}
\emph{The antiferromagnetic phase: bipartite canted reference and its saddle.} In the AS phase both
order parameters are on, so the reference is a two-sublattice product state. We parametrise the two
sublattices by Bloch angles $\beta\pm\alpha$ in the plane,
$\langle\sigma^z\rangle_{A,B}=\sin(\beta\pm\alpha)$, $\langle\sigma^x\rangle_{A,B}=\cos(\beta\pm\alpha)$,
so the uniform and staggered components are $\sin\beta\cos\alpha$ and $\cos\beta\sin\alpha$.
With the longitudinal field $h_l=\varepsilon$ and the cavity field $h_x$, the leading variational
energy per site is
\begin{equation}
e^{(0)}(\alpha,\beta)=J(\sin^2\beta-\sin^2\alpha)-h_l\sin\beta\cos\alpha-h_x\cos\beta\cos\alpha .
\end{equation}
Stationarity gives, on the canted branch, $2J\cos\alpha=h_l\sin\beta+h_x\cos\beta$ and
$\tan2\beta=2h_lh_x/(4J^2+h_x^2-h_l^2)$, which collapse the energy to the compact
$e^{(0)}_{\mathrm{mat}}=-J(\cos^2\alpha+\cos^2\beta)$. The cavity is closed by the envelope theorem
(i): $\partial_{h_x}e^{(0)}=-\cos\alpha\cos\beta$, so $\partial_m\evar^{(0)}=0$ fixes the leading
mean field $\cos\alpha\cos\beta$, i.e.\ the self-consistent cavity \emph{field}
$h_*\equiv\lambda m=\lambda\cos\alpha\cos\beta$ (the convention of
\cref{eq:AS_1d}).
Eliminating $\alpha,\beta$ between these three conditions gives the closed forms
\begin{equation}
h_*^2=\frac{\varepsilon(4J-\lambda)}{\eta}-4J^2-\varepsilon^2,\qquad
\cos^2\alpha+\cos^2\beta=\frac{\varepsilon}{2J\eta},\qquad
\eta\equiv\sqrt{1-\tfrac{\lambda}{2J}},
\end{equation}
whence $e^{(0)}_{\mathrm{mat}}=-\varepsilon/2\eta$ and
$\evar^{(0)}_{\mathrm{AS}}=h_*^2/(2\lambda)+e^{(0)}_{\mathrm{mat}}=-J-(\varepsilon-2J\eta)^2/(2\lambda)$.
Everything is anchored at $d=\infty$ from the start: the angles and $h_*$ above are the
\emph{leading-order} self-consistent solution, and they are never corrected---the $O(1/d)$ shift
of $m$ itself is not computed at all, because by stationarity of the reference [envelope theorem
(i)] it would enter the energy only at $O(1/d^2)$. The $O(1/d)$ term is \cref{eq:bond_nlo}
evaluated \emph{on} that $d=\infty$ saddle.
The single-bond term \cref{eq:bond_nlo},
$e^{(1)}_{\mathrm{mat}}/d=-J^2(\cos^2\alpha+\cos^2\beta-1)^3/(4d\,h_x\cos\alpha\cos\beta)$, evaluated
at this saddle with $\cos^2\alpha+\cos^2\beta-1=(\varepsilon-2J\eta)/2J\eta$ and
$h_x\cos\alpha\cos\beta=h_*^2/\lambda$, yields the $O(1/d)$ correction and reproduces
\cref{eq:AS_1d},
\begin{equation}
\evar_{\mathrm{AS}}=-J-\frac{(\varepsilon-2J\eta)^2}{2\lambda}-\frac{\lambda(\varepsilon-2J\eta)^3}{32\,d\,J\eta^3 h_*^2}+O(1/d^2).
\end{equation}
The two edges of the phase are where each order vanishes: $h_*=0$ gives the AN--AS line
$\varepsilon=2J\eta$, and $\cos\alpha=1$ (antiferromagnetic order off) gives the AS--PS line
$\varepsilon=(2J+\lambda)\eta$, i.e.\ \cref{eq:AS_lines}. At $d\to\infty$ these match the
mean-field result of Zhang \textit{et al.}~\cite{Zhang2014}, as quoted in
\cref{sec:results_1d}. (The full rational arithmetic and the $1/d$ coefficients, including the vanishing of the connected
multi-bond ($b\ge2$) clusters at this order, are checked by the scripts in the reproducibility
package; the single-bond power-counting (leaf) argument and the global minimality of the AS branch
between the two edges are the analytic statements established above.)

The \emph{order} of the AS--PS line in the physical dimensions cannot be read from this series:
the $d=\infty$ bifurcation is second order, while in $d\le3$ the first order is forced
non-perturbatively by the divergent $(d{+}1)$-Ising susceptibility (\cref{app:lp}), invisible to the
$1/d$ series at any order. The NLO branch crossing is a valid $1/d$ estimate of where the boundary
sits, not an order indicator. This is because at the $d=\infty$ critical point the AS and PS states
coincide, the staggered order vanishing there, so $\evar_{\mathrm{AS}}=\evar_{\mathrm{PS}}$ at that
point at every order in $1/d$. What fixes the order is not this equality but whether the two branches
meet there tangentially (second order) or with a kink (first order), and a finite $1/d$ truncation can
render a true tangency as a spurious kink (the bare hypercubic transverse-field Ising model does the
same). What decides the order is the curvature $\mathcal{S}_{\mathrm{AS}}$, taken up next.

\emph{The order at large dimension.} Above the upper critical dimension of the matter point
($d{+}1>4$, i.e.\ $d\ge4$) the susceptibility is finite, second order is possible, and---being analytic
in $1/d$---the series is the relevant tool at large $d$. The order is the sign of the \emph{binding}
antiferromagnetic-branch curvature $\mathcal{S}_{\mathrm{AS}}=1-\lambda_*\xmat^R(0)$ at the AS--PS
critical point, the full self-consistent response. It is the AS branch, not the polarised one, that
sets this bound, because the transition is asymmetric, as mean-field transitions of this kind always
are: the matter susceptibility is finite but larger on the antiferromagnetic (AS) side than on the
polarised (PS) side, $\xmat^R(0)>\chi_{\mathrm{reg}}$. A second-order transition needs both branches
stable, so the constraint comes from the less stable, larger-susceptibility side. At $d=\infty$ the
staggered angle vanishes,
$\cos^2\beta_c=\lambda/(2J)$ and $\sin\beta_c=\eta$; the critical cavity field equals the cavity mean
field, $h_c=h_*=\lambda^{3/2}/\sqrt{2J}$, while the \emph{conjugate magnetisation} is
$m_c=\langle\sigma^x\rangle=\cos\beta_c=\sqrt{\lambda/(2J)}$ (so $\lambda_*=h_c/m_c=\lambda$, the
self-consistency check). Relaxing only the uniform Bloch angle $\beta$ gives the regular polarised-side
response $\chi_{\mathrm{reg}}=\mathrm d\langle\sigma^x\rangle/\mathrm dh=\tan^2\beta_c/4J=\eta^2/(2\lambda)$
and $1-\lambda_*\chi_{\mathrm{reg}}=1-\tfrac12\eta^2=\tfrac12+\lambda/4J$; but this is \emph{not} the
binding curvature, because on the AS branch the \emph{staggered} tilt $\alpha$ relaxes with the field as
well. Including it gives the full response and
$\mathcal{S}_{\mathrm{AS}}^{(\infty)}=1-\lambda_*\xmat^R(0)=2\lambda(2J-\lambda)/[J(2J+\lambda)]$
(\cref{eq:Sinfty}), smaller than the regular value by the gap $(3\lambda-2J)^2/[4J(2J+\lambda)]\ge0$,
positive in the interior but vanishing at both ends ($\lambda\to0$ and $\lambda\to2J$). The $O(1/d)$
correction $\mathcal{S}^{(1)}<0$ then drives the marginal ends negative, giving two tricritical points
(a tricritical line in the $(d,\varepsilon)$ plane, $\lambda_{\mathrm{tri}}^{+}\approx2J-0.75J/d$ on the
zero-field side; the quadruple-point side is truncation-sensitive, the corner independently first
order). As an asymptotic expansion the series does not reach the physical dimensions: for $d\le3$ the
$(d{+}1)$-Ising susceptibility diverges and the onset is first order (\cref{app:lp}).

\section{The Larkin--Pikin criterion: three cases}
\label{app:lp}

At the bare-matter critical field $h_c$, write $m_c=\mumat(h_c)$ and split
$\emat=e_{\mathrm{reg}}+e_{\mathrm{sing}}$ with $\chi_{\mathrm{reg}}=-e_{\mathrm{reg}}''(h_c)$
finite. The point $m_c$ is stationary at $\lambda_*=h_c/m_c$ (\cref{sec:anchor}); expanding
$\Delta\evar(\delta)=\evar(m_c+\delta;\lambda_*)-\evar(m_c;\lambda_*)$ with $u=\lambda_*\delta$
(here $\delta=m-m_c$ is a local expansion variable, not the quadruple-point detuning of
\cref{sec:results_qp}),
\begin{equation}
\Delta\evar=\tfrac{K_2}{2}\,\delta^2+e_{\mathrm{sing}}(\lambda_*\delta)+O(\delta^3),\qquad
K_2=\lambda_*\big(1-\lambda_*\chi_{\mathrm{reg}}\big).
\end{equation}
Three universality regimes follow from the singular part:
\begin{itemize}
\item \emph{Case A ($\alpha>0$, power law).} $e_{\mathrm{sing}}=-A|u|^{2-\alpha}$ with $A>0$:
since $|\delta|^{2-\alpha}/\delta^2\to\infty$, the singular term dominates and is negative, so
$m_c$ is a local \emph{maximum}---first order, robust.
\item \emph{Case B ($\alpha=0$, log).} $e_{\mathrm{sing}}=B\,u^2\ln|u|$ with $B>0$ (Onsager
$2$D-Ising, or the upper-critical-dimension Ising log): $\Delta\evar=\delta^2[\tfrac{K_2}{2}+B\lambda_*^2\ln\lambda_*+B\lambda_*^2\ln|\delta|]$,
and the $\ln|\delta|\to-\infty$ term dominates, so $m_c$ is again a maximum---first order, but with
an exponentially narrow \emph{local} window $|\delta^*|\sim\exp(-K_2/(2B\lambda_*^2))$ about $m_c$.
This bounds the first-order \emph{character} from below; the size of the actual jump is set by the
global Maxwell construction (\cref{sec:completeness}) and is generically larger (in the classical
compressible problem the analogous equation of state and jump were computed within the
renormalisation group~\cite{BrunoSak1980}). The window $|\delta^*|$ is a half-width in the order
parameter $\delta=m-m_c$, not the order-parameter jump; at a critical point where $\xmat$ itself
diverges as the point is approached---the quadruple point of \cref{sec:results_qp},
$\xmat\sim1/\lambda_*$---$K_2$ and $B$ scale together and the window stays $O(1)$ rather than
shrinking. The verdict is
sharper than the classical compressible-magnet lore, where the logarithmic ($\alpha=0$) case is
precisely the \emph{marginal} one and the order is decided by subleading
couplings~\cite{Sak1974,BergmanHalperin1976}. There is no contradiction: there the elastic field
is a fluctuating, spatially varying degree of freedom that renormalises the transition, while
here the ``strain'' is a single global mode whose fluctuations are $1/N$-suppressed---the
Larkin--Pikin energetics is evaluated exactly at the saddle of \cref{sec:decoupling}, and the
bare logarithm already decides.
\item \emph{Case C ($\alpha<0$, or the bounded mean-field $\alpha=0$ jump above the upper critical
dimension; finite $\chi$).} For $\alpha<0$, $|\delta|^{2-\alpha}/\delta^2\to0$ and the regular
term dominates (the bounded $\alpha=0$ jump is itself $\propto\delta^2$ and merely renormalises $K_2$,
with the same conclusion); $m_c$ is then a minimum iff $\lambda_*\chi_{\mathrm{reg}}<1$, i.e.
\begin{equation}
\xmat(h_c)<\frac{m_c}{h_c}=\frac1{\lambda_*}\qquad\text{(quantitative Larkin--Pikin condition).}
\end{equation}
\end{itemize}
The sharp statement is therefore that a \emph{divergent} $\xmat$ (power \emph{or} log; Cases A,
B) forces first order, whereas finite $\xmat$ (Case C) permits second order. Geometrically, in
the $(h,\langle\sigma^z\rangle)$ plane the condition is that the response tangent at $h_c$ be
flatter than the secant from the origin, which a divergent tangent trivially violates. For the
field-driven antiferromagnetic QCP the universality is $(d{+}1)$-Ising: $d=1$ ($2$D-Ising, log; Case B),
$d=2$ ($3$D-Ising, $\alpha\approx0.11$; Case A), $d=3$ ($4$D-Ising, log; Case
B)\footnote{At $d=3$ the matter sits \emph{exactly} at the Ising upper critical dimension
($d{+}1=4$): the exponents are mean-field ($\alpha=0$) but carry multiplicative logarithmic
corrections~\cite{LarkinKhmelnitskii1969,WegnerRiedel1973}. Because the cavity-conjugate (uniform)
susceptibility is energy-like---the field $h$ is the thermal direction of the $\mathbb{Z}_2$
transition---it inherits the specific-heat correction,
$\xmat\sim|\ln|h-h_c||^{(4-n)/(n+8)}=|\ln|h-h_c||^{1/3}$ for $n=1$, i.e.\
$e_{\mathrm{sing}}\sim-u^2|\ln|u||^{1/3}$. This is still a genuine divergence---Case~B, first
order---but a \emph{weaker}, fractional-power-of-log one than the full Onsager logarithm
$\xmat\sim\ln|t|$ at $d=1$; the first-order window is correspondingly thinner,
$|\delta^*|\sim\exp\{-[K_2/(2B\lambda_*^2)]^3\}$. The Larkin--Pikin verdict is unchanged.}, all first
order, while above the upper critical dimension the mean-field $\xmat$ is finite (Case C). The
analysis is local; its global completion is the completeness argument of \cref{sec:completeness}.

\section{Closed forms and series for the non-Ising magnets}
\label{app:noning}

The three magnets of \cref{sec:results_beyond} rest on three different matter inputs---a Bethe
ansatz, a free-fermion band structure, and a high-field series. This appendix records the working
equations behind each, at the same level of detail as the Dicke--Ising appendices above; all three
are implemented in the reproducibility package.

\subsection{Heisenberg chain in a field: the Bethe working equations}
\label{app:bethe}

The curves $\mu_{\mathrm{B}}(h)$ and $e_{\mathrm{B}}(h)$ of \cref{sec:results_iso} come from the
Bethe solution of the chain~\cite{Bethe1931} in its zero-temperature dressed-energy form, the
route by which Griffiths first computed the magnetisation curve~\cite{Griffiths1964}. One
convention flag: our field couples to $\sigma^z=2S^z$, so the field of the spin-$\tfrac12$
literature is $2h$ and our $m=\langle\sigma^z\rangle$ is twice theirs ($J=1$ throughout). The
ground state fills real rapidities $\theta\in[-B,B]$; with
$a_n(\theta)=\frac{1}{2\pi}\frac{n}{\theta^2+n^2/4}$, the root density $\rho$ and the dressed
energy $\epsilon_{\mathrm{dr}}$ obey the Fredholm equations
\begin{align}
\epsilon_{\mathrm{dr}}(\theta)+\int_{-B}^{B}\!\mathrm{d}\theta'\,a_2(\theta-\theta')\,\epsilon_{\mathrm{dr}}(\theta')
&=\epsilon_{\mathrm{dr},0}(\theta)\equiv 2h-\frac{1}{2(\theta^2+\tfrac14)},\nonumber\\
\rho(\theta)+\int_{-B}^{B}\!\mathrm{d}\theta'\,a_2(\theta-\theta')\,\rho(\theta')&=a_1(\theta),
\label{eq:bethe_fredholm}
\end{align}
with the Fermi point $B(h)$ fixed by $\epsilon_{\mathrm{dr}}(\pm B)=0$, $\epsilon_{\mathrm{dr}}<0$ inside; then
\begin{equation}
\mu_{\mathrm{B}}(h)=1-2\int_{-B}^{B}\!\rho\,\mathrm{d}\theta,\qquad
e_{\mathrm{B}}(h)=\tfrac14-h+\int_{-B}^{B}\!\epsilon_{\mathrm{dr},0}\,\rho\,\mathrm{d}\theta
=\tfrac14-h+\int_{-B}^{B}\!a_1\,\epsilon_{\mathrm{dr}}\,\mathrm{d}\theta,
\label{eq:bethe_obs}
\end{equation}
the two energy forms equal by symmetry of the kernel. The anchors of
\cref{sec:results_iso} sit at the two ends of the Fermi interval. At $h=0$ it is the whole line
and \cref{eq:bethe_fredholm} closes by Fourier transform, $\rho(\theta)=1/(2\cosh\pi\theta)$ and
$\epsilon_{\mathrm{dr}}(\theta)=h-\pi/(2\cosh\pi\theta)$, giving $e_{\mathrm{B}}(0)=\tfrac14-\ln2$ and, by
Wiener--Hopf analysis of the $B\to\infty$ boundary, $\xmat(0)=4/\pi^2$---a limit approached with
the marginal $1/\ln(1/h)$ corrections discussed in \cref{fig:isocurves}. At $h\geq1$ the interval
has closed ($B=0$): the chain saturates, $\mu_{\mathrm{B}}=1$ and $e_{\mathrm{B}}(h)=\tfrac14-h$,
so \cref{eq:dheis_tr} returns
$\lambda_M=2\,[\,1+e_{\mathrm{B}}(1)-e_{\mathrm{B}}(0)\,]=2\ln2$. Numerically we
discretise \cref{eq:bethe_fredholm} by a Gauss--Legendre Nystr\"om scheme ($\leq900$ nodes) and
locate $B(h)$ by bracketed root finding on $\epsilon_{\mathrm{dr}}(B)=0$; below $h\sim10^{-3}$, where
$B\simeq\ln(\pi/h)/\pi$ grows and every Fermi-interval quantity is $O(h)$, we solve instead for
the exponentially small differences from the closed-form $B=\infty$ solution above, preserving
relative precision down to $h\sim10^{-40}$. The solver reproduces $e_{\mathrm{B}}(0)$ and
$\lambda_M$ to $10^{-11}$, $\xmat(0)$ to $\sim10^{-7}$ (extrapolating the log-corrected
$\chi(h)$ in powers of $1/\ln(1/h)$), and the tabulated $\mu_{\mathrm{B}}(h)$ behind
\cref{fig:isocurves,fig:dheis} to $|\Delta m|<2\times10^{-8}$.

\subsection{Compass chain: elliptic closed forms}
\label{app:compass}

The Jordan--Wigner bands of the compass chain in the field $h$ are
$\omega_\pm(k)=\sqrt{|A_k|^2+4h^2}\pm|A_k|$ with $A_k=J_1+J_2e^{ik}$, and the band sum
integrates to complete elliptic integrals ($K$, $E$; parameter convention):
\begin{equation}
\emat(h) = -\frac{2}{\pi}\,\sqrt{1+h^2}\;E\!\left(\frac{1-\Delta^2}{1+h^2}\right),
\qquad
\mumat(h) = \frac{2}{\pi}\,\frac{h}{\sqrt{1+h^2}}\;K\!\left(\frac{1-\Delta^2}{1+h^2}\right).
\label{eq:compass_closed}
\end{equation}
Expanding in $h^2$,
\begin{equation}
\xmat(0)=\frac{2}{\pi}\,K(1-\Delta^2)
\;\xrightarrow[\Delta\to0]{}\;\frac{2}{\pi}\Bigl[\ln\frac{1}{\Delta}+\ln 4\Bigr],
\qquad
c_4\equiv\mumat'''(0)=-\frac{6}{\pi}\,\frac{E(1-\Delta^2)}{\Delta^2}<0
\quad\text{for all }\Delta>0,
\label{eq:compass_chi_c4}
\end{equation}
so the additive constant of the logarithm in \cref{fig:compass}b is exactly $\ln4$, and the
quartic Landau coefficient
\begin{equation}
a_4=-\frac{c_4}{24}\,\lambda^4=\frac{\lambda^4}{4\pi}\,\frac{E(1-\Delta^2)}{\Delta^2}>0,
\qquad
a_4\big|_{\lambda=\lambda_c=\pi/2K}=\frac{\pi^3\,E(1-\Delta^2)}{64\,K(1-\Delta^2)^4\,\Delta^2},
\label{eq:compass_a4}
\end{equation}
is strictly positive: the onset is continuous at every bond asymmetry, with
$c_4\simeq-6/\pi\Delta^2$ diverging at the frustrated point and $c_4\to-3$ in the dimer limit
$\Delta\to1$. Moreover
$\mumat''(h)=-\tfrac{6h}{\pi}\int_0^{\pi/2}\!a_u\,(a_u+h^2)^{-5/2}\,\mathrm{d}u<0$ with
$a_u=1-(1-\Delta^2)\sin^2u$, so $\mumat$ is strictly concave and the stationarity curve
$\lambda(m)=\mumat^{-1}(m)/m$ is strictly monotone---fold-free---for all $\Delta>0$.

\subsection{Triangular lattice: the He--Hamer--Oitmaa Pad\'e input}
\label{app:hepade}

The triangular-lattice transverse-field Ising model is not exactly solvable; its matter input for
\cref{sec:results_triangular} is the high-field series of He, Hamer, and
Oitmaa~\cite{HeHamerOitmaa1990}, resummed as a $[7/7]$ Pad\'e approximant for $\mumat(h)$ in
$x=J/h$. The antiferromagnet is obtained from the ferromagnetic series by $x\to-x$, a parameter
continuation of the series that requires no bipartiteness. The self-consistent construction and
its validation against the published $L_y=6$ iDMRG of Saadatmand \emph{et
al.}~\cite{Saadatmand2018} (agreement to better than $1\%$ across the polarised phase,
\cref{fig:tri_mu}) are given in \cref{sec:results_triangular}; the digitised
reference data and the Pad\'e scripts are part of the reproducibility package.

\section{Infinite-system DMRG: full chain and quadruple point}
\label{app:idmrg}

The order of the AS--PS transition near the quadruple point (QP), where $\lambda\to0$ disables
ordinary perturbation theory, is determined directly by infinite-system DMRG (iDMRG/VUMPS)~\cite{ZaunerStauber2018}, using the
MPSKit.jl library~\cite{MPSKit}, on the self-consistent independent-set chain \cref{eq:IS}. This appendix records the strict-blockade
encoding, the bare-transition exponent determinations, and the dressed first-order jump quoted in
\cref{sec:results_qp}; it also fixes the methods behind the full-chain runs of
\cref{sec:results_tri,sec:results_1d}. Away from the corner nothing exotic is needed: the bare
spin-$\tfrac12$ chain is solved by ordinary two-site-unit-cell VUMPS at modest bond dimension
($\chi=8$, validated against $\chi=32$), scanning the longitudinal field $h$; the stationarity
curves of \cref{fig:b5ferro,fig:extremum_af,fig:c2preempt} are these bare data re-plotted as
$\lambda(m)=h/m$, their folds and the $\varepsilon=1.5,1.8$ jumps read off by the equal-energy
rule, and the near-corner jumps ($\varepsilon=1.85$--$1.99$) come from the same chain with the
self-consistent branch following of the final subsection of this appendix. At the corner itself the strict blockade
takes over. The result has two layers: the \emph{bare} matter
undergoes a single second-order transition, while the \emph{dressed} (self-consistent) one is first
order by the Larkin--Pikin mechanism of \cref{app:lp}.

\subsection{The strict-blockade encoding}

Reaching the QP requires solving the matter directly in its strict independent-set
(Rydberg-blockade) manifold, where no two neighbouring spins are up---and that manifold is
genuinely awkward for matrix-product methods. It is not a tensor product of on-site spaces, so no
local basis spans exactly the allowed states; a pure energetic penalty on the full spin chain
enforces the constraint only approximately while injecting a large scale that degrades
convergence; and the kinetic term must not leave the manifold, so every flip carries projectors
onto empty neighbours---an operator that straddles any cut of the chain. The encoding splits the
constraint in two. Two physical sites form one cell, and only the three states
$\{\ket{00},\ket{01},\ket{10}\}$ are kept: the \emph{intra}-cell violation $\ket{11}$ is removed
from the Hilbert space exactly, with no penalty at all. What remains is the \emph{inter}-cell
violation---the right site of one cell and the left site of the next both occupied---suppressed
by the single boundary penalty $V\,n_2^{(k)}n_1^{(k+1)}$ with $V=10$, two orders of magnitude
above the ray scales $|\delta|,h_z\lesssim0.1$;
the measured leakage $\langle n_2^{(k)}n_1^{(k+1)}\rangle$ falls from $\sim10^{-13}$ at $\chi=16$
to $\sim10^{-22}$ at $\chi=128$, so the constraint holds to far below any scale of the problem.
The blockaded flip assembles from the same pieces: within a cell, the flip of one site carries
the projector onto the other site being empty ($\ket{00}\leftrightarrow\ket{10}$; the would-be
$\ket{11}$ matrix element simply does not exist in the three-state basis), and the projector onto
the neighbouring cell's boundary site makes the full PXP term a two-cell operator. The staggered
($\mathbb{Z}_2$) order lives \emph{inside} the cell---$\ket{10}$ versus $\ket{01}$---so the
matrix-product state keeps a one-cell unit cell, translation invariant from cell to cell, with
$m_{\mathrm{stag}}=|\langle n_1\rangle-\langle n_2\rangle|/2$; the cavity order parameter is the
expectation of the operator the field couples to, $m=\langle\tilde X\rangle$ per physical site
(the blockaded flip above). Ground states are found with VUMPS; the bare-exponent determinations
of the next subsection run up a warm-started bond-dimension ladder $\chi=16\ldots128$---the
finite-entanglement scaling that controls the thermodynamic limit, with the leakage quoted above
falling along it. Parametrising the ray near the QP by an angle $\phi$ measured from the
polarised $\delta>0$ axis---for the bare scans of the next subsection $\phi$ lives in the
$(\delta,h_z)$ plane, while the dressed rays of \cref{sec:results_qp} are quoted as
$r=\lambda/(2|\delta|)$ in the $(\delta,\lambda)$ plane; the table at the end of this appendix
collects the conventions---the staggered magnetisation vanishes identically on the PS side and
rises on the AS side.

\subsection{The bare transition is second order}

\begin{figure}[!htbp]
\centering
\includegraphics{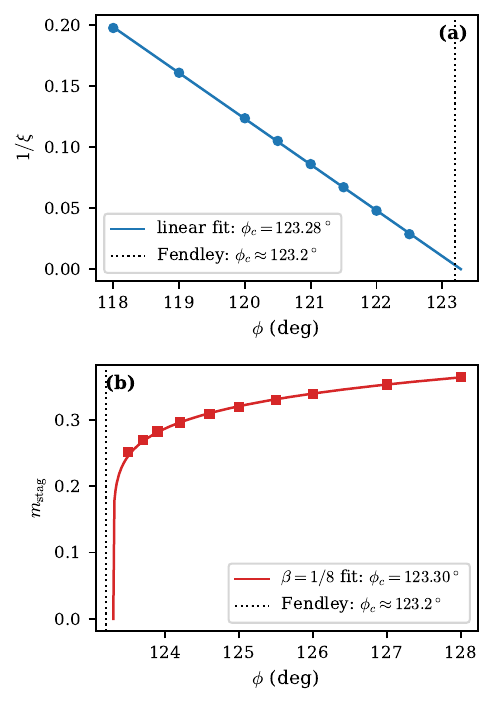}
\caption{The bare independent-set transition at the quadruple point (strict-blockade iDMRG,
$\rho=0.1$, $\chi=128$). \textbf{(a)} PS side: the inverse correlation length vanishes linearly
($\nu=1$); the fit gives $\phi_c=123.28^\circ$ ($R^2>0.9999$), with Fendley's
$(U/w)_c=-1.308$, i.e.\ $\phi_c\approx123.2^\circ$, dotted in both panels (the small offset of
the fitted intercept is finite-$\chi$ drift: the fit gives $123.35/123.30/123.28^\circ$ at
$\chi=32/64/128$, drifting toward the Fendley value). \textbf{(b)} AS side: the staggered
magnetisation follows the $2$D-Ising form $B\,(\phi-\phi_c)^{1/8}$ ($\beta=1/8$), the
independent fit giving $\phi_c=123.30^\circ$---the two sides agree on the critical angle to
$0.02^\circ$. Finite-$\chi$ check: between $\chi=64$ and $128$, $1/\xi$ moves by $0.1\%$ at the
far end and $2.5\%$ at the point nearest $\phi_c$; deep in the ordered phase $m_{\mathrm{stag}}$
is $\chi$-independent to $10^{-6}$ ($\chi=16$--$128$).}
\label{fig:bareqp}
\end{figure}

The bare independent-set chain \cref{eq:IS} undergoes a
single second-order transition in the $2$D-Ising universality class along the whole line,
and three independent determinations agree on it (\cref{fig:bareqp}). Our iDMRG/VUMPS gives $\phi_c\approx123.3^\circ$
with $\nu=1$---from the correlation length $\xi\sim|\phi-\phi_c|^{-1}$ diverging out of the PS
side (a linear $1/\xi$ fit, $R^2>0.9999$)---and $\beta=1/8$, from the staggered magnetisation
$\sim(\phi-\phi_c)^{1/8}$ on the AS side, the two fits agreeing on $\phi_c$ to $0.02^\circ$;
the hard-boson analysis of Fendley \textit{et al.}~\cite{fendley2004competing} places the
critical point at $(U/w)_c=-1.308$ (a numerical determination they quote---the $V=0$ line is not
integrable---i.e.\ $\phi_c\approx123.2^\circ$); and Chepiga and Mila~\cite{chepiga2019dmrg} give $\beta/\nu\approx0.128$,
$c\approx0.513$ on the hard-boson Ising line. The bare exponents are therefore $\beta=1/8$, $c=1/2$ ($2$D-Ising)---an unambiguous second-order \emph{bare} transition. Along these rays the uniform susceptibility of the bare chain is cleanly
unimodal---rising to the single critical peak and falling beyond it---the numerical check, at the
quadruple point, of the single-hump assumption of \cref{sec:completeness}.

\subsection{The dressed jump and its convergence}

\begin{figure}[!htbp]
\centering
\includegraphics{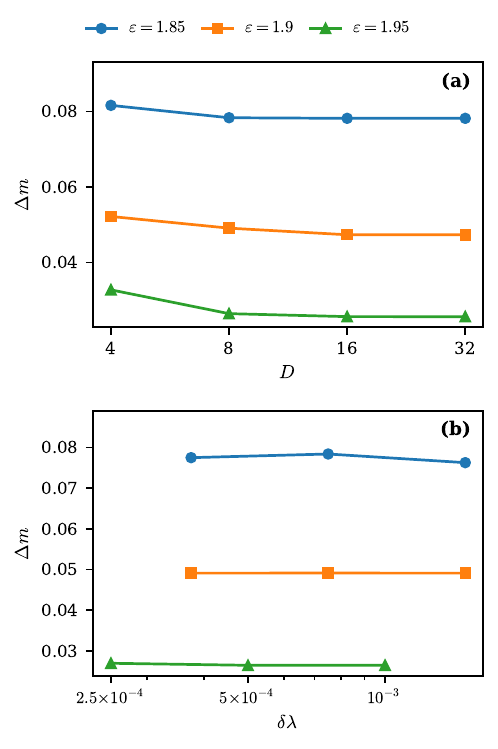}
\caption{Convergence of the near-corner AS--PS jumps ($J=1$; $\varepsilon=1.85,\,1.90,\,1.95$; $\Delta m$ in the full convention
$m=\langle\sigma^z\rangle$, $=\langle X\rangle$ in the rotated numerics frame of \cref{app:1d};
Maxwell estimator). \textbf{(a)} $\Delta m$ versus bond dimension $\chi$ at
fixed (half) coupling step: between $\chi=8$ and $\chi=32$ the jumps change by
$\lesssim2\times10^{-3}$ ($\chi=16$ and $\chi=32$ agree to
$10^{-4}$), while $\chi=4$ drifts, by up to $28\%$ toward the corner. \textbf{(b)} $\Delta m$
versus the coupling step at $\chi=8$ (full, half, quarter of the scan grid; quoted as a step in
$\lambda=g^2/2$): variations $\lesssim2\times10^{-3}$.
Both coexisting branches are gapped at the jump, which is why small $\chi$ suffices; the binding
resolution is the coupling grid.}
\label{fig:dmconv}
\end{figure}

\begin{figure}[!htbp]
\centering
\includegraphics{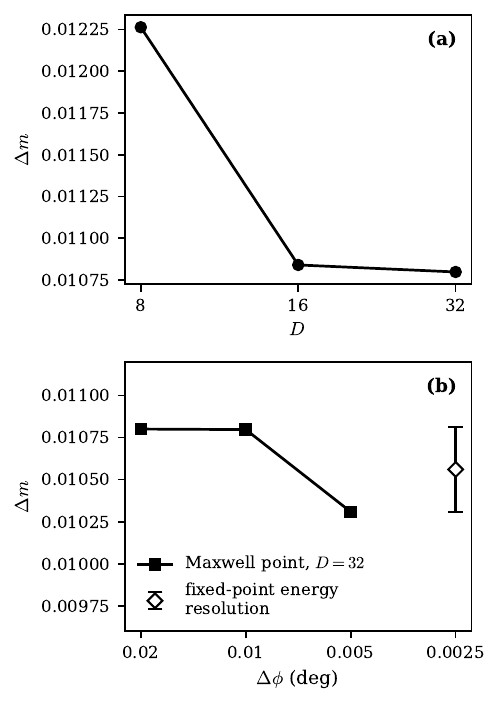}
\caption{Convergence of the scale-invariant QP-ray jump (strict blockade, $\rho=0.1$;
$\Delta m$ in the full convention $m=\langle\sigma^z\rangle$, $=\langle\tilde X\rangle$ in the
rotated numerics frame of \cref{app:1d}; Maxwell estimator).
\textbf{(a)} $\Delta m$ versus bond dimension at fixed $0.01^\circ$ angular grid:
$\chi=16$ and $32$ agree to $4\times10^{-5}$. \textbf{(b)} $\Delta m$ versus the angular step at
$\chi=32$: $0.0108,\,0.0108,\,0.0103$ for steps $0.02^\circ,\,0.01^\circ,\,0.005^\circ$, and the
finest grid ($0.0025^\circ$, six bistable points; open diamond) plateaus at
$\Delta m=0.0103$--$0.0108$, the error bar spanning the bistable points degenerate within the
fixed-point energy resolution: $\Delta m\approx0.0105(5)$, the $\approx0.011$ quoted in the
text.}
\label{fig:qpconv}
\end{figure}

The dressed (self-consistent) transition is, by
contrast, first order (\cref{app:lp}; \cref{sec:completeness}). The iDMRG resolves the first-order jump
directly---continuing both the polarised and antiferromagnetic branches metastably across the coexistence
window (the metastable-branch protocol of Ref.~\cite{leibig2025}), each seeded from the other
branch (near the corner, from the neighbouring grid point) and
iterated to its self-consistent fixed point, the iteration
accelerated by Anderson mixing~\cite{Anderson1965} and stopped once $m$ moves by less than
$10^{-4}$ ($10^{-5}$ nearest the corner)---and it shrinks toward the QP but stays finite,
growing away from it: $\Delta m\approx0.40$ at $\varepsilon=1.5$ and $0.11$ at $\varepsilon=1.8$
along the full chain, and $\approx0.011$ at the QP itself. This last value is the scale-invariant
hard-boson limit: the effective Hamiltonian \cref{eq:IS} is linear in $\delta$ and $h_z$, so the jump
depends only on the ray $r=\lambda/(2|\delta|)$, not on the distance $\rho$ from the corner---a finite,
$\rho$-independent QP jump. The correspondence with the fixed-$\varepsilon$ scans is the detuning: a
constant-$\varepsilon$ scan is the horizontal line $\delta=\varepsilon-2|J|$ in the
$(\delta,\lambda)$ plane and crosses the AS--PS line at $\lambda_M$, i.e.\ at the ray
$r=\lambda_M/(2|\delta|)$. The Maxwell points of \cref{fig:trikrit} read
$r=1.27,\,1.34,\,1.44,\,1.53$ for $\varepsilon=1.85\ldots1.99$---drifting onto the asymptotic ray
$r_{c2}\approx1.5$ as the strict manifold takes over---and their jumps approach the scale-invariant
ray value $\Delta m\approx0.011$, the floor seen in \cref{fig:qpjumps}(b). The jumps are converged in both control parameters
(\cref{fig:dmconv}): between bond dimensions $\chi=8$ and $32$ they change by
$\lesssim2\times10^{-3}$ ($\chi=16$ and $32$ agree to $10^{-4}$), and quartering the coupling step
changes them by $\lesssim2\times10^{-3}$ ($\varepsilon=1.85$--$1.95$). The binding resolution near the
corner is the coupling grid, not the bond dimension: both coexisting branches are \emph{gapped} at
the jump (correlation lengths $\xi\approx3$--$7$ on the antiferromagnetic branch, $\xi\approx10$--$21$
on the polarised branch at $\chi=8$), which is why moderate $\chi$ suffices---$\chi=4$ does
not (the jump drifts by up to $28\%$ approaching the corner and the $\varepsilon=1.99$ window is
not resolved at all)---while the coexistence window closes faster than any fixed grid. The
scale-invariant ray jump at the corner converges the same way (\cref{fig:qpconv}):
$\Delta m=0.0123,\,0.0108,\,0.0108$ at $\chi=8,16,32$ on the $0.01^\circ$ grid, and
$0.0108,\,0.0108,\,0.0103$ for angular steps $0.02^\circ,0.01^\circ,0.005^\circ$ at
$\chi=32$; a further halving of the step ($0.0025^\circ$, six bistable points) plateaus at
$\Delta m=0.0103$--$0.0108$, the spread now set by the energy resolution of the fixed-point
iteration itself. The corner jump is thus bounded away from zero: refining both the bond dimension and the grid
leaves it at $\Delta m\approx0.0105(5)$ with no downward trend, so what the data establish is a
nonzero, saturating floor; the precise third digit is not needed for the first-order verdict. Only
the window nearest the corner ($\varepsilon=1.99$) requires a finer grid, where a step of
$2.5\times10^{-4}$ resolves the coexistence window directly (five bistable grid points), the
hysteresis-loop split of \cref{fig:qpjumps}a. Rays are quoted throughout in the standard
$(\delta,\lambda)$ plane; the same physical
AS--PS ray appears at different angles in other planes, all linked by $h_z=\lambda m$:
\begin{center}
\begin{tabular}{lll}
\hline
plane & angle & ray \\
\hline
$(\delta,\lambda)$ & $\approx108^\circ$ & \emph{dressed}, $r=\lambda/(2|\delta|)\approx1.5$ (standard)\\
$(\delta,h_z)$ & $\phi_c\approx123^\circ$ & \emph{bare} critical ($\delta_c\approx-0.055$, $h_z^c\approx0.083$)\\
$(\delta,g^2)$ & $\phi\approx99^\circ$ & \emph{dressed}, $g^2=2\lambda$ (the only ray quoted outside the standard $(\delta,\lambda)$ plane, for \cref{fig:qpjumps}a)\\
\hline
\end{tabular}
\end{center}
dressing by $h_z=\lambda m_c$ carries the bare ray onto the dressed one. The
Larkin--Pikin window of \cref{app:lp} is a guarantee, not an estimate: the local analysis is a
\emph{lower bound} on the first-order character, establishing a barrier in the order parameter
$\delta=m-m_c$ around the critical point---hence the first order---while the discontinuity $\Delta m$
is fixed by the global Maxwell construction, and neither vanishes at the corner. The window is
\emph{not} exponentially thin there: the divergence $\xmat\sim1/\lambda_*$ makes $K_2\sim\lambda_*$
and the log amplitude $B\sim1/\lambda_*$ scale together, so the exponent $K_2/2B\lambda_*^2$ stays
$O(1)$ and the exponential factor does not drive $\Delta m\to0$. The ``effectively second-order''
appearance at coarse resolution is discussed in \cref{sec:results_qp}.

\section{The Dicke--Ising model at infinite dimension as a multi-mode Dicke model}
\label{app:dinf_dicke}

The $d\to\infty$ limit is the exact anchor of this paper: it fixes the diagram, and the $1/d$ expansion of
\cref{sec:results} is built on it. It is usually read as ``mean-field Ising matter in a cavity''. Here we
make a sharper statement: at $d=\infty$ the Dicke--Ising model \emph{is}, exactly, a multi-mode Dicke model
of the photon and the sublattice magnons, and its complete phase diagram and excitation spectrum follow in
closed form. The construction recovers, on the normal vacuum, the effective Dicke mapping of
Ref.~\cite{Schellenberger2024}; our contribution is the \emph{superradiant} ($m>0$) sector and, with it, the
closed-form polariton that softens at the antiferromagnetic-superradiant--polarised-superradiant (AS--PS)
transition.

\subsection{The collective Hamiltonian}

On a bipartite lattice the rescaled antiferromagnetic Ising coupling $J\to J/d$ becomes, as $d\to\infty$,
an all-to-all coupling between the two sublattice collective spins. Writing the sublattice Pauli sums
$\hat\Sigma^{A,B}_\mu=\sum_{i\in A,B}\sigma^\mu_i$ (each of length $N/2$) and
$\hat\Sigma_\mu=\hat\Sigma^A_\mu+\hat\Sigma^B_\mu$, and keeping the paper-wide normalisation in which the
cavity couples to $\hat S_z=\tfrac12\sum_i\sigma^z_i$ (\cref{eq:H_general}) with $\lambda=g^2/2\omega_c$, the
model is a two-collective-spin Dicke--Lipkin--Meshkov--Glick model (rotated as in \cref{app:1d} so the cavity
axis is $\sigma^x$ and the Ising/field axis is $\sigma^z$),
\begin{equation}
\hat H_{\infty}=\omega_c\hat a^\dagger\hat a
+\frac{g}{2\sqrt N}(\hat a+\hat a^\dagger)\,\hat\Sigma_x
+\frac{4J}{N}\,\hat\Sigma^A_z\hat\Sigma^B_z
-\varepsilon\,\hat\Sigma_z .
\label{eq:Hinf}
\end{equation}
Integrating out the photon at its coherent displacement leaves a pure-matter energy density
\begin{equation}
\evar=J\,Z_AZ_B-\tfrac{\varepsilon}{2}\,(Z_A+Z_B)-\tfrac{\lambda}{2}\,m^2 ,
\qquad m=\tfrac12(X_A+X_B),
\label{eq:dinf_efunc}
\end{equation}
the $d=\infty$ specialisation of the functional $\evar=\tfrac\lambda2 m^2+\emat(\lambda m)$ of
\cref{eq:functional}. Here $X_s=\langle\sigma^x\rangle_s$, $Z_s=\langle\sigma^z\rangle_s$ are the per-spin
sublattice magnetisations and the last, superradiant, term is the collective $\hat\Sigma_x^2$ of the
Lipkin--Meshkov--Glick form (the $-\tfrac\lambda2 m^2$ penalty being $-\tfrac{g^2}{4\omega_c N}\hat\Sigma_x^2$).
Two order parameters distinguish the four phases, the superradiant $m=\langle\hat\Sigma_x\rangle/N$ and the
staggered $m_s=\langle\hat\Sigma^A_z-\hat\Sigma^B_z\rangle/N$: PN $(0,0)$, PS $(\neq0,0)$, AN $(0,\neq0)$ and
AS $(\neq0,\neq0)$. Because $\hat H_\infty$ is fully collective, its ground state is the product coherent
state minimising \cref{eq:dinf_efunc}, exact as $N\to\infty$. (The saddle of \cref{eq:dinf_efunc} reproduces
the closed-form boundaries of \cref{eq:as_ps_line} below, the quadruple point $\varepsilon=2J$ at $\lambda=0$,
and the AS energy $\evar_{\mathrm{AS}}=-J-(\varepsilon-2J\eta)^2/2\lambda$ of \cref{sec:results}.)

A feature of the limit, used throughout: \emph{every} boundary is second order. The two tricritical fields
of \cref{sec:results_tri}, $\varepsilon_{\mathrm{tri}}^{\mathrm{ferro}}=|J|/d$ and
$\varepsilon_{\mathrm{tri}}^{\mathrm{AF}}=2|J|/\sqrt{8d-3}$, both vanish as $d\to\infty$, so for $\varepsilon>0$
there is no tricritical point and no first-order superradiant line. The first-order folds, the tricritical
points and the AN$\to$PS preemption are all $1/d$ effects (\cref{sec:lp,sec:results_1d}). Since every
$d=\infty$ boundary is a soft mode, the Gaussian (magnon) sector of \cref{eq:Hinf} reproduces the whole
diagram.

\subsection{Canted mean field and the AS phase}

In the rotated frame the AS ground state is the canted product state with sublattice Bloch angles
$\beta\pm\alpha$ measured from the cavity ($\sigma^x$) axis: $\langle\sigma^x\rangle_{A,B}=\cos(\beta\pm\alpha)$,
$\langle\sigma^z\rangle_{A,B}=\sin(\beta\pm\alpha)$, so that $m=\cos\alpha\cos\beta$ and
$m_s=\cos\beta\sin\alpha$ (the uniform cavity magnetisation and the staggered Ising magnetisation
respectively). With $\eta\equiv\sqrt{1-\lambda/2J}$ the saddle of \cref{eq:dinf_efunc} gives the two AS edges:
the AN--AS line at $\varepsilon=2J\eta$ (where $m\to0$) and, where the staggered order switches off
($\alpha\to0$), the
\begin{equation}
\text{AS--PS line:}\qquad \varepsilon_c=(2J+\lambda)\,\eta ,
\qquad \cos^2\beta_c=\frac{\lambda}{2J},\quad m_c=\sqrt{\frac{\lambda}{2J}} .
\label{eq:as_ps_line}
\end{equation}
At $d=\infty$ the AS--PS transition is \emph{second order}: the order parameter $m_s$ vanishes continuously
along \cref{eq:as_ps_line}, with a soft mode (below) and no fold, in agreement with the mean-field diagram of
Zhang \emph{et al.}~\cite{Zhang2014}. Whether this
continuity survives at finite $d$, where the $(d{+}1)$-Ising susceptibility diverges, is the
Larkin--Pikin question taken up in \cref{sec:lp} (the binding antiferromagnetic-branch curvature
$1-\lambda\,\xmat^R(0)$ below).

\subsection{Polariton spectrum: the photon kept explicit}

Rather than eliminate the photon, we keep it and read the polaritons directly. The cavity couples to the
uniform ($q=0$) matter response, whose closed form on the AS branch has two collective poles,
\begin{equation}
\xmat^R(\omega)=\frac{P}{\omega_L^2-\omega^2}+\frac{A}{\Delta^2-\omega^2},
\label{eq:dinf_chi}
\end{equation}
an \emph{optical} (Larmor) magnon at a finite frequency $\omega_L$, and a \emph{staggered} magnon at $\Delta$
that softens, $\Delta\to0$, along the AS--PS line \cref{eq:as_ps_line}. Crucially the cavity (uniform $q=0$)
couples to the staggered ($q=\pi$) magnon \emph{only} through the canting, so its weight $A\propto m_s^2\to0$
as the transition is approached; the static response $\xmat^R(0)$ stays finite and the binding curvature
$1-\lambda\,\xmat^R(0)=2\lambda(2J-\lambda)/[J(2J+\lambda)]$ stays positive for $0<\lambda<2J$ (vanishing
only at the line's quadruple-point and zero-field ends), so $\lambda\,\xmat^R(0)<1$ throughout (no Larkin--Pikin divergence). The closed pole positions and residues
$(\omega_L^2,\Delta^2(\varepsilon),A,P)$ are given, and verified, in the reproducibility script.

The dressed photon propagator has poles at $\Omega^2=\omega_c^2[1-\lambda\,\xmat^R(\Omega)]$; with
\cref{eq:dinf_chi} this is a cubic in $W=\Omega^2$ whose constant term is
$\propto\Delta^2\,\omega_L^2\,[1-\lambda\,\xmat^R(0)]$, giving three polariton branches
$\Omega_1<\Omega_2<\Omega_3$: a lower branch $\Omega_1$, the optical magnon, and a photon-like upper
polariton. Because the constant term $\propto\Delta^2$, the lower branch softens with the staggered magnon,
\begin{equation}
\boxed{\;\Omega_1=\sqrt{z}\,\Delta,\qquad z\in(0,1)\ \text{a finite constant},\;}
\label{eq:dinf_softpol}
\end{equation}
so $\Omega_1\to0$ at the AS--PS line as $|\varepsilon^2-\varepsilon_c^2|^{1/2}\sim|t|^{1/2}$ (the closed
$z$ is in the script). Setting $\Omega_1=0$ requires $\Delta^2=0$, which is exactly
$\varepsilon=\varepsilon_c=(2J+\lambda)\eta$: the spectral and thermodynamic AS--PS lines coincide. The
softening is driven by the staggered pole $\Delta\to0$, \emph{not} by a uniform susceptibility divergence
($\lambda\,\xmat^R(0)<1$ throughout): a positive static curvature ($1-\lambda\,\xmat^R(0)>0$, no
Larkin--Pikin) does \emph{not} keep the cavity mode gapped---the lower polariton softens through the
vanishing photon weight, the exactly solvable instance of the mechanism discussed in \cref{sec:spectral}.

The lower polariton's photon weight vanishes with the same canting factor, $Z_{\mathrm{phot}}\sim\Delta^2\sim
|t|\to0$: as the AS--PS line is approached the lower polariton sheds its photon content and becomes the bare
staggered magnon. The cavity sees a soft mode that simultaneously softens and decouples.

\subsection{Reduction to the normal phase and attribution}

Setting the canting to zero ($\alpha\to0$, hence $m_s\to0$ and $A\to0$) decouples the staggered pole from the
cavity and the dressed-photon cubic factorises into $(\Delta^2-W)$ times a photon$+$uniform-magnon quadratic.
The quadratic is the effective Dicke model of the antiferromagnetic \emph{normal} phase: its soft point
reproduces the AN--AS onset $g_c^2=4\varepsilon_A\varepsilon_B/(\varepsilon_A+\varepsilon_B)$ with
$\varepsilon_{A,B}=2J\mp\varepsilon$ (the molecular-field flip gaps, already rescaled $J\to J/d$, so the
$A$ gap closes at the quadruple point $\varepsilon=2J$), the three-boson secular condition of the normal-phase mapping
[\cite{Schellenberger2024}; obtained for the antiferromagnet in the present author's earlier,
unpublished analysis]. That normal-phase reading already shows the cavity couples only to the
uniform ($k=0$) magnon and not to the staggered ($k=\pi$) one. What the canted ($m>0$) extension adds is
precisely how the staggered mode re-enters: through the broken-symmetry mixing, with weight $\propto m_s^2$,
producing the softening lower polariton \cref{eq:dinf_softpol} at the AS--PS line. The normal-phase argument
that the staggered mode ``cannot soften via the Dicke mechanism'' therefore concerns the direct AN$\to$PS
transition, which is first order; the AS--PS transition, inside the superradiant phase, is second order and
carries the soft polariton above.

\subsection{Scope}

This is the $d=\infty$ saddle spectrum; the $O(1/N)$ fluctuation corrections are not included. In the physical
dimensions the $(d{+}1)$-Ising susceptibility diverges and forces the AS--PS transition first order through
the cavity Larkin--Pikin mechanism (\cref{sec:lp}); the soft polariton \cref{eq:dinf_softpol} is the
$d=\infty$ object, where the transition is continuous. The fate of the corresponding mode at finite $d$,
where the cavity meets a gapless two-particle continuum rather than a sharp pole, is left open
(\cref{sec:spectral,sec:conclusions}).

\bibliographystyle{apsrev4-2}
\makeatletter
\immediate\write\@auxout{\string\citation{apsrev42Control}}
\makeatother
\bibliography{../refs,control}

\end{document}